\documentclass[preprint,10pt]{elsarticle}

\usepackage{amssymb}
\journal{Annals of Physics}

\RequirePackage[T1]{fontenc}

\usepackage{pdflscape}
\RequirePackage{graphicx}
\RequirePackage{mathptmx}      % use Times fonts if available on your TeX system
\RequirePackage{flushend}
\newcommand{\mathsym}[1]{{}}
\newcommand{\unicode}[1]{{}}
\usepackage{textcomp}
\usepackage{eurosym}
\usepackage{amsfonts}
\usepackage{array}
\usepackage{bm}

\usepackage{graphicx}
\usepackage{subfigure}
\usepackage{bm, array}
\usepackage[T1]{fontenc}
\def\be{\begin{equation}}
\def\ee{\end{equation}}
\def\beq{\begin{eqnarray}}
\def\eeq{\end{eqnarray}}

\newcommand{\ben}{\begin{eqnarray}}
\newcommand{\een}{\end{eqnarray}}

\def\R{{}^3\!R}

\usepackage{color}
\usepackage{enumerate}
\usepackage{soul}
\usepackage[normalem]{ulem}
\usepackage{color}
\usepackage{color}
\usepackage{mathtools, cuted}
\def\ex{e_1{}^1}

\def\e{\mbox{\rm e}}

\newcommand{\CC}{\Lambda}

\newcommand{\mincir}{\raise
-3.truept\hbox{\rlap{\hbox{$\sim$}}\raise4.truept\hbox{$<$}\ }}
\newcommand{\magcir}{\raise
-3.truept\hbox{\rlap{\hbox{$\sim$}}\raise4.truept\hbox{$>$}\ }}

\newcommand{\pM}{P_m}

\def\bea{\begin{eqnarray}}
\def\eea{\end{eqnarray}}

\def\be{\begin{equation}}
\def\ee{\end{equation}}
\def\e{\mathbf{e}}

\def\ex{e_1{}^1}

\def\R{{}^3\R}
\def\S{{}^3\S_+}

\def\R{{}^3\R}
\def\S{{}^3\S_+}

\PassOptionsToPackage{linktocpage}{hyperref}
\def\case#1/#2{\textstyle\frac{#1}{#2}}
\def\e{\mathbf{e}}

\def\ex{e_1{}^1}

\def\R{{}^3\R}

\def\be{\begin{equation}}
\def\ee{\end{equation}}
\def\bea{\begin{eqnarray}}
\def\eea{\end{eqnarray}}

\begin{document}

\begin{frontmatter}



\title{Static Spherically Symmetric Einstein-aether models: Integrability and the Modified Tolman-Oppenheimer-Volkoff approach}

\author[label1]{Genly Leon}
\address[label1]{Departamento  de  Matem\'aticas,  Universidad  Cat\'olica  del  Norte, Avda. Angamos  0610,  Casilla  1280  Antofagasta,  Chile.}
\author[label2]{A. Coley}
\address[label2]{Department of Mathematics and Statistics,
 Dalhousie University,  Halifax, Nova Scotia, Canada  B3H 3J5}
\author{Andronikos Paliathanasis}
\address[label3]{Institute of Systems Science, Durban University of Technology, \\
PO Box 1334, Durban 4000, Republic of South Africa.}
 
\begin{abstract}
We study the evolution of the dynamics and the existence of analytic solutions for the field equations in the Einstein-aether theory for a static spherically symmetric spacetime. In particular, we investigate if the gravitational field equations in the Einstein-aether model in the static spherically symmetric spacetime possesses the Painlev\`e property, so that an analytic explicit integration can be performed. We find that analytic solutions can be presented in terms of Laurent expansion only when the matter source consists of a perfect fluid with linear equation of state (EoS) $\mu =\mu _{0}+\left( \eta -1\right) p,~\eta >1$. In order to study the evolution of the dynamics for the field equations we apply the Tolman-Oppenheimer-Volkoff (TOV) approach. We find that the relativistic TOV equations are drastically modified in Einstein-aether theory, and we explore the physical implications of this modification. We study perfect fluid models with a scalar field with an exponential potential. We discuss all of the equilibrium points and discuss their physical properties.
\end{abstract}

%\begin{keyword}


%\end{keyword}

\end{frontmatter}

%\newpage
\tableofcontents

\section{Introduction}

Einstein-aether theory \cite{Jacobson:2000xp,Eling:2004dk,DJ,kann,Zlosnik:2006zu,CarrJ,Jacobson,Carroll:2004ai,Garfinkle:2011iw} is an effective field theory preserving locality and covariance, and consists of General Relativity (GR)  dynamically coupled to a  time-like unit  aether vector field, in which the local spacetime structure is  determined by both the  dynamical  aether vector field  and the geometrical metric tensor field \cite{Jacobson}. Since a preferred frame is specified by the aether at each spacetime point, Lorentz invariance is broken spontaneously. We note that every hypersurface-orthogonal Einstein-aether solution is a solution of the IR limit of (extended) Horava gravity \cite{Horava}; the relationship between Einstein-aether theory and  Horava-Lifschitz (HL) gravity is further clarified in \cite{TJab13}.

Exact solutions and the qualitative analysis of Einstein-aether cosmological models have been presented by a number of authors, \cite{Barrow:2012qy,Sandin:2012gq,Alhulaimi:2013sha,Coley:2015qqa,Latta:2016jix}, with an emphasis on the impact of Lorentz violation on the inflationary scenario \cite{CarrJ,kann,Zlosnik:2006zu}. More recent developments of the HL theory were reviewed in \cite{Wang:2017brl}, including  universal horizons, black holes and their thermodynamics, gauge-gravity duality, and the possible quantization of the theory. HL cosmological scenarios were tested  against new observational constraints in \cite{Nilsson:2018knn} using updated Cosmic Microwave Background, Type Ia supernovae, and Baryon Acoustic Oscillations cosmological data. It is known that cosmologically viable extended Einstein-aether theories are compatible with cosmological data \cite{Trinh:2018pcb}.

\subsection{Spherically symmetric models}

The Einstein-aether field equations (FE) are derived from  the Einstein-aether action \cite{Jacobson:2000xp,DJ}. In such a model there will be additional terms in the FE due to
the effects of the curvature of the  underlying spherically symmetric  geometry and from the stress tensor, $T_{ab}^{\text{ea}}$, for the aether field (which depends on a number of dimensionless parameters $c_i$). 
In addition, because a FLRW limit is not required by the analysis, the aether  field can be ``tilted'' with respect to the (perfect fluid) CMB rest frame, which provides additional terms in the matter stress tensor, $T_{ab}^{mat}$ (characterized by a ``hyperbolic tilt angle'', $\alpha$, measuring the relative boost). It is anticipated that this tilt will decay to the future \cite{tilt}.
For example, (and consistent with the local linear stability analysis presented in \cite{kann}), $\alpha \rightarrow 0$ at late times in a tilted (aether) Bianchi I cosmological model in which the  initial hyperbolic tilt angle $\alpha$ and its time derivative are sufficiently small  \cite{CarrJ}.

The  Einstein-aether parameters $c_i$ are dimensionless constants. In the case of spherical symmetry the aether is hypersurface orthogonal, \footnote{Hence, all spherically symmetric solutions of Einstein-aether theory are also solutions of Horava gravity. The converse is not  generally true, except in the case of solutions with a regular center \cite{Jacobson}.} so that the twist vanishes and $c_4$ can be set to zero without any loss of generality \cite{Jacobson}. The $c_i$ parameters also determine the effective Newtonian gravitational constant $G$ in the model, and consequently  $8 \pi G$ can be set to unity by an appropriate renormalization. The  spherically symmetric model can thus be characterized by two non-trivial constant parameters. In GR $c_i = 0$. In the qualitative analysis that follows we shall assume values for the non-GR parameters $c_i$ compatible with all current constraints \cite{Jacobson,Barausse:2011pu}.

There is a 3-parameter family of spherically symmetric solutions in the case of a static vacuum \cite{Eling:2006df}. Solutions involving a static metric coupled to a stationary aether  are called 
``stationary spherically symmetric'' solutions. The special case when the aether is parallel to the Killing vector is referred  as a ``static aether'', and such a static vacuum solution is explicitly  known \cite{Eling:2006df}.

Spherically symmetric static and stationary  Einstein-aether solutions in which there is an additional (to GR)  radial tilting  aether mode are of physical importance, and a number of black hole and time-independent solutions  are known \cite{Jacobson,Barausse:2011pu,Eling:2006df,Eling:2006ec}. In general, the dynamics of the cosmological scale factor and perturbations in non-rotating neutron star and
black hole solutions are very similar to those in GR. Although a fully nonlinear positive energy result has been demonstrated for spherically symmetric solutions at an instant of time symmetry \cite{Garfinkle:2011iw}, a thorough study of the fully nonlinear solutions has not yet been carried out.

\subsection{Background}

Recently spherically 
symmetric Einstein-aether cosmological models 
with a  perfect fluid matter source  have been studied \cite{paper1}, including  FRW models \cite{Campista:2018gfi}, Kantowski-Sachs models \cite{VanDenHoogen:2018anx,Latta:2016jix},  and spatially homogenous metrics \cite{Alhulaimi:2017ocb}. In all cases  the matter source was assumed to include a scalar field which is coupled  to the expansion of the aether field through a generalized exponential potential.  

In order to perform a dynamical systems analysis it is useful 
to introduce suitable normalized variables \cite{WE,Coley:2003mj,Copeland:1997et} to simplify the FEs,
which also facilities their numerical study. 
In particular, we derive the
equilibrium points of the algebraic-differential system by introducing
proper normalized variables \cite{Coley:2015qqa}. For every point the
stability and the physical variables can be given.

The static spherically symmetric model with perfect
fluid, first introduced in Section 6.1 of \cite{Coley:2015qqa}, has been investigated in \cite{Coley:2019tyx}
using appropriate dynamical variables inspired by \cite{Nilsson:2000zf}. In the reference \cite{Coley:2019tyx} two of us have presented 
asymptotic expansions for all of the equilibrium points in the finite region of the phase space. 
We have shown that the Minkowski spacetime can be given in explicit spherically symmetric form \cite{Nilsson:2000zf} irrespective of the aether parameters. We have shown that we can have non-regular self-similar perfect fluid solutions (like those in \cite{Tolman:1939jz,Oppenheimer:1939ne,Misner:1964zz})%
, self-similar plane-symmetric perfect fluid models, and Kasner's
plane-symmetric vacuum solutions \cite{kasner}. We discussed the existence of new solutions related with
naked singularities or with horizons. The line elements have been presented in explicit form. We also discussed the
dynamics at infinity and presented some numerical results supporting our
analytical results. Scalar field configurations in static spherically symmetric metrics were also included.  The physical consequences of the models were discussed in \cite{Nilsson:2000zf,Nilsson:2000zg,Carr:1999rv}, most of which are astrophysical rather than cosmological ones. Relativistic polytropic EoS were examined in \cite{Vernieri:2018sxd} in Horava gravity and in Einstein-aether theories with anisotropic fluid. 
Assuming viable analytic and exact interior solutions, there was found the equation of state (EoS) by a polynomial fit, not accurate enough to describe the relation between the density and the radial and tangential pressure, so that a correction of the relativistic polytropic EoS was needed to give a satisfactory fit. 

The present research is the follow-up of reference  \cite{Coley:2019tyx}. Here we continue the investigation of static spherically symmetric Einstein-aether models. Particularly, we investigate the integrability of the FE and we present the Modified Tolman-Oppenheimer-Volkoff approach. 

\subsection{Integrability and modified  Tolman-Oppenheimer-Volkoff formulation}

In order to prove that there  actually exist solutions, the integrability of the system should be established
 \cite{silent1,silent2,miritzis,helmi,cots,sinFR,sinFT}.
We first study the integrability 
of static spherically
symmetric spacetimes  in Einstein-aether gravity with
a matter source of the form of a perfect fluid
with EoS $\mu =\mu _{0}+\left( \eta -1\right) p,~\eta >1$. The governing
equations for such a spacetime form an algebraic-differential system of
first order which can be written as system of second-order
differential equations with fewer independent variables by utilizing
the ARS  (Ablowitz, Ramani and Segur) algorithm \cite{Abl}. We first show that the system passes
the singularity test, which implies that they are integrable.
Additionally,  we consider a polytropic EoS $p=q \mu^\Gamma, \Gamma=1+\frac{1}{n}, n>0$, and we extend the analysis by adding a scalar field with an exponential potential.

Specifically, we investigate if the gravitational FE in the Einstein-aether
model in the static spherically symmetric spacetime  posses the Painlev\`e property (so that an  analytic explicit integration can be performed). In order to perform such an analysis we apply the classical treatment for the singularity analysis which is summarized in the ARS algorithm. As far as  the dynamical system with only the perfect fluid is concerned, we show that the FE are integrable and we write the analytic solution in terms of mixed Laurent expansions. We show that if the scalar field is not present, and assuming that the perfect fluid has the EoS $\mu =\mu _{0}+\left( \eta -1\right) p,~\eta >1$, the system passes 
the singularity test, which implies that it is integrable (in a similar way to what has been done for Kantowski-Sachs Einstein-aether theories in \cite{Latta:2016jix}). That is, in the case of
the static spherically symmetric spacetime with a perfect fluid there  always exist a negative
resonance which means that the Laurent expansion is expressed in a Left and
Right Painlev\'{e} Series because we integrate over an annulus around the singularity which has two borders \cite{Andriopoulos}.
On the other hand, in the presence of a scalar field, or when the EoS of the perfect fluid is polytropic $p=q \mu^\Gamma, \Gamma=1+\frac{1}{n}, n>0$, the FE do not pass the Painlev\`e test. Therefore, they are not integrable.

It is worth mentioning a few technicalities. First, we have to choose coordinates that can make the calculations simpler. Furthermore,  since we know that the full system is, in general, non-integrable (including simpler models with realistic matter for neutron star solutions or for vacuum solutions in the context of modified gravity) we have to use techniques that do not involve actually solving the FE.  The local semi-tetrad splitting  \cite{Ganguly:2014qia} allows us to recast the FE into an autonomous system of covariantly defined quantities  \cite{Clarkson:2002jz, Clarkson:2007yp}.  This enables us to get ride of all the coordinate singularities that may appear because of badly defined coordinates. In addition, the autonomous systems can be simplified further by incorporating Killing symmetries of the spacetimes. 
For studying the integrability of the FE we combine the ARS method \cite{Latta:2016jix}, Lie/Noether symmetries \cite{Dimakis:2017kwx, Giacomini:2017yuk} and Cartan symmetries \cite{Karpathopoulos}. In this research we use the ARS method.

One special application of interest is to use dynamical system tools to  determine the conditions under which stable stars can form. By using the Tolman-Oppenheimer-Volkoff (TOV) approach \cite{Tolman:1939jz,Oppenheimer:1939ne,Misner:1964zz}, the relativistic TOV equations are drastically modified in Einstein-aether theory, and we can explore the physical implications of this modification. This approach to studying static stars  uses the mass and pressure as dependent variables of the Schwarzschild radial coordinate. The resulting equations are however not regular at the center of the star, and a regularizing method for analyzing the solution space is required. In the relativistic and Newtonian case the resulting equations of the regularizing procedure are polynomial, and therefore they are regular. However, for the non-relativistic case the main difficulty is that the resulting system is rational. However, we are able to find some analytical results that are confirmed by numerical integrations of 2D and 3D dynamical system in compact variables, to obtain a global picture of the solution space for a linear EoS, that can visualized in a geometrical way. This study can be extended to a wide class of EoS. For example, we study the case of a polytropic EoS: $p=q \mu^\Gamma, \Gamma=1+\frac{1}{n}, n>0$ in the static Einstein-aether theory, by means of the so-called $(\Sigma_1, \Sigma_2,Y)$-formulation given by the model \eqref{XXXeq:23}.  
For $\beta=1$ the model of \cite{Nilsson:2000zg} is recovered. In GR we find that the point $(\Sigma_1, \Sigma_2,Y)=\left(\frac{2}{3}, \frac{1}{9}, 1\right)$ corresponds to the so called self-similar Tolman solution discussed in \cite{Nilsson:2000zf}. In the Einstein-aether theory this solution is generalized to a 1-parameter set of solutions given by the equilibrium point $P_1: (\Sigma_1, \Sigma_2,Y)= \left(\frac{2}{3}, \frac{1}{9} (9-8\beta), 1\right)$. The solutions having $Y=1$ corresponds to the limit $\frac{p}{\mu}\rightarrow \infty$.

The paper is organized as follows. In Section \ref{einstein} we give the basic definitions of Einstein-aether gravity. In Section \ref{Section:2} we study static spherically symmetric spacetime with a perfect fluid with a linear EoS (Subsection \ref{Section:2.1}) and with a polytropic EoS  (Subsection \ref{model2}). For each choice of the EoS for the perfect fluid we provide i) a singularity analysis and integrability analysis of the FE; ii) a local stability analysis of the equilibrium points in the finite region of the phase space, using the $\Sigma_1, \Sigma_2, Y$ formulation, that takes advantage of the local semi-tetrad splitting, iii) a local stability analysis of the equilibrium points at infinity (when $\beta < 0$) using the dominant quantities to create compact variables; and finally iv) we present a dynamical systems analysis based on the Newtonian homology invariants, $(U,V,Y)$-formulation, that makes use of the TOV formulation of the FE. As we see, are obtained as a particular case the results for relativistic and Newtonian stellar models. For the TOV formulation of the Einstein-aether modification, the main difficulty is that although  the non-regular nature of the $m, p$ equations at the center is overcome using Newtonian homology-like invariants, the resulting  equations are rational (includes radicals) and are more difficult to analyze as compared with the local semi-tetrad splitting formulation. Finally, we investigate in Section \ref{sf} a stationary comoving aether with a perfect fluid with both linear and polytropic EoS and a scalar field with exponential potential in a static metric. We find that the models do not have the Painlev\`e property. Therefore, they are not integrable. We investigate the stability of the equilibrium points using the local semi-tetrad splitting formulation due to their advantages over the Newtonian homology-like invariants formulation in the Subsections \ref{Section:3.1.2} (for the linear EoS) and \ref{Section:3.2.2} (for polytropic EoS). Finally in Section \ref{conclusions} we discuss our results
and draw our conclusions.  In the Appendices \ref{App_A} and \ref{App_A2} we present a detailed stability analysis of the equilibrium  points when the matter content consists on  a perfect fluid and a scalar field with an exponential potential.

\section{Einstein-aether Gravity}

\label{einstein}
The Einstein-aether Action Integral is given by \cite{Jacobson,Carroll:2004ai}
\begin{equation}
S=S_{GR}+S_{u}+S_{m},  \label{action}
\end{equation}%
where~$S_{GR}=\int d^{4}x\sqrt{-g}\left( \frac{R}{2}\right) $ is the
Einstein-Hilbert term,~$S_{m}$ is the term which corresponds to the matter
source and

\begin{equation}
S_{u}=\int d^{4}x\sqrt{-g}\left( -K^{ab}{}_{cd}\nabla _{a}u^{c}\nabla
_{b}u^{d}+\Lambda \left( u^{c}u_{c}+1\right) \right)
\end{equation}%
corresponds to the aether field. $\Lambda $ is a Lagrange multiplier
enforcing the time-like constraint on the aether \cite{Garfinkle:2011iw},
for which we have introduced the coupling \cite{Jacobson}
\begin{equation}
K^{ab}{}_{cd}\equiv c_{1}g^{ab}g_{cd}+c_{2}\delta _{c}^{a}\delta
_{d}^{b}+c_{3}\delta _{d}^{a}\delta _{c}^{b}+c_{4}u^{a}u^{b}g_{cd},  \label{aeLagrangian}
\end{equation}%
which depends upon four dimensionless coefficients $c _{i}$. Finally, $%
u^{a}$ is the normalized time-like vector (observer) in which $u^{a}u_{a}=-1$.

Variation with respect to the metric tensor in (\ref{action}) provides
the gravitational FE%
\begin{equation}
{G_{ab}}=T_{ab}^{\text{ea} }+T_{ab}^{\text{mat}}  \label{EFE2}
\end{equation}%
in which $G_{ab}$ is the Einstein tensor, $T_{ab}^{\text{mat}}$ corresponds to $S_{m}$, %
 and $T_{ab}^{\text{ea} }$ is the aether tensor \cite{Garfinkle:2007bk}%
\begin{align}
{T_{ab}^{\text{ea} }}& =2c_{1}(\nabla _{a}u^{c}\nabla _{b}u_{c}-\nabla
^{c}u_{a}\nabla _{c}u_{b})-2[\nabla _{c}(u_{(a}J^{c}{}_{b)})+\nabla
_{c}(u^{c}J_{(ab)})-\nabla _{c}(u_{(a}J_{b)}{}^{c})]  \notag \\
& -2c_{4}\dot{u}_{a}\dot{u}_{b}+2\Lambda u_{a}u_{b}+g_{ab}\mathcal{L}%
_{u},  \label{aestress}
\end{align}%
where
\begin{equation}
{{J^{a}}_{m}}=-{{K^{ab}}_{mn}}{\nabla _{b}}{u^{n}}~~\text{and}~~{\dot{u}_{a}}%
={u^{b}}{\nabla _{b}}{u_{a}},  \label{a}
\end{equation}
and variation with respect to the vector field $u^{a}$ and the
Lagrange multiplier gives us
\begin{subequations}
\label{FE1}
\begin{eqnarray}
\Lambda {u_{b}} &=&{\nabla _{a}}{{J^{a}}_{b}}+c _{4}\dot{u}_{a}\nabla
_{b}u^{a}  \label{evolveu} \\
{u^{a}}{u_{a}} &=&-1,  \label{unit}
\end{eqnarray}%
where from (\ref{evolveu}) we derive the Lagrange multiplier to be
\end{subequations}
\begin{equation}
\Lambda =-u^{b}\nabla _{a}J_{b}^{a}-c_{4}\dot{u}_{a}\dot{u}^{a},
\label{definition:lambda}
\end{equation}
and the compatibility conditions
\begin{equation}
0=h^{bc}\nabla _{a}J_{b}^{a}+c_{4}h^{bc}\dot{u}_{a}\nabla _{b}u^{a}.
\label{restrictionaether}
\end{equation}
 We shall use the equation \eqref{definition:lambda} as a
definition for the Lagrange multiplier, whereas the second equation %
\eqref{restrictionaether} leads to a set of restrictions that the aether
vector must satisfy.

The energy momentum tensor of a matter source of the form of a
perfect fluid (which includes a scalar field) with energy density $\mu $, and pressure $p$, in the 1+3
decomposition with respect to $u^{a}$, is
\begin{equation}
{T_{ab}^{\text{mat}}}\equiv \mu u_{a}u_{b}+p(g_{ab}+u_{a}u_{b}),
\end{equation}%
where $h_{ab}=(g_{ab}+u_{a}u_{b})$ is the projective tensor where $%
h^{ab}u_{b}=0$, $\mu$ is the density and $p$ is the pressure of the perfect fluid, respectively.

For simplicity, in the following we introduce the constants redefinition $c_{\theta }=c_{2}+(c_{1}+c_{3})/3,  c_{\sigma
}=c_{1}+c_{3},\ c_{\omega }=c_{1}-c_{3}, c_{a}=c_{4}-c_{1}, \beta=c_a+1$. Concerning the model parameters in flat space, the theory has additional degrees of freedom compared with GR. That is, the theory presents two spin-2 polarizations, as in GR, but also one spin-0 and two spin-1 polarizations. The squared propagation speeds on flat space are respectively given by  \cite{Jacobson:2004ts}:
$s_2^2=-\frac{1}{c_\sigma-1}, s_1^2=-\frac{c_\sigma c_\omega-c_\sigma-c_\omega}{2 (\beta -1) (c_\sigma-1)}$, and $s_0^2=-\frac{(\beta +1)
   (3 c_\theta+2 c_\sigma)}{3 (\beta -1) (3
   c_\theta+2)}$. Stability at the classical and quantum levels requires all $s_{i}^2$ to be positive \cite{Garfinkle:2011iw,Jacobson:2004ts}. Ultra high energy cosmic ray observations requires 
$s_i^2>1-\mathcal{O}(10^{-15}),  i=0,1,2$, to prevent cosmic rays from losing energy into gravitational modes via a Cherenkov-like cascade \cite{Elliott:2005va}. 
Additionally, from the constraints on the PPN parameters, it follows that $|\alpha_1|\lesssim 10^{-4}$ and $|\alpha_2|\lesssim 10^{-5}$, where 
$\alpha_1$, and $\alpha_2$ are defined explicitly in \cite{Foster:2005dk}. Combining all of the above restrictions we find  
	$c_\omega\approx \mathcal{O}(10^{-15}), \beta \approx 1+\mathcal{O}(10^{-4}), c_\sigma \approx 3 c_\theta+\mathcal{O}(10^{-4})$. However, these bounds change if we consider a static spherically symmetric spacetime with curvature, or if we change the matter content to include a perfect fluid and a scalar field. In general, in the following we assume no bounds on the model parameters.  
  
\section{Static spherically symmetric spacetime with a perfect fluid}
\label{Section:2}

We consider a static spherically symmetric spacetime with line element 
\begin{equation}
ds^{2}=-N^{2}(r)dt^{2}+\frac{dr^{2}}{r^2}+K^{-1}(r)(d\vartheta ^{2}+\sin
^{2}\vartheta d\varphi ^{2}),  \label{met2}
\end{equation}
where we have fixed the spatial gauge to have $\ex(r)\equiv r$, and $x\equiv \frac{1}{2} r \frac{d\ln(K)}{dr},~y \equiv r  \frac{d\ln(N)}{dr}$. 

The FE are \cite{Coley:2015qqa}:
\begin{subequations}
\label{system_12}
\begin{align}
&r \frac{dx}{dr}=\frac{\mu +3p}{2\beta }+2(\beta-1)y^2+3xy+K
\label{eq.01},\\
&r \frac{dy}{dr}=\frac{\mu +3p}{2\beta }+2xy-y^{2}  \label{eq.02}
,\\
&r \frac{dp}{dr}=-y\left( \mu +p\right)   \label{eq.03}
,\\
&r \frac{dK}{dr}=2xK.  \label{eq.04}
\end{align}%
\end{subequations}
 Furthermore, there exists the constraint
equation
\begin{equation}
x^{2}=(\beta-1)y^2+2xy+p+K.  \label{eq.05}
\end{equation}

From (\ref{eq.03}) and (\ref{eq.04}) we have that $y=-\frac{r p^{\prime }}{\mu
+p}~\ ,~x=\frac{1}{2}\frac{r K^{\prime }}{K}$, where the prime ${}^{\prime }$ denotes a  derivative with respect $r$. By substituting into (\ref{eq.01}%
), and~(\ref{eq.02}) we find a system of two second-order ordinary differential
equations,%
\begin{subequations}
\begin{align}
&r^2\left[\frac{p^{\prime \prime }}{\mu +p}-\frac{p^{\prime }\mu^{\prime }}{\left( \mu
+p\right)^2}-\frac{p^{\prime }K^{\prime }}{K\left( \mu
+p\right) }-\frac{2\left( p^{\prime }\right) ^{2}}{\left( \mu +p\right) ^{2}}\right]%
+\frac{r p^{\prime }}{\mu +p}+\frac{\mu +3p}{2\beta }=0.  \label{eq.07}\\
& r^2 \left[\frac{K^{\prime \prime }}{2K}-\frac{1}{2}\left( \frac{K^{\prime }}{K}\right)
^{2}+\frac{3}{2}\frac{p^{\prime }K^{\prime }}{K\left( \mu +p\right) }-\frac{%
2\left(\beta-1\right)\left( p^{\prime }\right) ^{2}}{\left( \mu +p\right) ^{2}}\right]+\frac{r K^{\prime }}{2 K}-K-\frac{\mu
+3p}{2\beta }=0,  \label{eq.08}
\end{align}
\end{subequations}%
while (\ref{eq.05}) becomes%
\begin{equation}
r^2\left[\left(\beta-1\right)\left( \frac{p^{\prime }}{\mu +p}\right) ^{2}+\frac{p^{\prime
}K^{\prime }}{K\left( \mu +p\right) }-\frac{1}{4}\left( \frac{K^{\prime }}{K}%
\right) ^{2}\right]+p+K=0.  \label{eq.10}
\end{equation}

In order to apply Tolman-Oppenheimer-Volkoff (TOV) approach 
(recall that we use units where $8  \pi G=1$)
we use a coordinate change to write the line element as 
\begin{equation}
\label{TOV-coords}
ds^{2}=-N(\rho)^2 dt^{2}+\frac{d{\rho}^{2}}{1-\frac{2 m(\rho)}{\rho}}+\rho^2 (d\vartheta ^{2}+\sin
^{2}\vartheta d\varphi ^{2}),  
\end{equation}%
and define
\begin{eqnarray}
K=\rho^{-2}, \quad 
\frac{d r}{r}=\left(1-\frac{2m(\rho)}{\rho}\right)^{-\frac{1}{2}} {d \rho} \label{EQ_54},
\end{eqnarray}
where $m(\rho)$ denotes the mass up to the radius $\rho$, with range $0\leq \rho <\infty$, 
which in the relativistic case is defined by 
\begin{equation}
m(\rho)=\frac{1}{2} \int_{0}^{\rho} s^2 \mu(s) d s,
\end{equation}
such that (in the relativistic case)
\begin{equation}
\label{EQ59a}
m'(\rho)=\frac{\mu \rho ^2}{2}.
\end{equation}

Now we will calculate $m'(\rho)$, $p'(\rho)$ and $\Omega'(\rho)$ in the Einstein-aether theory and compare with the relativistic TOV equations. By definition 
\begin{subequations}
\begin{align}
&x=-\frac{1}{\rho}\sqrt{1-\frac{2 m(\rho)}{\rho}}, \label{EQ_57}\\
&y =- \frac{1}{\mu +p}\sqrt{1-\frac{2 m(\rho)}{\rho}}\frac{dp}{d \rho}.  \label{EQ_58}
\end{align}
\end{subequations}
\\
Equation \eqref{EQ_57} leads to the definition 
\begin{equation}
\label{EQ_59}
m(\rho )= \frac{1}{2} \rho \left(1-\rho ^2 x^2\right).
\end{equation}
Taking derivative with respect to $\rho$ in both sides of \eqref{EQ_59} and using Eqs. \eqref{eq.01} and \eqref{EQ_54} we obtain
\begin{equation}
\label{EQ59}
m'(\rho )=\frac{2 (\beta -1) \rho  (\rho -2 m) p'(\rho )^2}{(\mu +p)^2}+\frac{3 (\rho -2 m) p'(\rho )}{\mu+p}+\frac{3 m}{\rho }+\frac{\rho ^2 (\mu +3 p)}{2 \beta }.
\end{equation}
On the other hand, the identity 
 \begin{equation}
x^{2}=(\beta-1)y^2+2xy+p+K,  
\end{equation} 
can be written as
\begin{equation}
\frac{(\rho -2 m) p'(\rho ) \left((\beta -1) \rho  p'(\rho )+2 (\mu+p)\right)}{\rho ^2 (\mu +p)^2}+p+\frac{1}{\rho ^2}=\frac{\rho -2 m}{\rho ^3}.  \label{EQ61}
\end{equation}

Equations \eqref{EQ59}, \eqref{EQ61} have two solutions for $p'(\rho )$ and $m'(\rho)$. We consider the branch 
\begin{subequations}
\label{modified-TOV-eqs}
\begin{small}
\begin{align}
& p'(\rho )= \frac{(\mu +p) \left(\sqrt{(2 m-\rho ) \left(2 \beta  m+(\beta -1) p
   \rho ^3-\rho \right)}+2 m-\rho \right)}{(\beta -1) \rho  (\rho -2 m)}, \label{modified-TOV-eqs(a)}\\
&	m'(\rho )= -\frac{-2 \beta  \rho +2 \beta  \left((\beta +1) m+\sqrt{(2 m-\rho ) \left(2 \beta  m+(\beta -1) p \rho ^3-\rho \right)}\right)+(\beta
   -1) \rho ^3 ((4 \beta -3) p-\mu )}{2 (\beta -1) \beta  \rho }, \label{modified-TOV-eqs(b)}
	\end{align}\end{small}
and the evolution equation for the gravitational potential $\Omega$, related with the lapse function by $N=e^{\Omega}$, is 
	\begin{align}
	& \Omega'(\rho)= -\frac{\left(\sqrt{(2 m-\rho ) \left(2 \beta  m+(\beta -1) p
   \rho ^3-\rho \right)}+2 m-\rho \right)}{(\beta -1) \rho  (\rho -2 m)}. \label{modified-TOV-eqs(c)}\end{align}
	\end{subequations}
\\
Taking the limit $\beta\rightarrow 1$ in the  Eqs. \eqref{modified-TOV-eqs} we obtain the proper general relativistic equations:
\begin{equation}
\label{eq-122a}
p'(\rho )=-\frac{\left(p \rho ^3+2 m \right) (p +\mu  )}{2 \rho  (\rho -2 m )}, \quad
 m'(\rho )=\frac{\mu  \rho ^2}{2}, \quad \Omega'(\rho )=\frac{\left(p \rho ^3+2 m \right)}{2 \rho  (\rho -2 m )}.
	\end{equation}
 We have to consider the reality conditions:
$$\rho >0,  \beta \leq 1,  p>0,  m>0,  \rho >2 m,$$ or 
$$\rho >0, 0<p<\frac{1}{(\beta -1) \rho ^2},  0<m\leq \frac{\rho ^3 (p-\beta  p)+\rho
   }{2 \beta }.$$

For the analysis we have to specify the equation of state, at least on implicit form $p=p(s), \mu=\mu(s)$.

\subsection{Model 1: Linear equation of state}
\label{Section:2.1}
For the equation of state for the perfect fluid we consider that $\mu
(r)=\mu_{0}+\left( \eta -1\right) p\left( r\right) $, with $\mu_{0}\geq 0, \eta >1$.

\subsubsection{Singularity analysis and integrability}

In this section we use the line element \eqref{met2}. Then, by introducing the radial rescaling 
$r=e^{\ell}$, such that $\ell\rightarrow -\infty$ as $r\rightarrow 0$ and $\ell\rightarrow \infty$ as $r\rightarrow \infty$, we obtain the equations  
\begin{subequations}
\begin{align}
&\frac{p^{* * }}{\mu +p}-\frac{p^{* }\mu^{*}}{\left( \mu
+p\right)^2}-\frac{p^{* }K^{*}}{K\left( \mu
+p\right) }-\frac{2\left( p^{*}\right) ^{2}}{\left( \mu +p\right) ^{2}}%
+\frac{\mu +3p}{2\beta }=0,  \label{eq.07b}\\
&\frac{K^{* * }}{2K}-\frac{1}{2}\left( \frac{K^{* }}{K}\right)
^{2}+\frac{3}{2}\frac{p^{* }K^{* }}{K\left( \mu +p\right) }-\frac{%
2\left(\beta-1\right)\left( p^{* }\right) ^{2}}{\left( \mu +p\right) ^{2}}-K-\frac{\mu
+3p}{2\beta }=0,  \label{eq.08b}
\end{align}
\end{subequations}%
while (\ref{eq.05}) becomes%
\begin{equation}
\left(\beta-1\right)\left( \frac{p^{*}}{\mu +p}\right) ^{2}+\frac{p^{*
}K^{* }}{K\left( \mu +p\right) }-\frac{1}{4}\left( \frac{K^{* }}{K}%
\right) ^{2}+p+K=0  \label{eq.10b}
\end{equation}
where ${}^{* }$ means derivative with respect to $\ell$. 

 Next, we apply the ARS algorithm and we find that the dominant terms are%
\footnote{%
We have taken the position~${\ell}_{0}$ of the singularity to be at zero.}~$p_{d}=k{\ell}^{-2}~,~K=K_{0}{\ell}^{-2},~$where%
\begin{equation}
k=\frac{4}{\eta ^{2}}\beta ~,~K_{0}=-\frac{%
4 (2 \beta -1)-4\eta -\eta ^{2}}{\eta ^{2}}.
\end{equation}

For the resonances we have that
\begin{equation}
s_{1}=-1~,~s_{2}=\frac{2}{\eta}\left( 2+\eta\right) ~,~s_{\pm}=\frac{1}{2}%
\left( -1-\frac{2}{\eta}\pm\frac{\sqrt{\Delta\left( \eta,\beta\right) }}{\eta%
}\right),
\end{equation}
where%
\begin{equation}
\Delta\left( \eta,\beta\right) =4 (16 \beta -7)-28\eta-7\eta^{2}
\end{equation}
\\
Furthermore, as for $\eta >1$ the resonance $s_{-}$ is negative, the Laurent
expansion is a Left-Right Painlev\'{e} Series. Importantly, for the
singularity analysis to succeed the resonances must be rational numbers.

Consider the special case $\eta =4,~$with $\beta=\frac{11}{2}$, which gives
the resonances~$s_{1}=-1~,~s_{2}=3~,~s_{+}=\frac{1}{2}~,~s_{-}=-2.~$ Hence
the solution is given by%
\begin{equation}
p\left( {\ell}\right) =\sum_{\alpha_{-} =-1}^{-\infty }p_{\alpha_{-} }{\ell}^{-2+\frac{\alpha_{-}}{2}%
}+k{\ell}^{-2}+\sum_{\alpha_{+} =1}^{+\infty }p_{\alpha_{+} }{\ell}^{-2+\frac{\alpha_{+} }{2}},
\label{ss.1}
\end{equation}%
\begin{equation}
K\left( {\ell}\right) =\sum_{\alpha_{-} =-1}^{-\infty }K_{\alpha_{-} }{\ell}^{-2+\frac{\alpha_{-} }{2}%
}+K_{0}{\ell}^{-2}+\sum_{\alpha_{+} =1}^{+\infty }K_{\alpha_{+} }{\ell}^{-2+\frac{\alpha_{+} }{2}}.
\label{ss.2}
\end{equation}

For $\mu _{0}=0$, after we replace the solution (\ref{ss.1}), (\ref{ss.2})
in (\ref{eq.07}) we find that the coefficient constants are%
\begin{equation}
K_{1}=\frac{17}{11}\left( p_{1}\right),~p_{2}=-\frac{135}{88}\left(
p_{1}\right) ^{2},~K_{2}=-\frac{999}{242}\left( p_{1}\right) ^{2},
\label{ss.3}
\end{equation}%
\begin{equation}
p_{3}\simeq 2.35\left( p_{1}\right) ^{3},~K_{3}\simeq 10.358\left(
p_{1}\right) ^{3},  \label{ss.4}
\end{equation}%
\begin{equation}
p_{4}\simeq -3.64\left( p_{1}\right) ^{4},~~K_{4}\simeq -26\left(
p_{1}\right)^{4}, p_{5}\simeq 5.77\left( p_{1}\right) ^{5}, \label{ss.5}
\end{equation}%
\begin{equation}
K_{5}\simeq 63.7\left( p_{1}\right) ^{5},~p_{6}\simeq 132.57\left(
p_{1}\right) ^{6}+\frac{11}{12}K_{6},~\text{etc.},
\end{equation}%
where $p_{1},K_{6}$ are the two constants of integration. The third constant
of integration is the position of the singularity while the fourth constant
of integration appears in the next coefficients. Furthermore, from (\ref{eq.10b})
we have the constraint for the constants of integration,~$z_{6}\simeq
154.84\left( p_{1}\right) ^{6}$.

For $\mu_{0}\neq0$, we find the coefficients (\ref{ss.3}), (\ref{ss.4}),
\begin{equation}
p_{4}\simeq-3.64\left( p_{1}\right) ^{4}-\frac{5}{18}\mu_{0}
,~K_{4}\simeq-26\left( p_{1}\right) ^{4}-\frac{5}{99}\mu_{0},
\end{equation}%
\begin{equation}
p_{5}\simeq5.77\left( p_{1}\right) ^{5}+\frac{5}{154}p_{1}%
\mu_{0},~K_{5}=63.7\left( p_{1}\right) ^{5}+\frac{983}{5082}p_{1}\mu_{0},
\end{equation}%
\begin{equation}
p_{6}\simeq132.57\left( p_{1}\right) ^{6}+\frac{11}{12}K_{6}+\frac {7633}{%
15246}\left( p_{1}\right) ^{2}\mu_{0},~\text{etc.},
\end{equation}
where again $p_{1},K_{6}$ are two constants of integration.

From \ this analysis we performed the consistency test and we showed that
the FE (\ref{eq.07b})-(\ref{eq.10b}) pass the singularity
analysis and form an integrable system for values of the free parameters for
which the resonances are rational numbers.

\subsubsection{Equilibrium points in the finite region of the phase space}
\label{DSSection_2.1.2}

In this section we consider the line element \eqref{met2}, with the definition ($N=e^{\Omega}$). 
For the dynamical systems formulation we introduce the phase space variables
\begin{subequations}
\begin{align}
& Y=\frac{p}{p+\mu}, \Sigma_1=\frac{y}{\theta}, \Sigma_2=\frac{K}{\theta^2}, 
 P=\frac{p}{\theta^2}, \theta=y-x,\\
& P=1-\beta\Sigma_1^2-\Sigma_2. \label{XAeqr}
\end{align}
\end{subequations}
Assuming that the energy density $\mu$ and the pressure $p$ are both non-negative we obtain $0\leq Y \leq 1$, where we have attached the boundaries $Y=0$ and $Y=1$.
On the other hand,  $\Sigma_2\geq 0$ by definition.

The relation between the gravitational potential $\Omega$, and the matter field is given by 
\begin{align}
\frac{d \Omega}{d p}=-\frac{1}{\mu + p}, \quad \mu =\frac{\mu_0 (1-Y)}{1-\eta  Y}, \quad p=\frac{\mu_{0} Y}{1-\eta  Y}. 
\end{align}
Hence,
\begin{align}
e^\Omega=e^{c_1} (\mu_{0}+\eta  p)^{-1/\eta }
= \alpha\left(1-\eta Y\right)^{\frac{1}{\eta}}, 
\end{align}
where $\alpha$ is a freely specifiable  constant corresponding to the freedom of scaling the time coordinate in the line element. 
This in turn reflects the freedom in specifying the value of the gravitational potential at some particular value of $r$. Matching an interior solution with the exterior Schwarzschild solution at the value where the pressure vanishes, however, fixes this constant.
Using the variables $\Sigma_1, \Sigma_2, Y$ we obtain the dynamical system 
\begin{subequations}
\label{XAeq:23}
\begin{align}
&\frac{d\Sigma_1}{d\lambda}= -\frac{\beta   {\Sigma_1}^2+ {\Sigma_2}+2 Y \left(\beta   {\Sigma_1} \left(-2 \beta   {\Sigma_1}^2+ {\Sigma_1}- {\Sigma_2}+2\right)+ {\Sigma_2}-1\right)-1}{2 \beta },\\
&\frac{d\Sigma_2}{d\lambda}=	2  {\Sigma_2} Y \left(2 \beta   {\Sigma_1}^2+ {\Sigma_2}-1\right),\\
&\frac{dY}{d\lambda}= {\Sigma_1} Y (\eta  Y-1),
\end{align}
\end{subequations}
where we have defined the new radial coordinate $\lambda$ through
\begin{equation}
\frac{d f}{d\lambda}=\frac{Y}{\theta} r \frac{df}{dr}.
\end{equation}

We have the additional equation 
\begin{equation}
\frac{d P}{d\lambda}=P\left[\Sigma_{1} (Y (4 \beta  \Sigma_{1}-2)-1)+2 \Sigma_{2} Y\right],
\end{equation}
which implies that $P=0$ is an invariant subset. Thus, since $\Sigma_2\geq 0$, we have from \eqref{XAeqr} that $P\geq 0$ for $0\leq Y \leq 1$.
Hence, system \eqref{XAeq:23} defines a flow on the (invariant subset) phase space 
\begin{equation}
\Big\{(\Sigma_1, \Sigma_2,Y)\in\mathbb{R}^3: 0\leq Y\leq \frac{1}{\eta}, \quad \beta\Sigma_1^2+\Sigma_2\leq 1, \quad \Sigma_2\geq 0\Big\}.
\end{equation}

Finally, from the definitions of $x, y$ and $\theta$ we obtain the auxiliary equations
\begin{align}
\label{AuX_1}
& \frac{d \theta}{d \lambda}=Y \left(\Sigma_{1}-2 \beta  \Sigma_{1}^2-\Sigma_{2}\right) \theta, \quad \frac{d \Omega}{d \lambda}=\Sigma_1 Y, \quad \frac{d r}{d \lambda}=Y(1-\Sigma_1)r.
\end{align}
The first equation implies that the sign of $\theta$ is invariant. Thus, we can choose $\theta>0$ and then, the direction of the radial variable $\lambda$ is well defined. 
The last two equations lead
\begin{equation}
\label{AuX_2}
\frac{d \Omega}{d \ln r}=\frac{\Sigma_1}{1-\Sigma_1}, 
\end{equation}
which relates the variable $\Sigma_1$ and the logarithmic derivative of the gravitational potential  \cite{Nilsson:2000zg}.
The third equation implies that the function $r$, defined by \begin{equation}
\label{eqr}
\frac{dr}{r d\lambda}=Y(1-\Sigma_1),
\end{equation} is monotone for $y\neq 0, r\neq 0$, and $\Sigma_1\neq 1$. Additionally, the function $Z=\frac{Y}{1-Y},\quad \frac{d Z}{d \lambda}=-\frac{\Sigma_1 }{n+1} Z$ is monotonic for $\Sigma_1\neq 0, Y\neq 0$. Together with other monotone functions defined on the boundary subsets, this implies that all of the attractors for $\beta\geq 0$ (when the phase space is compact) are equilibrium points on the boundary sets $Y=0, Y=1$, $\Sigma_2=0$ and $P:=1-\beta\Sigma_1^2-\Sigma_2=0$ which are invariant for the flow of \eqref{XAeq:23}. 

The equilibrium points and the corresponding eigenvalues of the dynamical system \eqref{XAeq:23} are the following: 
\begin{enumerate}
\item $L_1$: $(\Sigma_1, \Sigma_2,Y)=\left( {\Sigma_1^*},1-\beta   {\Sigma_1^*}^2,0\right)$. Exists for $\beta\leq 0, \Sigma_1^*\in\mathbb{R}$, or  $\beta> 0, -\sqrt{\frac{1}{\beta}}\leq \Sigma_1^* \leq \sqrt{\frac{1}{\beta}}$. The eigenvalues are $0,- {\Sigma_1^*},- {\Sigma_1^*}$. Nonhyperbolic with a 2D stable (resp. unstable) manifold for $\Sigma_1^*>0$ (resp. $\Sigma_1^*<0$). 

\item $L_2$: $(\Sigma_1, \Sigma_2,Y)=(0,1,Y_0)$. Exists for $-1\leq Y_0\leq 1$. The eigenvalues are $0,-Y_0,2 Y_0$. It is a saddle.

\item $P_1$: $(\Sigma_1, \Sigma_2,Y)=\left(\frac{2}{\eta +2},1-\frac{8 \beta }{(\eta +2)^2},\frac{1}{\eta }\right)$. Exists for $\eta \geq 1, 0\leq \beta \leq \frac{1}{8} (\eta +2)^2$. The eigenvalues are $\frac{2}{\eta +2},-\frac{\eta+2 +\sqrt{64 \beta -7 (\eta +2)^2}}{2 \eta  (\eta +2)},-\frac{\eta+2 -\sqrt{64 \beta -7 (\eta +2)^2}}{2 \eta  (\eta +2)}$. It is a saddle. 

\item $P_2$: $(\Sigma_1, \Sigma_2,Y)=\left(\frac{1}{\sqrt{\beta }},0,\frac{1}{\eta }\right)$. Exists for $\beta>0$. The eigenvalues are $\frac{2}{\eta },\frac{-\eta +4 \sqrt{\beta }-2}{\sqrt{\beta } \eta },\frac{1}{\sqrt{\beta }}$. It is a source for $\eta >1, \beta >\frac{1}{16} \left(\eta+2\right)^2$, saddle otherwise. 

\item $P_3$: $\left(-\frac{1}{\sqrt{\beta }},0,\frac{1}{\eta }\right)$. Exists for $\beta>0$. The eigenvalues are $\frac{2}{\eta },-\frac{1}{\sqrt{\beta }},\frac{\eta +4 \sqrt{\beta }+2}{\sqrt{\beta } \eta }$.  It is a saddle.
	
	\item $P_4$: $(\Sigma_1, \Sigma_2,Y)=\left(\frac{\eta +2}{4 \beta },0,\frac{1}{\eta }\right)$. Exists for $\beta <0,  \eta \geq 1$ or $\eta \geq 1,  \beta \geq \frac{1}{16} (\eta +2)^2$. The eigenvalues are $\frac{\eta +2}{4 \beta },\frac{(\eta +2)^2-8 \beta }{4 \beta  \eta },\frac{(\eta +2)^2-16 \beta }{8 \beta  \eta }$. It is nonhyperbolic for $\beta=\frac{1}{16} \left(\eta +2\right)^2$. Is is a sink for $\beta <0, \eta >1$. 
	
\end{enumerate}

%%%%%%%%%%%
We have the useful equations
\begin{equation}
 r^2 =\frac{\left({\Sigma_2}-({\Sigma_1}-1)^2\right)
   (1-\eta  Y) \left(1-\beta  {\Sigma_1}^2-{\Sigma_2}\right)}{\mu_{0}
   {\Sigma_2} \left(Y \left({\Sigma
   _2}-({\Sigma_1}-1)^2\right)+1\right)}, \quad \mathcal{M}(r):=\frac{m}{r}=\frac{\Sigma_2-(1-\Sigma_1)^2}{2 \Sigma_2},
\end{equation}
where $\mathcal{M}(r)$ denotes the Misner-Sharp mass \cite{Misner:1964je}. 

From  these expressions we obtain the $m-r$-relation 
\begin{equation}
\label{RM0} 
r^2=\frac{2}{\mu_0} \frac{\mathcal{M}}{D},\quad 
m^2=\frac{2}{\mu_0} \frac{\mathcal{M}^3}{D}, \quad D=\frac{\left(Y \left({\Sigma_2}-(1-{\Sigma_1})^2\right)+1\right)}{(1-\eta  Y) \left(1-\beta \Sigma_1^2-\Sigma_2\right)}.
\end{equation}
Using the equations \eqref{RM0}, we can define the compact variables 
\begin{equation}
\label{RM0compact}
R_{\text{comp}}=\frac{\sqrt{\mu_0}r}{\sqrt{1+ \mu_0 r^2\left(1+{\mathcal{M}}^2\right)}}, \quad M_{\text{comp}}=\frac{\sqrt{\mu_0}\mathcal{M}r}{\sqrt{1+ \mu_0 r^2\left(1+{\mathcal{M}}^2\right)}}, 
\end{equation}
which depends only of the phase-space variables $(\Sigma_1,\Sigma_2,Y)$. 
 Evaluating numerically the expressions $R_{\text{comp}}, M_{\text{comp}}$ at the orbits of the system \eqref{XAeq:23}, we can see whether the resulting model leads to finite radius and finite mass.

\begin{table}[!t]
\centering
\scalebox{0.87}{
\begin{tabular}{|c|c|c|c|c|}
\hline
Labels & $(\Sigma_1,\Sigma_2,Y)$ &   Eigenvalues & Stability   \\\hline
$L_1$ & $\left({\Sigma_1^*},1-\beta   {\Sigma_1^*}^2,0\right)$ & $0,- {\Sigma_1^*},- {\Sigma_1^*}$ & Normally hyperbolic.  \\
 &&& Stable for $\Sigma_1^*>0$.   \\
 &&& Unstable for  $\Sigma_1^*<0$. \\\hline
$L_2$ & $(0,1,Y_0)$& $0,-Y_0,2 Y_0$ & Saddle.   \\\hline
$P_1$ & $\left(\frac{2}{\eta +2},1-\frac{8 \beta }{(\eta +2)^2},\frac{1}{\eta }\right)$& $\frac{2}{\eta +2},-\frac{\eta+2 +\sqrt{64 \beta -7 (\eta +2)^2}}{2 \eta  (\eta +2)},-\frac{\eta+2 -\sqrt{64 \beta -7 (\eta +2)^2}}{2 \eta  (\eta +2)}$ & Saddle. \\\hline
$P_2$ & $\left(\frac{1}{\sqrt{\beta }},0,\frac{1}{\eta }\right)$  & $\frac{2}{\eta },\frac{-\eta +4 \sqrt{\beta }-2}{\sqrt{\beta } \eta },\frac{1}{\sqrt{\beta }}$ & Source for \\
 &&& $\eta >1, \beta >\frac{1}{16} \left(\eta +2\right)^2$,  \\
 &&& saddle otherwise.\\\hline 
$P_3$ & $\left(-\frac{1}{\sqrt{\beta }},0,\frac{1}{\eta }\right)$ & $\frac{2}{\eta },-\frac{1}{\sqrt{\beta }},\frac{\eta +4 \sqrt{\beta }+2}{\sqrt{\beta } \eta }$ & Saddle. \\\hline
$P_4$ & $\left(\frac{\eta +2}{4 \beta },0,\frac{1}{\eta }\right)$ & $\frac{\eta +2}{4 \beta },\frac{(\eta +2)^2-8 \beta }{4 \beta  \eta },\frac{(\eta +2)^2-16 \beta }{8 \beta  \eta }$& Sink for \\
 &&& $\beta <0, \eta >1$. 
 \\\hline
\end{tabular}}
\caption{\label{Tab1}  Equilibrium points and the corresponding eigenvalues of the dynamical system \eqref{XAeq:23}.}
\end{table}

In table \ref{Tab1} we present the equilibrium points and the corresponding eigenvalues of the dynamical system \eqref{XAeq:23}.

\begin{figure}[!t]
	\subfigure[\label{fig:Syst3102c} $\beta=1, \eta=1$.]{\includegraphics[width=0.4\textwidth]{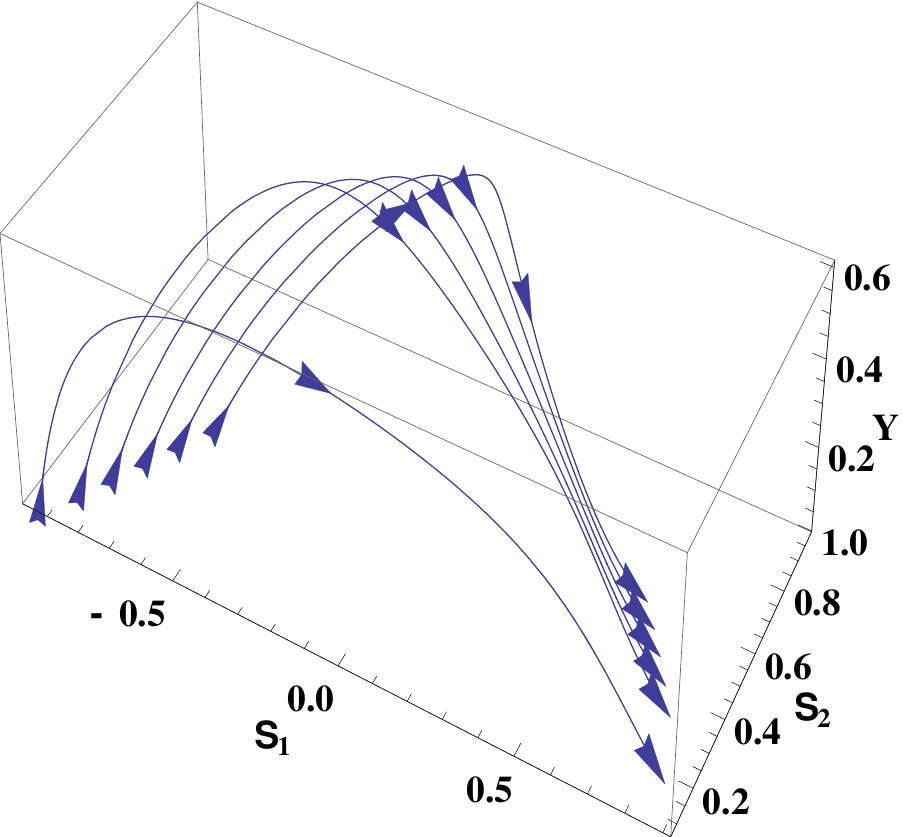}} \hspace{2cm} 	
	\subfigure[\label{fig:Syst3103c} $\beta=1.25,\eta=1$.]{\includegraphics[width=0.4\textwidth]{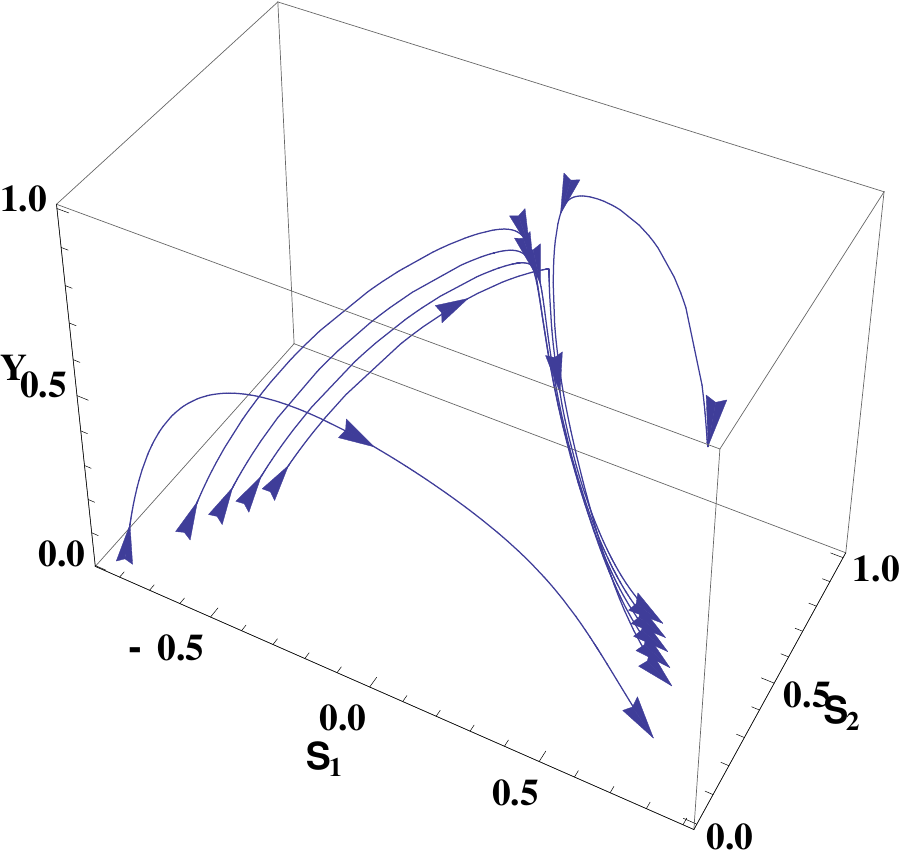}}\\
	\subfigure[\label{fig:Syst3102a} $\beta=1, \eta=1$. ]{\includegraphics[width=0.4\textwidth,height=0.35\textwidth]{FIGURE1C}} \hspace{2cm} 
	\subfigure[\label{fig:Syst3103a} $\beta=1.25, \eta=1$. ]{\includegraphics[width=0.4\textwidth, height=0.35\textwidth]{FIGURE1D}}\\
	\subfigure[\label{fig:Syst3102b}  Evolution of the compact variables 
$R_{\text{comp}},  M_{\text{comp}}$ for $\beta=1, \eta=1$.]{\includegraphics[width=0.4\textwidth,height=0.35\textwidth]{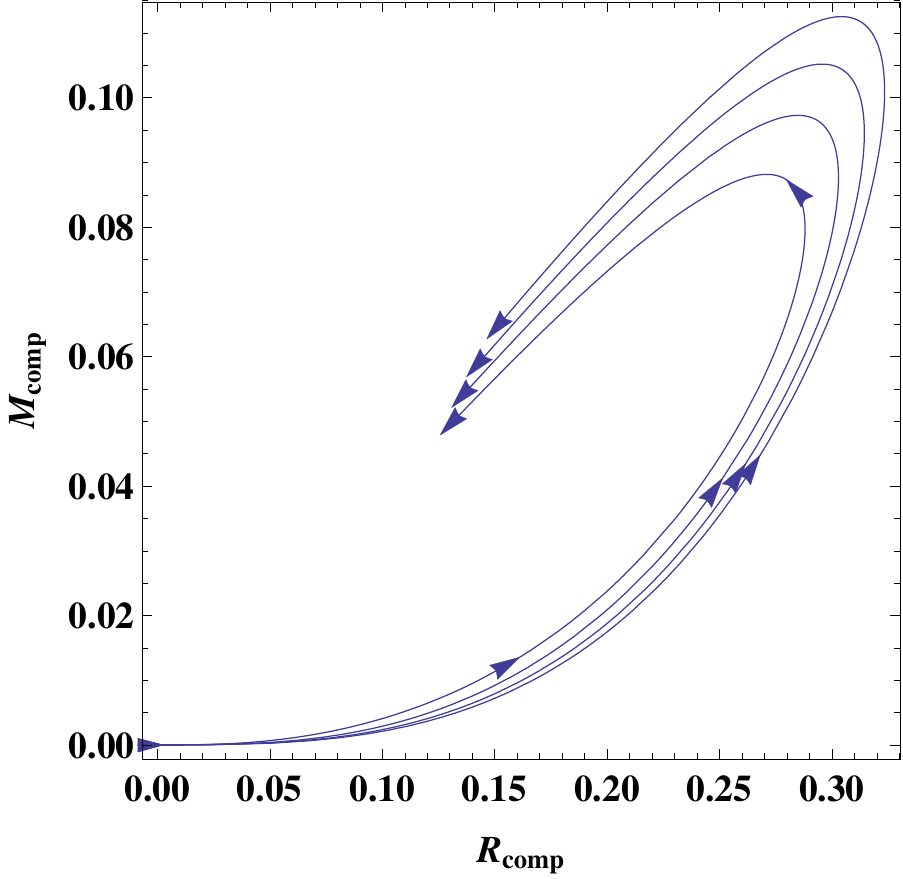}} \hspace{2cm}
	\subfigure[\label{fig:Syst3103b}  Evolution of the compact variables 
$R_{\text{comp}},  M_{\text{comp}}$ for $\beta=1.25, \eta=1$.]{\includegraphics[width=0.4\textwidth,height=0.35\textwidth]{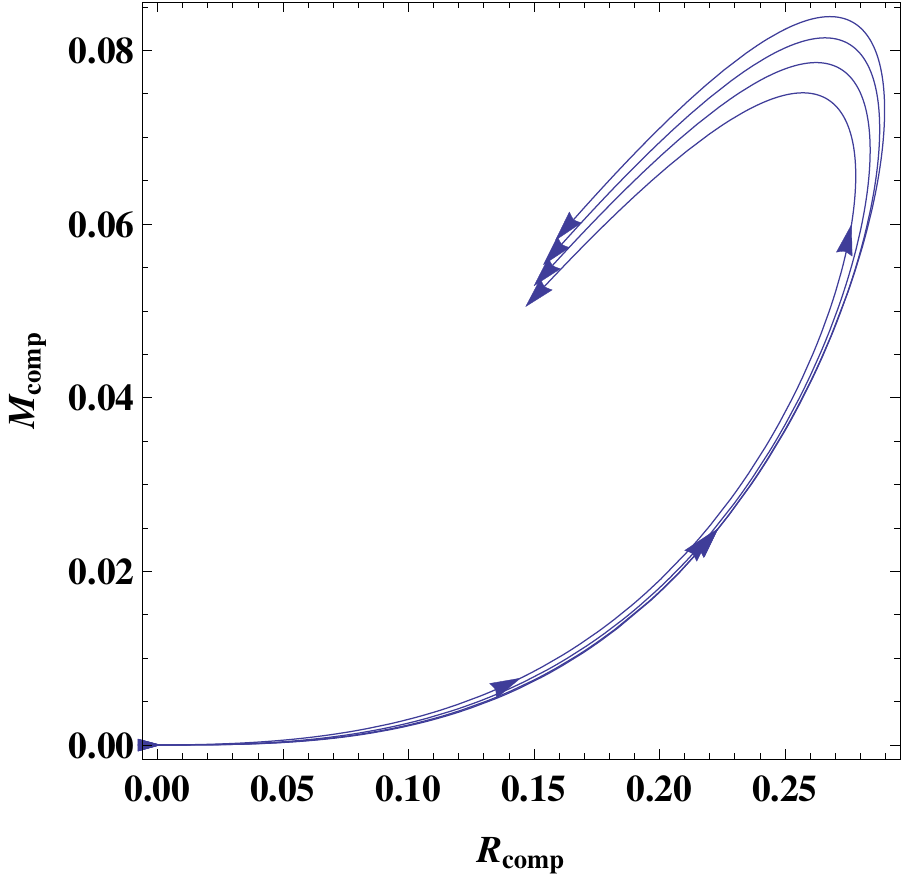}}	
	\caption{\label{fig:SystFIG1} Dynamics of the system \eqref{XAeq:23} for some choices of the parameters.}
	\end{figure}

In the Figure \eqref{fig:Syst3102c} are shown some orbits of the system \eqref{XAeq:23} on the plane $\Sigma_2=0$ for $\beta=1, \eta=1$. 
In the Figure \eqref{fig:Syst3103c} are shown some orbits of the system \eqref{XAeq:23} on the plane $\Sigma_2=0$ for $\beta=1.25, \eta=1$. In the Figure \eqref{fig:Syst3102a} are shown some projections of the orbits of the system \eqref{XAeq:23} on the plane $\Sigma_2=0$ for $\beta=1, \eta=1$. 
In the Figure \eqref{fig:Syst3103a} are shown some projections of the orbits of the system \eqref{XAeq:23} on the plane $\Sigma_2=0$ for $\beta=1.25, \eta=1$. 
In the Figure \ref{fig:Syst3102b} it is presented how the compact variables 
$R_{\text{comp}}, M_{\text{comp}}$, defined by \eqref{RM2compact} varies along the orbits of \eqref{XAeq:23}.
In the Figure \ref{fig:Syst3103b} it is presented how the compact variables 
$R_{\text{comp}}, M_{\text{comp}}$, defined by \eqref{RM2compact} varies along the orbits of \eqref{XAeq:23}.
The last two plots shows that the typical behavior for this choice of parameters is to obtain regions of finite radius and finite mass $(R_{\text{comp}}, M_{\text{comp}})$.

\subsubsection{Equilibrium points at infinity.}\label{SECT_2.1.3}
In this section we cover the case $\beta<0$, which gives an unbounded phase space in which  $\Sigma_1, \Sigma_2$ can take infinite values. Due to in $\Sigma_2\leq 1 -\beta\Sigma_1^2, \beta<0$, $1 -\beta\Sigma_1^2$ is the dominant quantity, and we define
\begin{equation}
\label{INFProy}
\Pi_1=  \frac{\sqrt{|\beta|} \Sigma_1}{\sqrt{1+|\beta| \Sigma_1^2}}, \quad
\Pi_2= \frac{\Sigma_2}{1+|\beta| \Sigma_1^2}.  
\end{equation}
and we introduce the derivative
	\begin{equation}
	\label{INFProy2}
\frac{d f}{d\bar{\lambda}}=\sqrt{1+|\beta| \Sigma_1^2}\frac{df}{d\lambda},
\end{equation}
to obtain the dynamical system
\begin{subequations}
\begin{align}
& \frac{d\Pi_1}{d\bar{\lambda}}=Y \left({\Pi_1} \left(\sqrt{1-{\Pi_1}^2} ({\Pi_2}-2) \sqrt{\left| \beta \right|
   }-{\Pi_1} {\Pi_2}+{\Pi_1}\right)+{\Pi_2}-1\right)-\frac{1}{2} \left({\Pi_1}^2-1\right) ({\Pi_2}-1),\\
& \frac{d\Pi_2}{d\bar{\lambda}}=({\Pi_2}-1) {\Pi_2} \left(2 \sqrt{1-{\Pi_1}^2} Y \sqrt{\left| \beta \right| }-{\Pi_1}-2 {\Pi_1}
   Y\right),\\
& \frac{dY}{d\bar{\lambda}}={\Pi_1} Y (\eta  Y-1),
	\end{align}
	\end{subequations}

defined in the compact phase space
\begin{equation}
\Big\{(\Pi_1, \Pi_2, Y)\in \left[-1,1\right]\times [0,1]\times [0,1]\Big\}.
\end{equation}

Now we discuss about the fixed points at infinity of the system \eqref{XAeq:23} in the above compact coordinates:
\begin{enumerate}
\item[$I_1$]: $(\Pi_1, \Pi_2 ,Y)=(-1,0,0)$, with eigenvalues $-1,-1,1$. Therefore, it is a saddle.  

\item[$I_2$]: $(\Pi_1, \Pi_2 ,Y)=(1,0,0)$, with  eigenvalues $-1, 1, 1$. Therefore, it is a saddle.  

\item[$I_3$]: $(\Pi_1, \Pi_2 ,Y)=\left(-1,0,\frac{1}{\eta }\right)$, with eigenvalues $-\frac{\eta +2}{\eta },-1,\frac{\infty  \sqrt{\left| \beta \right| }}{\eta }$. Therefore, it is a saddle.  

\item[$I_4$]: $(\Pi_1, \Pi_2 ,Y)=\left(1,0,\frac{1}{\eta }\right)$, with eigenvalues $1,\frac{\eta +2}{\eta },\frac{\infty  \sqrt{\left| \beta \right| }}{\eta }$. Therefore, it is a source.  

\item[$I_5$]: $(\Pi_1, \Pi_2 ,Y)=\left(-1,1,\frac{1}{\eta }\right)$, with eigenvalues $-1,\frac{\eta +2}{\eta },\frac{\infty  \sqrt{\left| \beta \right| }}{\eta }$. Therefore, it is a saddle.  

\item[$I_6$]: $(\Pi_1, \Pi_2 ,Y)=\left(1,1,\frac{1}{\eta }\right)$, with eigenvalues $-\frac{\eta +2}{\eta },1,\frac{\infty  \sqrt{\left| \beta \right| }}{\eta }$.  Therefore, it is a saddle.  
\end{enumerate}
There are not physical relevant equilibrium points at infinity, apart of the point $I_4$ that has $\Sigma_1\rightarrow \infty$, $\Sigma_2$ finite and $Y=\frac{1}{\eta}$, which is a source (it exists only for $\beta<0$). 
This solution corresponds to $y$ finite and $x\rightarrow y$. That is, $\frac{d p}{d\rho}=\frac{1}{\mu+p}, \frac{p}{p+\mu}=\frac{1}{\eta},$ where $\rho$ is the radial coordinate. Then  $\frac{dp}{d\rho}=\frac{1}{\eta p}\implies p(\rho )= \frac{\sqrt{\eta  p_{c}^2+2 \rho }}{\sqrt{\eta }},$ where $p_c$ is the central pressure (at $\rho=0$).  As $\rho\rightarrow \infty$ the pressure is increasing as the solutions move away this point.   

\subsubsection{Dynamical systems analysis based on the Newtonian homology invariants}
\label{SECTUVY}

In this section we use dynamical system tools to  determine conditions under which stable stars can form. We start with the set of equations \eqref{modified-TOV-eqs},  which determine the star's structure and the geometry in the static spherically symmetric Einstein-aether theory for a perfect fluid. Equation \eqref{modified-TOV-eqs(a)} is a modification of the well-known Tolman-Oppenheimer-Volkoff equation for relativistic star models. The idea is to construct a 3D dynamical system in compact variables and obtain a global picture of the solution space for a linear equation of state, that can be visualized in a geometrical way. This study can be extended to a wide class of equations of state. We therefore introduce the $(U,V,Y)$- formulation as follows: we define
\begin{align}
U=\frac{\mu  r^3}{2 m+\mu  r^3}, V=\frac{\mu  m}{\mu  m+p r}
\end{align}
that are obtained by a compactification of the so- called Newtonian homology invariants 
\begin{equation}
u=\frac{r^3 \mu}{2 m}, v=\frac{m \mu}{r p},
\end{equation}
where $\mu$ and $p$ are the density and pressure of the background fluid.

The relation between the variables $\{U, V, Y\}$ and $\{Q, S, C\}$ used in paper \cite{Coley:2019tyx} are: 
\begin{subequations}
\label{TOV-vars}
\begin{align}
& U=\frac{C (\eta -1)-\eta +Q^2+\beta  (\eta -1) S^2}{C (\eta -2)-\eta +2 Q^2-2 Q S+S^2 (\beta  (\eta -1)+1)},\\
& V=\frac{\left(C-(Q-S)^2\right) \left(C (\eta -1)-\eta
   +Q^2+\beta  (\eta -1) S^2\right)}{\left(C+(Q-S)^2\right) \left(C-Q^2+\beta  S^2\right)+\eta  \left(C-(Q-S)^2\right) \left(C+\beta  S^2-1\right)},\\
& Y=\frac{C-Q^2+\beta  S^2}{\eta  \left(C+\beta  S^2-1\right)}.
\end{align}
\end{subequations}
Furthermore, the relation between the variables ${U, V}$ and $\Sigma_1, \Sigma_2$ are 
\begin{align}
&U=\frac{(Y-1) \left(\beta  {\Sigma_1}^2+{\Sigma_2}-1\right)}{-\beta  {\Sigma_1}^2-{\Sigma_2}+Y ({\Sigma_1} ((\beta -1) {\Sigma_1}+2)+2 ({\Sigma_2}-1))+1},\\
&V=	\frac{(Y-1) \left(({\Sigma_1}-1)^2-{\Sigma_2}\right)}{({\Sigma_1}-1)^2 (Y-1)+{\Sigma_2} (Y+1)}.
\end{align}

To use the dynamical system tools, we have to define 
a new independent variable by
\begin{equation}
\label{EQ_2.38}
\frac{d r}{r d{\lambda}}=(1 - U)  (1 - Y)^2  (1 - V - Y - V Y). 
\end{equation}
Under the conditions $\mu\geq 0, p\geq 0$, it follows that the variables $U,V,Y$ are compact. The new variable ${\lambda}$ has the same range as $\Omega=\ln(r)$, i.e., ${\lambda}\rightarrow -\infty$ as $r\rightarrow  0$ and ${\lambda}\rightarrow +\infty$ as $r\rightarrow  +\infty$. 
\\Assumptions of the positive energy density, pressure and mass, with the condition $\eta\geq 1$, leads to the inequalities:
\begin{equation}
0<U<1, 0<V<1, 0<Y<\frac{1}{\eta}.
\end{equation}
\\
Since we are dealing with relativistic models the condition $2 m/\rho<1$ has to be satisfied, so that we have  
\begin{equation}
\eta \geq 1, 0<U<1, 0<Y<\frac{1}{\eta }, 0<V<\frac{1-Y}{1+Y} \implies F:=1 - V - Y - V Y>0.
\end{equation}

As a first step before to studying the full system we consider the case $\beta=1$, which corresponds to the equations of GR \eqref{eq-122a}. We compare our results with the results found in \cite{Nilsson:2000zg,Heinzle:2003ud}. 
 
\paragraph{Relativistic case ($\beta=1$).}

Using the equations \eqref{eq-122a}, and considering the time derivative  \eqref{EQ_2.38},  we obtain the dynamical system 
\begin{subequations}
\label{Relativistic_1}
\begin{align}
&\frac{dU}{d\lambda}=(1-U) U \Big\{U \left(V \left(Y \left(\eta +4 Y^2-2 (\eta +1) Y-5\right)+4\right)+4 (Y-1)^3\right)\nonumber \\
& \;\;\;\;\;\;\;\;\;\;\; +(1-Y) \left(V \left(3 Y^2-\eta  Y+Y-3\right)+3
   (Y-1)^2\right)\Big\},\\
&\frac{dV}{d\lambda}=(V-1) V \Big\{U \left(V (Y (-\eta +2 Y (\eta +Y-1)-4)+3)+2 (Y-1)^3\right) \nonumber \\
& \;\;\;\;\;\;\;\;\;\;\; +(1-Y) \left(V (Y (\eta +Y)-2)+(Y-1)^2\right)\Big\},\\
&\frac{dY}{d\lambda}=	V (1-Y) Y (U (2 Y-1)-Y+1)  (\eta  Y-1), \label{eqY}
\end{align}
defined on the phase space
\begin{equation}
\label{phase-space:68}
\Big\{(U,V,Y)\in\mathbb{R}^3: 0\leq U \leq 1, 0\leq Y\leq\frac{1}{\eta}, 0\leq V\leq \frac{1-Y}{1+Y}\Big\}.
\end{equation}
\end{subequations}
We have included the boundary subsets $U=0, U=1$, together with the ``static surface'' $(1+Y)V-(1-Y)=0$. From \eqref{eqY} follows that the variable $Y$ is monotonically decreasing and all the orbits ends at the Newtonian boundary subset $Y=0$.

The equilibrium points and curves of equilibrium points of \eqref{Relativistic_1} are summarized below. 
\begin{enumerate}
\item $L_1$, exists for $0\leq Y_0\leq\frac{1}{\eta}$ (analogous to line $L_1$ characterized in table II of \cite{Nilsson:2000zg}). It is a source. 
\item $L_2$, exists for $0\leq Y_0\leq\frac{1}{\eta}$ (analogous to line $L_2$ characterized in table II of \cite{Nilsson:2000zg}). it is a saddle.
\item $L_3$, exists for $0\leq U_0\leq 1, \eta=1$ (analogous to line $L_3$ studied in table II of \cite{Nilsson:2000zg}). It is non-hyperbolic. 
\item $L_4$, exists for $0\leq Y_0\leq\frac{1}{\eta}$ (analogous to line $L_4$ characterized in table II \cite{Nilsson:2000zg}). It is a saddle.

\item The system  \eqref{Relativistic_1} admits the line of equilibrium points $L_5$ (which is the analogue of $L_5$ characterized in table II of  \cite{Nilsson:2000zg}). It is non-hyperbolic. From the assumption $0\leq V \leq (1-Y)/(1+Y)$ the static solutions are forced to be away the line $L_5$.  

\item $L_6$, exists for $0\leq U_0\leq 1$ (analogous to line $L_6$ studied in \cite{Nilsson:2000zg}). The center manifold of this line is tangent to the eigenvector $\left(1,0,0\right)^{T}$, such that the line is normally hyperbolic. Due to the negativity of the non-zero eigenvalues it is a sink.

\item  $M_1$, always exists. It is the analogous to the point $P_1$ presented in table II of  \cite{Nilsson:2000zg} by choosing $n=0$. It is a saddle. 

\item  $M_2$, always exists.  This point is the analogous to the point $P_2$ showed in table II of  \cite{Nilsson:2000zg}. It is a special point of the line $L_6$ when $U_0=0$. It is a sink.

\item $M_3$ is the analogous to the point $P_3$ showed in table II of  \cite{Nilsson:2000zg}. It is a saddle. This point is of little interest for the static solutions.  

 \item $M_5$, always exists.  It is the analogous to the point $P_5$ showed in table II of  \cite{Nilsson:2000zg} (it is contained in the attractor line $L_6$ described below). It is nonhyperbolic.

\item  The points $M_6, M_7$, and  $M_8$  only exists for linear equations of state as we have considered in our model. They are always saddle.

\end{enumerate}

 There is no equilibrium point of  \eqref{Relativistic_1} analogous  to $P_4$, presented in table II of  \cite{Nilsson:2000zg}. It is possible to extend this analysis to polytropic EoS as we show in Section \ref{model2}. 

Finally, there are three equilibrium points with $V=1, Y=\frac{1}{\eta}$, which are of little interest for the static solutions, because they are not over  the ``static surface''. Strictly speaking since the last inequality defining the phase space \eqref{phase-space:68} is not fulfilled at these points they do not exist. Therefore, we omit any discussion of them.

\begin{table}[!t]
\centering
\scalebox{0.9}{
\begin{tabular}{|c|c|c|c|c|}
\hline
Labels & $(U,V,Y)$  &   Eigenvalues & Stability \\\hline
$L_1$ & $(1,0, Y_0)$   & $0,(1-Y_{0})^3,(1- Y_{0})^3$ & Source. \\\hline
$L_2$ & $\left(\frac{3}{4}, 0, Y_0\right)$   & $0,\frac{1}{2} (1-Y_0)^3, -\frac{3}{4} (1-Y_0)^3$ & Saddle.  \\\hline 
$L_3$ & $\left(U_0,0,1\right)$  & $0,0,0$ & Nonhyperbolic.  \\\hline
$L_4$ & $(0, 0, Y_0)$   & $0,3 (1-Y_{0})^3,-(1-Y_{0})^3$ & Saddle.   \\\hline
$L_5$ & $\left(0, V_0, 1\right)$   & $0,0,0$ & Nonhyperbolic.    \\\hline
$L_6$ & $\left(U_0, 1,0\right)$   & $0, -1+U_{0}, -1+U_{0}$ & Sink. \\\hline
 $M_1$ & $\left(0, \frac{1}{2}, 0\right)$, $(0,0)$ & $\frac{3}{2},-\frac{1}{2},\frac{1}{2}$ & Saddle.  \\\hline
 $M_2$ & $\left(0, 1, 0\right)$  & $0, -1, -1$ &  Sink.  \\\hline
 $M_3$ & $\left(1,1,1\right)$    & $1-\eta , \eta -1, \eta -1$ & Saddle. \\\hline
 $M_5$ & $\left(1,1,0\right)$  &  $0, 0, 0$ & Nonhyperbolic.  \\\hline
 $M_6$ & $\left(0, \frac{\eta -1}{\eta +1}, \frac{1}{\eta }\right)$   &  $-\frac{(\eta -1)^3}{\eta ^2 (\eta +1)},\frac{(\eta -1)^3}{\eta ^2 (\eta +1)},\frac{2 (\eta -1)^3}{\eta ^3 (\eta +1)}$ & Saddle.  \\\hline
$M_7$ & $\left(1,  \frac{\eta -1}{\eta +1}, \frac{1}{\eta } \right)$, $\left(\frac{2}{\sqrt{5}},\frac{1}{\sqrt{5}}\right)$ & $\frac{(\eta -1)^2}{\eta ^2 (\eta +1)},\frac{(\eta -1)^2}{\eta ^2 (\eta +1)},-\frac{2 (\eta -1)^3}{\eta ^3 (\eta +1)}$ & Saddle.  \\\hline
$M_8$ & $\left(\frac{1}{2}, \frac{2 (\eta -1)^2}{3 \eta ^2-2},  \frac{1}{\eta }\right)$  & $\frac{(\eta -1)^3}{\eta  \left(3 \eta ^2-2\right)},-\frac{(\eta -1)^3 \left(\eta +2-\Delta\right)}{4 \eta ^2 \left(3 \eta ^2-2\right)},-\frac{(\eta -1)^3 \left(\eta +2+\Delta\right)}{4 \eta ^2 \left(3 \eta ^2-2\right)}$ & Saddle. \\\hline
\end{tabular}}
\caption{\label{Tab2}  Equilibrium points and the corresponding eigenvalues of the dynamical system defined by equations\eqref{Relativistic_1}.  We use the notation $\Delta=\sqrt{36-7 \eta ^2-28 \eta
   }$.}
\end{table}
%%%%%%%%%Figure%%%%%%%%%%%%%%%%%%%%%%%%%%%%%%
\begin{figure*}[!t]
	\subfigure[\label{fig:Syst387a}  $\beta=1, \eta=2$.]{\includegraphics[width=0.9\textwidth]{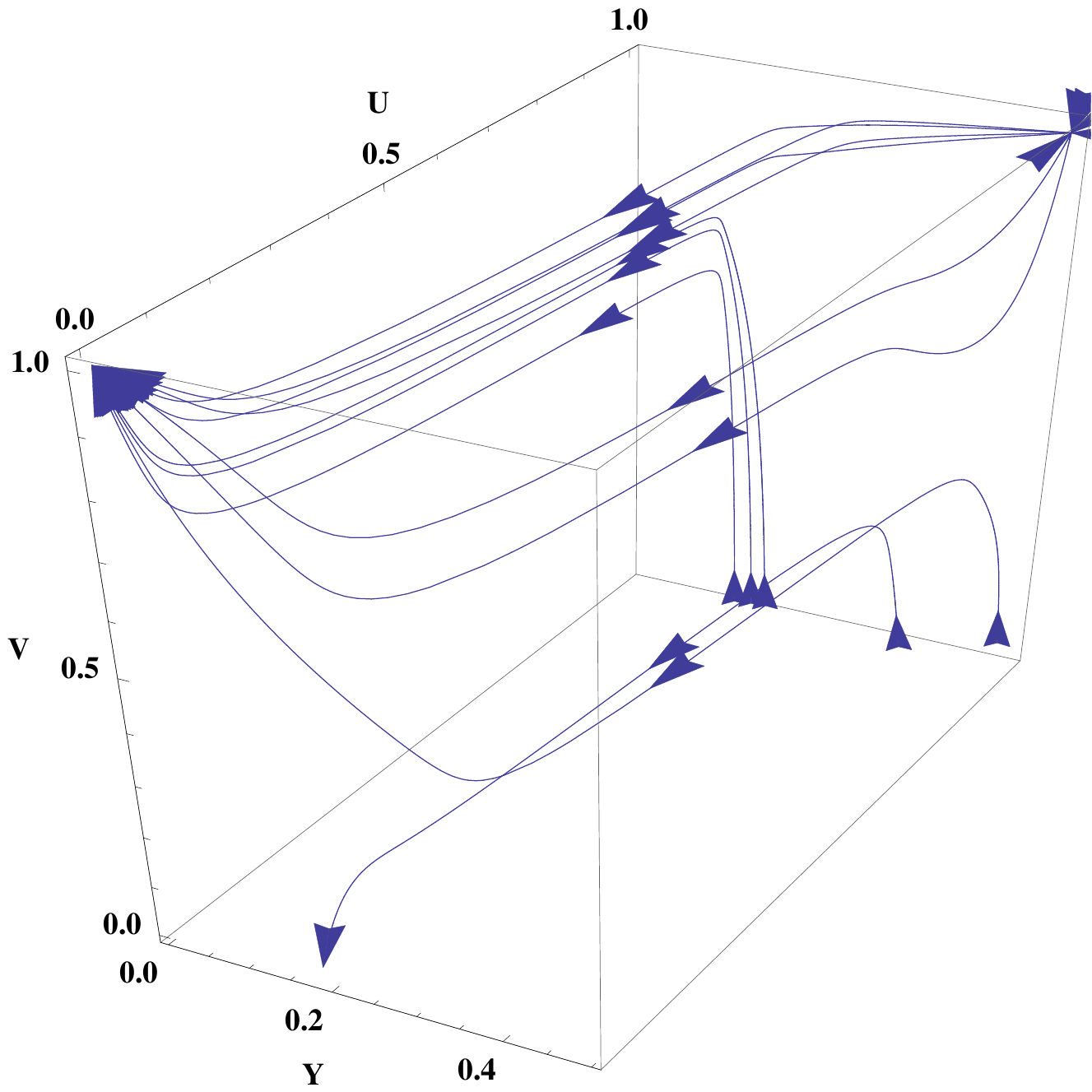}}	
	\subfigure[\label{fig:Syst387b}  $\beta=1, \eta=1$.]{\includegraphics[width=0.9\textwidth]{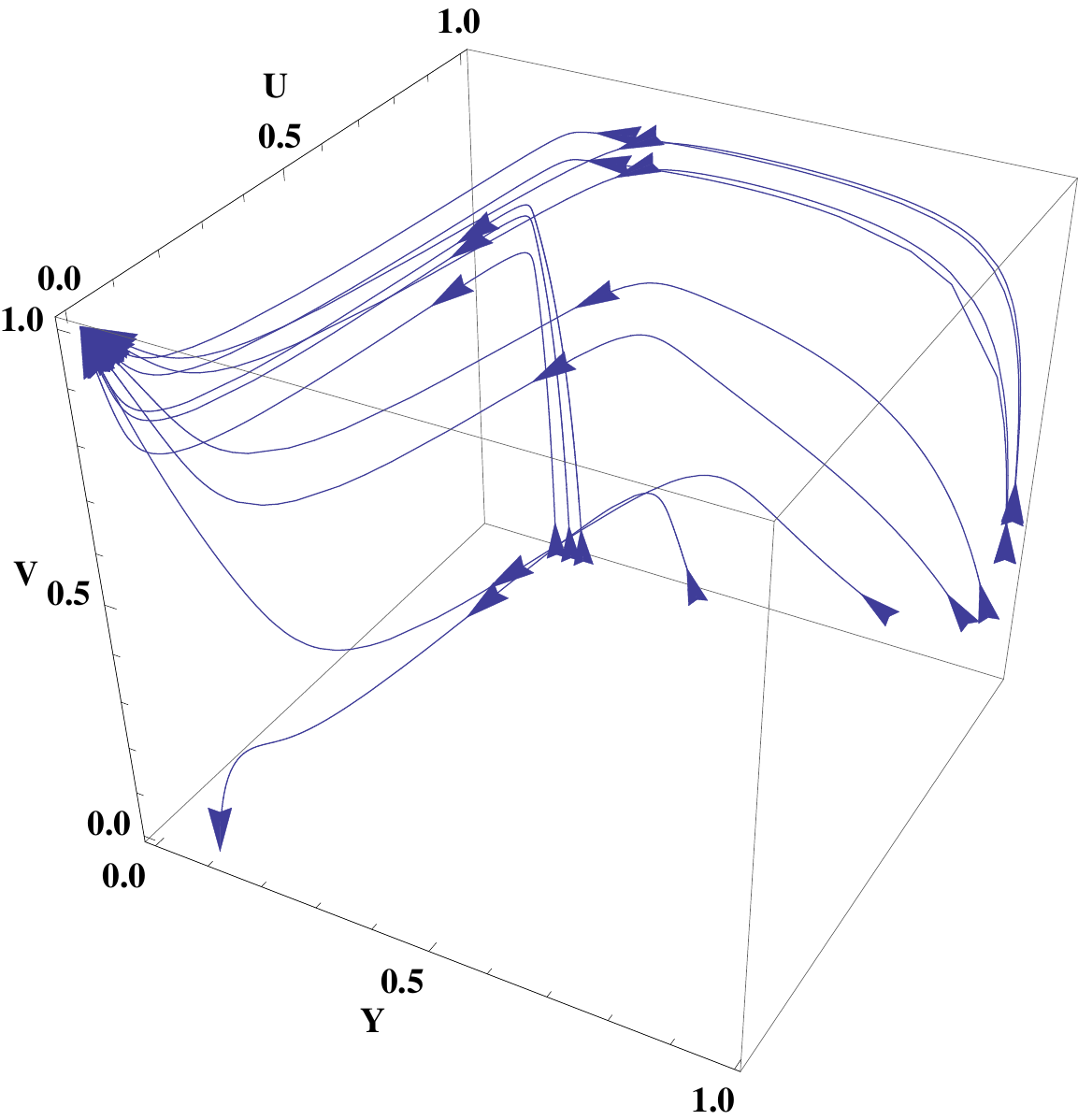}} 
								\caption{\label{fig:Syst387} Evolution of the compact variables 
$R_{\text{comp}},  M_{\text{comp}}$, defined by \eqref{RM1compact}, and some orbits of the system \eqref{Relativistic_1}.}
			\end{figure*}
%%%%%%%%%%%%%%%%%%%%%%%%%%%%%%%	

We have the useful equations
\begin{align}
& r^2 =\frac{2 U V Y (1-\eta  Y)}{\mu_0 (1-U) (1-V) (1-Y)^2}, \mu =\frac{\mu_0 (1-Y)}{1-\eta  Y}, m=\frac{r  V Y}{(1-V) (1-Y)}, p= \frac{\mu_{0} Y}{1-\eta  Y}, \\
& \mathcal{M}(r):=\frac{m(r)}{r}=\frac{V Y}{(1-V) (1-Y)}.
\end{align}
where $\mathcal{M}(r)$ defines the  Misner-Sharp mass \cite{Misner:1964je}.

From  these expressions we obtain the $m-r$-relation 
\begin{equation}
\label{RM1}
r^2=\frac{2}{\mu_0} \frac{\mathcal{M}}{D}\quad 
m^2=\frac{2}{\mu_0} \frac{\mathcal{M}^3}{D},  \quad D=\frac{(1-U)(1-Y)}{U(1-\eta Y)}.
\end{equation}
Using the equations \eqref{RM1}, we can define the compact variables 
\begin{equation}
\label{RM1compact}
R_\text{comp}=\frac{\sqrt{\mu_0}r}{\sqrt{1+ \mu_0 r^2\left(1+{\mathcal{M}}^2\right)}}, \quad M_\text{comp}=\frac{\sqrt{\mu_0}\mathcal{M}r}{\sqrt{1+ \mu_0 r^2\left(1+{\mathcal{M}}^2\right)}}, 
\end{equation}
which depends only of the phase-space variables $(U,V,Y)$. Evaluating numerically the expressions $R_\text{comp}, M_\text{comp}$ at the orbits of the system \eqref{Relativistic_1}, we can see whether the resulting model leads to finite radius and finite mass. 

In the table \ref{Tab2} is summarized the information of the equilibrium points and the corresponding eigenvalues of the dynamical system defined by equations \eqref{Relativistic_1}.
In the Figure \ref{fig:Syst387} it is presented how the compact variables 
$R_{\text{comp}}, M_{\text{comp}}$, defined by \eqref{RM1compact}, vary along the orbits of \eqref{Relativistic_1}, and the attractor solutions has finite mass and finite radius.

\paragraph{The Newtonian subset ($Y=0$).}

We now consider the ``Newtonian subset'' $Y=0$, which is of great importance for relativistic stars. The stability analysis below is restricted to the dynamics onto the invariant set $Y=0$, and the stability along the $Y$-axis is not taken into account. 
 
The evolution equations are given by
\begin{equation}
\label{system_69A}
\frac{d U}{d{\lambda}}=-(U-1) U (V-1) (4U-3),\\
\quad \frac{d V}{d{\lambda}}=(V-1) V (U (3 V-2)-2 V+1),
\end{equation}
defined on the phase plane
\begin{equation}
\Big\{(U,V,Y)\in\mathbb{R}^3: 0\leq U \leq 1, 0\leq V\leq 1\Big\}.
\end{equation}

The equilibrium (curves of equilibrium) points in this invariant set are:
\begin{enumerate}
\item $L_1(0):=(U,V)=\left(1, 0\right)$, which is an endpoint of the line $L_1$ with $Y_0=0$ described in the previous section. The eigenvalues are $1,1$, such that it is a source. 

\item $L_2(0):= (U,V)=\left(\frac{3}{4}, 0\right)$, which is an endpoint of the line $L_2$ with $Y_0=0$ described in the previous section. The eigenvalues are $-\frac{3}{4},\frac{1}{2}$, such that it is a saddle.

\item $L_4(0):=(U,V)=\left(0, 0\right)$, which is an endpoint of the line $L_4$ with $Y_0=0$ described in the previous section. The eigenvalues are $3,-1$, such that it is a saddle. 
 
\item $L_6:= (U,V)=(U_0,1), U_0\in[0,1]$. The eigenvalues are $U_0-1,0$. It it a sink.   

\item $M_1:=(U,V)=\left(0, \frac{1}{2}\right)$. The eigenvalues are $\frac{3}{2},\frac{1}{2}$, such that it is a source. 

\item $M_2:=(U,V)=\left(0, 1\right)$. The eigenvalues are $-1,0$, such that it is a non-hyperbolic equilibrium point (it is contained in the attractor line $L_6$ previously described). 

\item $M_5:= (U,V)=\left(1,1\right)$. 
 The eigenvalues are $0,0$. Hence,  it is a non-hyperbolic equilibrium point (it is contained in the attractor line $L_6$ previously described). 
\end{enumerate}

%%%%%%%%%Figure5%%%%%%%%%%%%%%%%%%%%%%%%%%%%%%
\begin{figure*}[!t]
\centering
\includegraphics[width=0.4\textwidth]{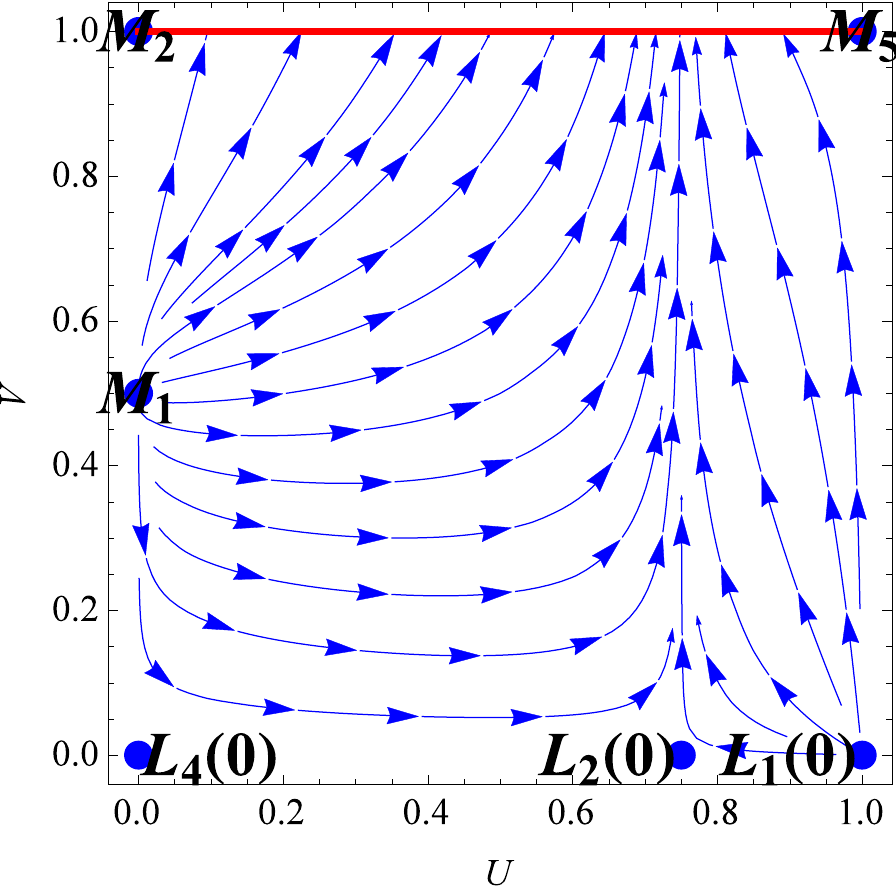}				\caption{\label{Fig5} Streamlines of the system \eqref{system_69A}. The solid (red) line denotes the line $L_6$. }
			\end{figure*}
%%%%%%%%%%%%%%%%%%%%%%%%%%%%%%%
In the Figure \ref{Fig5} are presented some streamlines of the system \eqref{system_69A}. The solid (red) line denotes the line $L_6$, which is a local attractor. 
The straight vertical line connecting $L_2(0)$ with the line $L_6$ is the so-called Tolman orbit (associated to the positive eigenvalue of $L_2(0)$, the so-called Tolman point).

\paragraph{High pressure regime ($Y=\eta^{-1}$).} 
In the limit $\beta=1, Y=\eta^{-1}$, the equations becomes 
\begin{subequations}
\label{system_Model_361}
\begin{align}
&\frac{dU}{d\lambda}= 
-\frac{(\eta -1) (U-1) U \left(\eta ^2 (U (5 V-4)-4 V+3)+\eta  (-2 U (V-4)+V-6)+(3-4 U) (V+1)\right)}{\eta ^3},\\
&\frac{dV}{d\lambda}=\frac{(\eta -1)^2 (2 U-1) (V-1) V (\eta  (V-1)+V+1)}{\eta ^3}. 
\end{align}
\end{subequations}

%%%%%%%%%Figure5%%%%%%%%%%%%%%%%%%%%%%%%%%%%%%
\begin{figure*}[!t]
\centering
\includegraphics[width=1\textwidth]{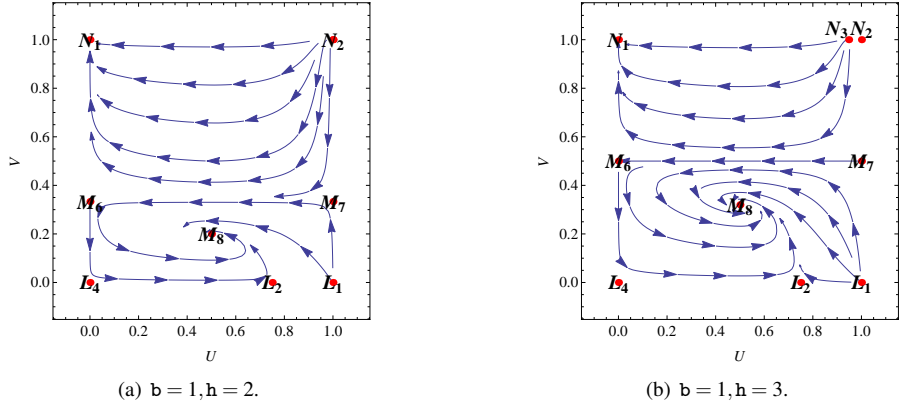}				\caption{\label{Model_361} Streamlines of the system \eqref{system_Model_361}. }
			\end{figure*}
%%%%%%%%%%%%%%%%%%%%%%%%%%%%%%%

The equilibrium points in this regime are 
\begin{enumerate}
\item $L_1(1/\eta)$: $(U,V)= (1, 0)$, with eigenvalues $\frac{(\eta -1)^3}{\eta ^3},\frac{(\eta -1)^3}{\eta ^3}$. Local source. 
\item $L_2(1/\eta)$:  $(U,V)=\left(\frac{3}{4},  0\right)$, with eigenvalues $\frac{(\eta -1)^3}{2 \eta ^3},-\frac{3 (\eta -1)^3}{4 \eta ^3}$. Saddle. 	
\item $L_4(1/\eta)$:  $(U,V)= (0, 0)$, with eigenvalues $\frac{3 (\eta -1)^3}{\eta ^3},-\frac{(\eta -1)^3}{\eta ^3}$. Saddle. 
\item $M_6$: $(U,V)=\left(0, \frac{\eta -1}{\eta +1}\right)$, with eigenvalues $\frac{2 (\eta -1)^3}{\eta ^3 (\eta +1)},-\frac{(\eta -1)^3}{\eta ^2 (\eta +1)}$. Exists for $\eta>1$.  Saddle.  

\item $M_7$: $(U,V)=\left( 1, \frac{\eta -1}{\eta +1}\right)$. The eigenvalues are $-\frac{2 (\eta -1)^3}{\eta ^3 (\eta +1)},\frac{(\eta -1)^2}{\eta ^2 (\eta +1)}$. Exists for $\eta>1$. Saddle. 
	
\item $M_8$: $(U,V)=\left(\frac{1}{2},  \frac{2 (\eta -1)^2}{3 \eta ^2-2}\right)$, with eigenvalues\\$-\frac{(\eta -1)^3 \left(\eta +\sqrt{36-7 \eta  (\eta +4)}+2\right)}{4 \eta ^2 \left(3 \eta ^2-2\right)},-\frac{(\eta -1)^3 \left(\eta -\sqrt{36-7 \eta  (\eta
   +4)}+2\right)}{4 \eta ^2 \left(3 \eta ^2-2\right)}$. Exists for $\eta>1$. It is an stable node for $1<\eta \leq \frac{2}{7} \left(4 \sqrt{7}-7\right)$. Stable spiral for $\eta >\frac{2}{7} \left(4 \sqrt{7}-7\right)$.
	
\item $S_1$:  $(U,V)= (0, 1)$, with eigenvalues $-\frac{(\eta -1)^2 (\eta +6)}{\eta ^3},-\frac{2 (\eta -1)^2}{\eta ^3}$. This point is over the static surface. It can be a local attractor on the invariant set $Y=1/\eta$. 

\item $S_2$: $(U,V)= (1, 1)$, with eigenvalues $\frac{2 (\eta -1)^2}{\eta ^3},-\frac{(\eta -2) (\eta -1)}{\eta ^3}$.  This point is over the static surface. It is a local source for $1<\eta <2$.  It is a saddle for $\eta>2$. 

\item $S_3$: $(U,V)=\left(\frac{(\eta -1) (\eta +6)}{\eta ^2+6 \eta -8}, 1\right)$, with eigenvalues $\frac{(\eta -2) (\eta -1)^2 (\eta +6)}{\eta ^3 (\eta  (\eta +6)-8)},\frac{2 (\eta -1)^2 (\eta  (\eta +4)-4)}{\eta ^3 (\eta  (\eta +6)-8)}$. Exists for $\eta \geq 2$.  This point is over the static surface. It is a local source for $\eta >2$.

\end{enumerate}

In the Figure \ref{Model_361} are presented some streamlines of the system \eqref{system_Model_361} which corresponds to the high pressure regime.
In this regime, the solution $M_8$, that has finite mass and radius, is a local attractor.     
The line $V=\frac{\eta -1}{\eta +1}$ acts as a separatrix. The orbits above this line are attracted by $S_1$.

\paragraph{Einstein-Aether modification.}

In this case  the dynamical system is given by 
\begin{subequations}
\label{TOVEA}
\begin{align}
&\frac{dU}{d\lambda}=\frac{(U-1) U (Y-1) \left(1 - V - Y - V Y\right)}{(\beta -1) \beta  V Y}\Big\{U \Big\{ \beta  (G+1)
   \left[(Y-1)^2 -V\right] \nonumber \\
& +V Y(\beta  (-4 \beta +\eta +(\eta -1) G+2)+Y (\beta  (8 \beta +G-9)+2)+1)\Big\} \nonumber \\
	& -\beta  \left(G \left(V (Y (\eta +Y-1)-1)+(Y-1)^2\right)+V (Y (\eta +(4 \beta -3) (Y-1))-1)+(Y-1)^2\right)\Big\},\\
&\frac{dV}{d\lambda}=-\frac{(V-1) (Y-1) \left(1 - V - Y - V Y\right)}{(\beta -1) \beta  Y} \Big\{\beta  (-G-1) \nonumber \\
& +U \left(V Y (\beta +\beta  (-(2 \beta +\eta +\eta  G))+Y (\beta  (6 \beta +G-7)+2)+1)+\beta  (G+1) (Y-1)^2\right)\nonumber \\
& +\beta  Y (-G (-\eta  V+V
   Y+Y-2)+V (2 \beta +\eta -2 \beta  Y+Y-2)-Y+2)\Big\},\\
&\frac{dY}{d\lambda}=-\frac{(G+1) (U-1) (Y-1)^2 \left(1 - V - Y - V Y\right) (\eta  Y-1)}{\beta -1},\label{eqY2}
\end{align}
\end{subequations}
where  
\begin{align}
& G:=-\sqrt{\frac{\left(p \rho ^3+\rho -2 \beta  m-\beta  p \rho ^3\right)}{\rho-2 m}}\nonumber \\
& = -\sqrt{\frac{(1-Y) (1-(2 \beta-1)  V Y-V-Y)-U \left(V (Y (-2 \beta +4 \beta  Y-3 Y+2)-1)+(Y-1)^2\right)}{(1-U) (1-Y)(1-V Y-V-Y)}}<0.
\end{align}
Observe that $\lim_{\beta\rightarrow 1}  G=-1$. In the limit $\beta\rightarrow 1$, we obtain an  equivalent dynamical system to \eqref{Relativistic_1}.

The dynamical system \eqref{TOVEA} is defined on the bounded phase space 
\begin{align}
&\Big\{(U,V,Y)\in\mathbb{R}^3: 0\leq U \leq 1, 0\leq Y\leq\frac{1}{\eta}, 0\leq V\leq \frac{1-Y}{1+Y},  \eta \geq 1, \nonumber \\
&  \beta \left(2 V Y (U (2
   Y-1)-Y+1)\right) <U \left(V \left(3 Y^2-2 Y+1\right)-(Y-1)^2\right)+(1-V) (Y-1)^2\Big\}.
\end{align}

From the equation \eqref{eqY2} we determine the invariant sets $ U=1, Y=1, Y=\frac{1}{\eta}$.  Using this fact, we can find  some relevant equilibrium points/lines of equilibrium points (see Table \ref{Tab2B}):

\begin{enumerate}
\item $L_1$: The line of equilibrium points $(U,V,Y)=\left(1, 0, Y_0\right)$, with eigenvalues\\$0,-\frac{(Y_0-1)^2 ((3 \beta -2) Y_0-1)}{\beta },-\frac{(Y_0-1)^2 ((3 \beta -2) Y_0-1)}{\beta }$. This a line that it is an attractor for $\eta >1, \beta <0, 0<Y_0<\frac{1}{\eta }$, or $\eta >1, \beta >\frac{\eta +2}{3}, \frac{1}{3 \beta -2}<Y_0<\frac{1}{\eta }$. It is a source for $\eta >1, 0<\beta \leq \frac{\eta +2}{3}, 0<Y_0<\frac{1}{\eta }$, or $\eta >1, \beta >\frac{\eta +2}{3}, 0<Y_0<\frac{1}{3 \beta -2}$.

\item $L_1(1/\eta)$: The equilibrium point $(U,V,Y)=\left(1, 0, \frac{1}{\eta}\right)$, with eigenvalues\\ $0,-\frac{(\eta -1)^2 (3 \beta -\eta -2)}{\beta  \eta ^3},-\frac{(\eta -1)^2 (3 \beta -\eta -2)}{\beta  \eta ^3}$. Nonhyperbolic with a 2D unstable manifold for $\eta >1, 0<\beta <\frac{\eta +2}{3}$. Nonhyperbolic with a 2D stable manifold for $\eta >1, \beta <0$, or $\eta >1, \beta >\frac{\eta +2}{3}$. 

\item $L_2$: The equilibrium point $(U,V,Y)=\left(\frac{3}{4}, 0,Y_0\right)$. Exists only for $\beta=1$. The eigenvalues are $0,-\frac{1}{2} (Y_0-1)^3,\frac{3}{4} (Y_0-1)^3$.    Nonhyperbolic and it behaves as a saddle. 

\item $L_2(0)$: The equilibrium point $(U,V,Y)=\left(\frac{3\beta}{1+3\beta}, 0,0\right)$. The eigenvalues are $0,-\frac{3}{3 \beta +1},\frac{2}{3 \beta +1}$.  Nonhyperbolic and it behaves as a saddle.

\item $L_2(1)$: The equilibrium point $(U,V,Y)=\left(\frac{3\beta}{1+3\beta}, 0,1\right)$. The eigenvalues are $0,0,0$. It is nonhyperbolic.

\item $L_3$: The line of equilibrium points $(U,V,Y)=\left(U_0, 0, 1\right)$. The eigenvalues are $0,0,0$. It is nonhyperbolic. 

\item $L_4$: The lines of equilibrium point $(U,V,Y)=\left(0, 0, Y_0\right)$, with eigenvalues $0,-3 (Y_0-1)^3,(Y_0-1)^3$. Nonhyperbolic for $Y_0=0$, saddle otherwise.

\item $L_5(1)$: The equilibrium point $(U,V,Y)=\left(0, 1, 1\right)$, with eigenvalues $0, 0, 0$. It is nonhyperbolic. 

\item $L_6$: The line of equilibrium points $(U,V,Y)=\left(U_0, 1, 0\right)$. The eigenvalues are\\$0,-\frac{2 \left(\sqrt{\beta }+1\right) (U_0-1)}{\beta -1},-\frac{2 \left(\sqrt{\beta }+1\right) (U_0-1)}{\beta -1}.$ This line of equilibrium points is nonhyperbolic with a 2D stable manifold for $0<U_0<1, 0\leq \beta <1$ (two negative eigenvalues) or $\beta <0, 0<U_0<1$ (two complex conjugated eigenvalues with negative real parts). This line of equilibrium points is nonhyperbolic with a 2D unstable manifold for $0<U_0<1, \beta >1$.

\item $M_1$: The equilibrium point $(U,V,Y)=\left(0, \frac{1}{2}, 0\right)$, with eigenvalues $\frac{3}{2},\frac{1}{2},-\frac{1}{2}$. Saddle 
 
\item $M_2$: The equilibrium point $(U,V,Y)=\left(0, 1, 0\right)$, with eigenvalues $\frac{2 \left(1+\sqrt{\beta }\right)}{\beta -1},0,\frac{2 \left(1+\sqrt{\beta }\right)}{\beta -1}$. Nonhyperbolic with a 2D stable manifold for $0<\beta <1$ (two negative real eigenvalues) or $\beta <0$ (two complex conjugated eigenvalues).

\item $M_3$: The equilibrium point $(U,V,Y)=\left(1, 1, 1\right)$, with eigenvalues $0, 0, 0$. 	It is nonhyperbolic.

\item $M_5$: The equilibrium point $(U,V,Y)=\left(1, 1, 0\right)$. The eigenvalues at the fixed points are $0,0,0$. Therefore, it is nonhyperbolic. However, taking $Y=\varepsilon\ll 1$, instead of exactly $Y=0$, we obtain the eigenvalues $0,-\frac{2 \varepsilon }{\beta },\frac{2 \varepsilon }{\beta }$, that is, the fixed point is indeed a saddle point. 
 
\item $M_6$: The equilibrium point $(U,V,Y)=\left(0, \frac{\eta-1}{\eta+1}, \frac{1}{\eta}\right)$, with eigenvalues $0,0,\frac{4 (\eta -1)^2}{\eta ^2 (\eta +1)}$.  Non hyperbolic with a 1D unstable manifold. 

\item $M_7$: The equilibrium point $(U,V,Y)=\left(1, \frac{\eta-1}{\eta+1}, \frac{1}{\eta}\right)$, with eigenvalues $0,0,\frac{2 (\eta -1) (4 \beta -\eta -2)}{\beta  \eta ^2 (\eta +1)}$.  Non hyperbolic with a 1D unstable manifold for $\beta <0, \eta >1$, or $\beta >\frac{3}{4}, 1<\eta <4 \beta -2$. Nonhyperbolic with a 1D stable manifold for $0<\beta \leq \frac{3}{4}, \eta >1$, or $\beta >\frac{3}{4}, \eta >4 \beta -2$.  
\end{enumerate}

Furthermore, there are four equilibrium points with $V=1, Y=\frac{1}{\eta}$, which are of little interest for the static solutions, because they are not over  the ``static surface''. Since the inequalities $0\leq V\leq \frac{1-Y}{1+Y},  \eta \geq 1,$ are not fulfilled at these points they do not exist. We omit the discussion of them.

Finally, we have found new lines of/equilibrium points in comparison with the previous case \eqref{Relativistic_1} (see table \ref{Tab2}), like: 
\begin{enumerate}

\item $M_9$: The line of equilibrium points $(U,V,Y)=\left(1, 1, Y_0\right)$. The eigenvalues are\\$0,\frac{2Y_0 ((2-4 \beta )Y_0+1)}{\beta },-\frac{2Y_0 ((2-4 \beta )Y_0+1)}{\beta}$. This line is nonhyperbolic and it behaves as a saddle.

\item $M_{10}$: The line of equilibrium points $(U,V,Y)=\left(1, V_0, \frac{1-V_0}{1+V_0}\right)$. This line of fixed points has the eigenvalues $\frac{2 (V_0-1) V_0 (4 \beta  (V_0-1)-V_0+3)}{\beta  (V_0+1)^2},0,0$. It is nonhyperbolic.  

\item $M_{11}$: The line of equilibrium points $(U,V,Y)=\left(1, V_0, \frac{1}{4 \beta -2}\right)$. The eigenvalues are $0,0,0$. It is nonhyperbolic.

\item $M_{12}$: The line of equilibrium points $(U,V,Y)=\left(U_0, \frac{\eta-1}{\eta+1}, \frac{1}{\eta}\right)$.  The eigenvalues are\\$0,0,-\frac{2 (\eta -1)^2 (U_0 (2 \beta  (\eta -3)+\eta +2)-2 \beta  (\eta -1))}{\beta  \eta ^3 (\eta +1)}$. It is nonhyperbolic.

\end{enumerate}

%%%%%%%%%Figure%%%%%%%%%%%%%%%%%%%%%%%%%%%%%%
\begin{figure*}[!t]
	\subfigure[\label{Fig5a}  $\beta=-0.1$.]{\includegraphics[width=0.4\textwidth]{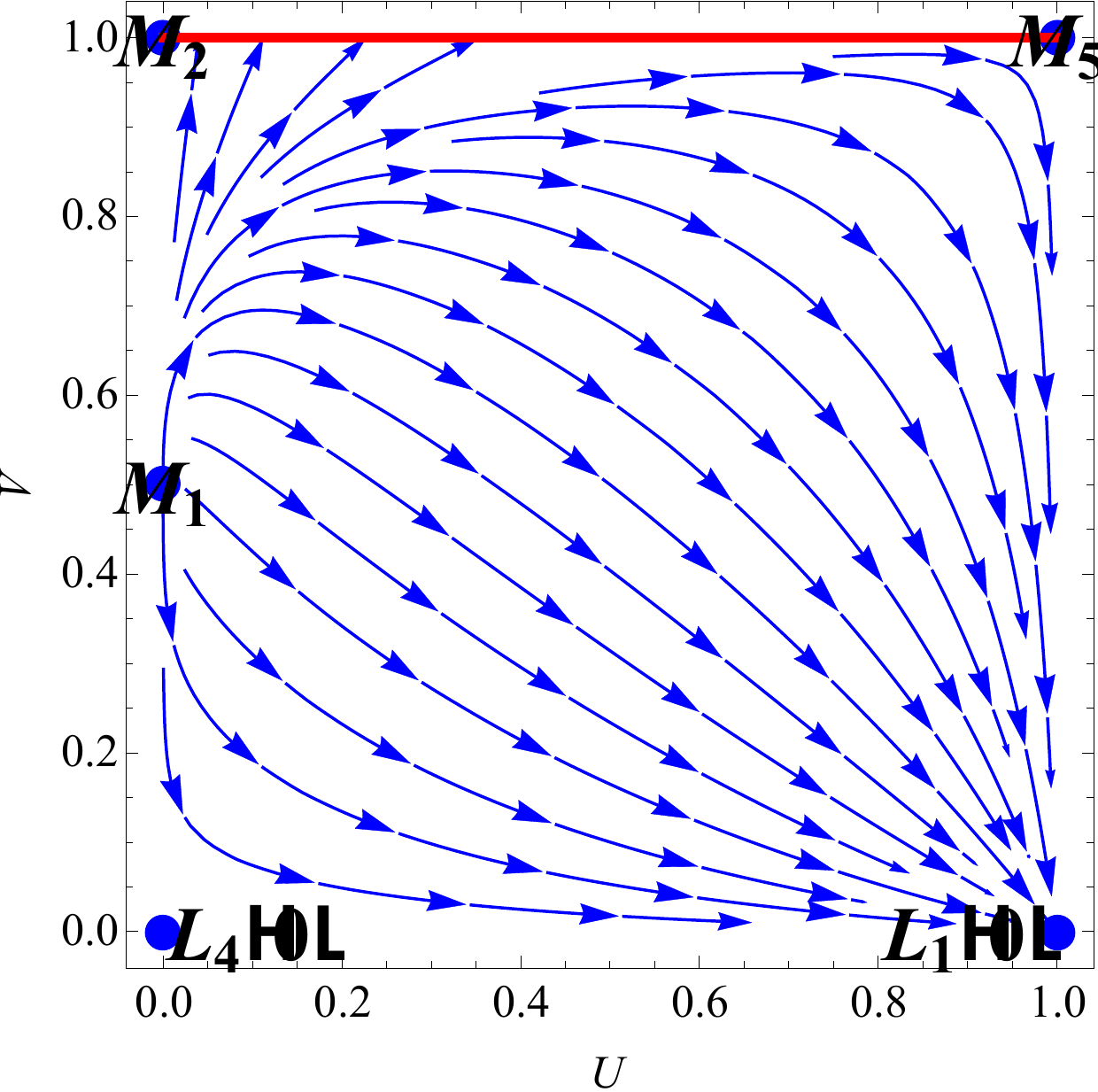}}	\hspace{2cm}
	\subfigure[\label{Fig5b}  $\beta=0.9$.]{\includegraphics[width=0.4\textwidth]{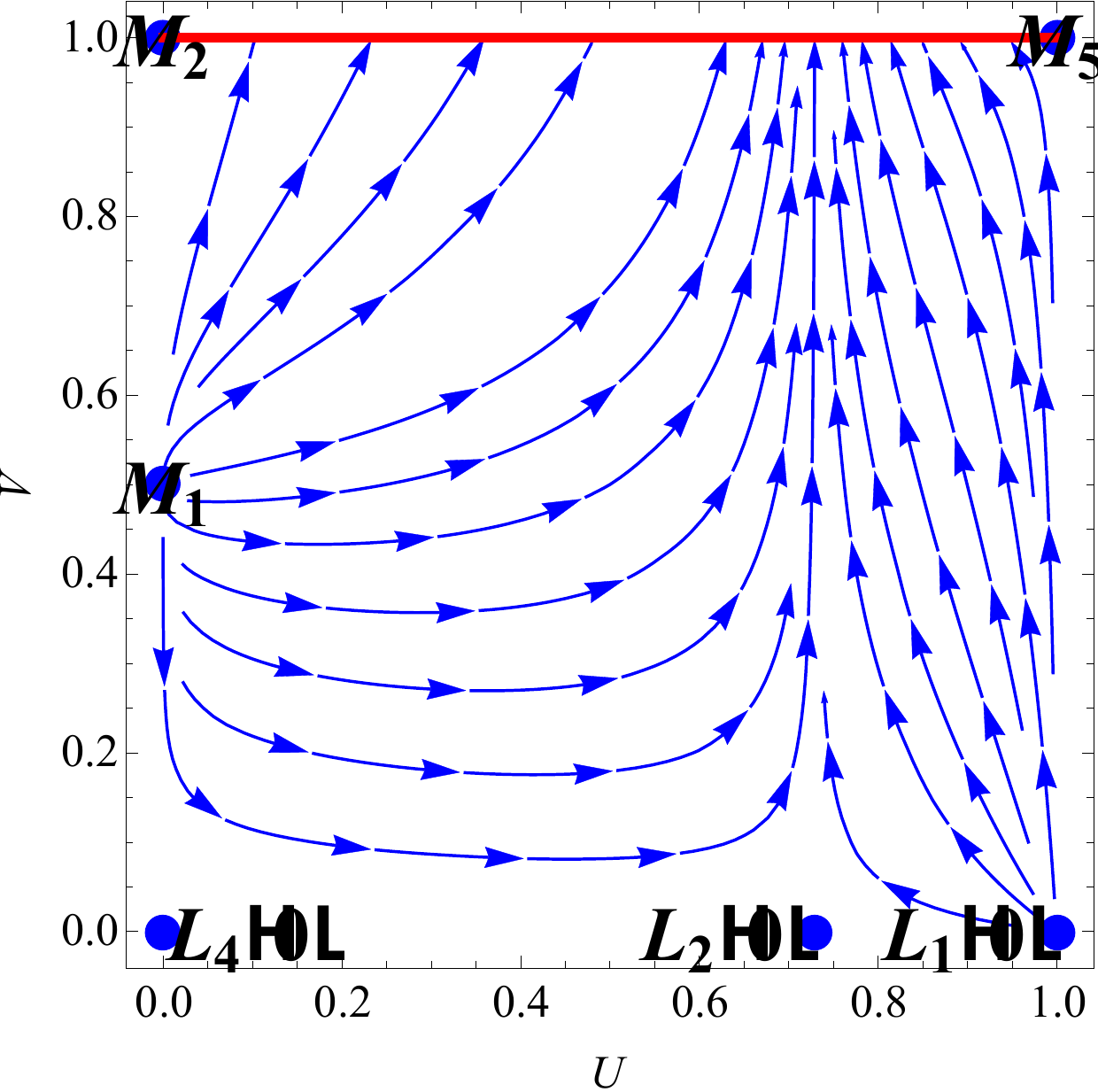}} 
								\caption{\label{Fig5ab} Streamlines of the system \eqref{system394}.}
			\end{figure*}
%%%%%%%%%%%%%%%%%%%%%%%%%%%%%%%

\begin{table}[!t]
\centering
\scalebox{0.75}{
\begin{tabular}{|c|c|c|c|c|}
\hline
Labels & $(U,V,Y)$  &   Eigenvalues & Stability  \\\hline
$L_1$ & $(1,0, Y_0)$ & $0,-\frac{(Y_0-1)^2 ((3 \beta -2) Y_0-1)}{\beta },-\frac{(Y_0-1)^2 ((3 \beta -2) Y_0-1)}{\beta }$ & Attractor for   \\
&&& $\eta >1, \beta <0, 0<Y_0<\frac{1}{\eta }$, or  \\
&&& $\eta >1, \beta >\frac{\eta +2}{3}, \frac{1}{3 \beta -2}<Y_0<\frac{1}{\eta }$.  \\
&&& Source for   \\
&&& $\eta >1, 0<\beta \leq \frac{\eta +2}{3}, 0<Y_0<\frac{1}{\eta }$, or  \\
&&& $\eta >1, \beta >\frac{\eta +2}{3}, 0<Y_0<\frac{1}{3 \beta -2}$.  \\\hline
$L_1(1/\eta)$& $\left(1, 0, \frac{1}{\eta}\right)$  & $0,-\frac{(\eta -1)^2 (3 \beta -\eta -2)}{\beta  \eta ^3},-\frac{(\eta -1)^2 (3 \beta -\eta -2)}{\beta  \eta ^3}$. & Nonhyperbolic.\\
&&&  2D unstable manifold for \\
&&& $\eta >1, 0<\beta <\frac{\eta +2}{3}$. \\
&&& 2D stable manifold for \\
&&& $\eta >1, \beta <0$,  \\
&&& or $\eta >1, \beta >\frac{\eta +2}{3}$. \\\hline
$L_2$ & $\left(\frac{3}{4}, 0,Y_0\right)$ $\beta=1$ & $0,-\frac{1}{2} (Y_0-1)^3,\frac{3}{4} (Y_0-1)^3$& Saddle.  \\
$L_2(0)$ & $\left(\frac{3\beta}{1+3\beta}, 0,0\right)$ & $0,-\frac{3}{3 \beta +1},\frac{2}{3 \beta +1}$& Saddle.  \\ 
$L_2(1)$ & $\left(\frac{3\beta}{1+3\beta}, 0,1\right)$ & $0,0,0$ & Nonhyperbolic.  \\\hline 
$L_3$ & $\left(U_0,0,1\right)$ & $0,0,0$ & Nonhyperbolic. \\\hline
$L_4$ & $(0, 0, Y_0)$  & $0,-3 (Y_0-1)^3,(Y_0-1)^3$ & Nonhyperbolic for $Y_0=0$,  \\
&&& saddle otherwise.  \\\hline
$L_5(1)$ & $\left(0, 1, 1\right)$   & $0, 0, 0$ & Nonhyperbolic.  \\\hline
$L_6$ & $\left(U_0, 1, 0\right)$  & $0,-\frac{2 \left(\sqrt{\beta }+1\right) (U_0-1)}{\beta -1},-\frac{2 \left(\sqrt{\beta }+1\right) (U_0-1)}{\beta -1}.$ & Nonhyperbolic. \\
&&& 2D stable manifold for \\
&&& $0<U_0<1, 0\leq \beta <1$, or  \\
&&& $\beta <0, 0<U_0<1$. \\
&&& 2D unstable manifold for \\
&&& $0<U_0<1, \beta >1$. \\\hline
 $M_1$ & $\left(0, \frac{1}{2}, 0\right)$   & $\frac{3}{2},-\frac{1}{2},\frac{1}{2}$ & Saddle.  \\\hline
 $M_2$ & $\left(0, 1, 0\right)$   &  $\frac{2 \left(1+\sqrt{\beta }\right)}{\beta -1},0,\frac{2 \left(1+\sqrt{\beta }\right)}{\beta -1}$ &  Nonhyperbolic. \\
&&& 2D stable manifold for \\
&&& $0<\beta <1$, or  $\beta <0$. \\\hline
 $M_3$ & $\left(1,1,1\right)$  &  $0, 0, 0$ & Nonhyperbolic.\\\hline
 $M_5$ & $\left(1,1,0\right)$  &  $0, 0, 0$ & Saddle. \\\hline
 $M_6$ & $\left(0, \frac{\eta -1}{\eta +1}, \frac{1}{\eta }\right)$   & $0,0,\frac{4 (\eta -1)^3}{\eta ^3 (\eta +1)}$ & Nonhyperbolic. \\
&&& 1D unstable manifold.  \\\hline
$M_7$ & $\left(1,  \frac{\eta -1}{\eta +1}, \frac{1}{\eta }\right)$ & $0,0,\frac{2 (\eta -1)^2 (4 \beta -\eta -2)}{\beta  \eta ^3 (\eta +1)}$. & Nonhyperbolic. \\
&&& 1D unstable manifold for  \\
&&& $\beta <0, \eta >1$, or  \\
&&& $\beta >\frac{3}{4}, 1<\eta <4 \beta -2$.  \\
&&& 1D stable manifold for  \\
&&& $0<\beta \leq \frac{3}{4}, \eta >1$, or  \\
&&& $\beta >\frac{3}{4}, \eta >4 \beta -2$.  \\\hline
$M_9$ & $\left(1, 1, Y_0\right)$	 & $0,\frac{2Y_0 ((2-4 \beta )Y_0+1)}{\beta },-\frac{2Y_0 ((2-4 \beta )Y_0+1)}{\beta}$& Saddle.\\\hline 
$M_{10}$ &  $\left(1, V_0, \frac{1-V_0}{1+V_0}\right)$	 & $0,0,\frac{2 (Y_0-1)^2 Y_0 ((4 \beta -2)  {Y_0}-1)}{\beta  (Y_0+1)}$ & Nonhyperbolic. \\\hline 
$M_{11}$& $\left(1, V_0, \frac{1}{2 (2 \beta -1)}\right)$, $\beta>\frac{3}{4}$ & $0,0,0$ & Nonhyperbolic.  	\\\hline
$M_{12}$& $\left(U_0, \frac{\eta-1}{\eta+1}, \frac{1}{\eta}\right)$ & $0,0,-\frac{2 (\eta -1)^2 (U_0 (2 \beta  (\eta -3)+\eta +2)-2 \beta  (\eta -1))}{\beta  \eta ^3 (\eta +1)}$ & Nonhyperbolic. 	\\\hline 
\end{tabular}}
\caption{\label{Tab2B}  Equilibrium points and the corresponding eigenvalues and stability conditions of the dynamical system \eqref{TOVEA}. We use the notation 
$\delta=\sqrt{10 {V_0}^2+16 \beta ^2
   ({V_0}-1)^2-24 \beta 
   ({V_0}-1)^2-18 {V_0}+9}$. }
\end{table}

\paragraph{Low pressure regime (The Newtonian subset).} 

We now consider the ``Newtonian subset'' $Y=0$, which is of great importance for relativistic stars.

The evolution equations are given by
\begin{subequations}
\label{system394}
\begin{align}
&\frac{d U}{d{\lambda}}=-\frac{(U-1) U (V-1) (U+3 (U-1) \beta )}{\beta },\\
&\frac{d V}{d{\lambda}}= \frac{(V-1) V (U (V-1)+(U-1) (2 V-1) \beta )}{\beta },
\end{align}
\end{subequations}
defined on the phase plane
\begin{equation}
\Big\{(U,V,Y)\in\mathbb{R}^3: 0\leq U \leq 1, 0\leq V\leq 1\Big\}.
\end{equation}

The stability analysis below is restricted to the dynamics onto the invariant set $Y=0$, and the stability along the $Y$-axis is not taken into account. 

The equilibrium (curves of equilibrium) points are in this invariant set are:
\begin{enumerate}
\item $M_1:=(U,V)=\left(0, \frac{1}{2}\right)$. The eigenvalues are $\frac{3}{2},\frac{1}{2}$, such that it is a source. 

\item $M_2:=(U,V)=\left(0, 1\right)$. The eigenvalues are $-1,0$, such that it is a non-hyperbolic equilibrium point (it is contained in the attractor line $L_6$ described below). 

\item $M_5:= (U,V)=\left(1,1\right)$. 
 The eigenvalues are $0,0$. Hence,  it is a non-hyperbolic equilibrium point (it is contained in the attractor line $L_6$ described below).

\item $L_1(0):=(U,V)=\left(1, 0\right)$. The eigenvalues are $\frac{1}{\beta },\frac{1}{\beta }$, such that it is a source for $\beta>0$ and a sink for $\beta<0$. 

\item $L_2(0):= (U,V)=\left(\frac{3\beta }{1+3\beta}, 0\right)$, which is an endpoint of the line $L_2$ described in the previous section. It exists for $\beta\geq 0$. The eigenvalues are $ -\frac{3}{3 \beta +1},\frac{2}{3 \beta +1}$, such that it is a saddle.

\item $L_4(0):=(U,V)=\left(0, 0\right)$, which is an endpoint of the line $L_4$ described in the previous section. The eigenvalues are $3,-1$, such that it is a saddle.

\item The line $L_6:= (U,V)=(U_0,1), U_0\in[0,1]$. The eigenvalues are $U_0-1,0$. This line is normally hyperbolic and due to the non-zero eigenvalue is negative it is a sink.
\end{enumerate}

We see  that the main difference with respect the case previously discussed is in that $L_1(0)$ can be an attractor for $\beta<0$, see figure \ref{Fig5a}. In the Figure \ref{Fig5b} it is presented the case for $\beta=0.9>0$, essentially the same behavior as for the model 
\eqref{system_69A}.

\subsection{Model 2: Polytropic equation of state}
\label{model2}

We use the line element 
\begin{equation}
ds^{2}=-e^{2\Omega(\lambda)}dt^{2}+r(\lambda)^2 \left[N(\lambda)^2d\lambda^{2}+d\vartheta ^{2}+\sin^{2} \vartheta d\phi ^{2}\right],  \label{XXXmet2}
\end{equation}%
where $\Omega(\lambda)$ is the gravitational potential, $r(\lambda)$ is the usual Schwarzschild radial parameter, and $N(\lambda)$ is a dimensionless (under scalar transformations) freely specifiable function, that defines the spatial radial coordinate $\lambda$. The choice $N=1$ corresponds to isotropic coordinates and the function $N$ can viewed as a relative gauge function with respect to the isotropic gauge \cite{Nilsson:2000zg}. 

The FE are \cite{Coley:2015qqa}:
\begin{subequations}
\label{XXXsystA}
\begin{align}
& {\mathbf{e}_{1}}(x) =3 x  y +2 (\beta -1) y ^2+\frac{\mu  +3 p }{2\beta}+K ,\label{XXXeq.01}\\
& {\mathbf{e}_{1}}(y)=	2 x  y -y ^2+\frac{\mu  +3
   p }{2\beta},\label{XXXeq.02}\\
& {\mathbf{e}_{1}}(p)=	-y  (\mu  +p ),\label{XXXeq.03}\\
& {\mathbf{e}_{1}}{K}=	2 x K ,\label{XXXeq.04}
\end{align}%
\end{subequations} 
where for any function $f(\lambda)$, we define ${\mathbf{e}_{1}} (f)\equiv \frac{f'(\lambda)}{N(\lambda) r(\lambda)}$, 
and $x=-\frac{{\mathbf{e}_{1}}(r)}{r}, \quad y= {\mathbf{e}_{1}}(\Omega), \quad K=\frac{1}{r^2}$. 
Furthermore there exists the constraint
equation%
\begin{equation}
2 x  y -x ^2+(\beta-1) y ^2+K +p =0.  \label{XXXeq.05}
\end{equation}%

Choosing a new radial coordinate $\tau$ such that
\begin{equation}
\frac{d\lambda}{d\tau}=\frac{1}{N(\lambda) r(\lambda)},
\end{equation}
the system \eqref{XXXsystA} becomes
\begin{subequations}
\begin{align}
&x'(\tau) =3 x   y +2(\beta-1)y ^2+\frac{\mu  +3 p }{2
 \beta}+K ,\label{XXXEq.01}\\
&Y'(\tau)=	2 x  y -y ^2+\frac{\mu  +3 p }{2 \beta},\label{XXXEq.02}\\
&p'(\tau)=	-y  (\mu  +p ),\label{XXXEq.03}\\
&K'(\tau)=	2 x K ,\label{XXXEq.04}
\end{align}%
\end{subequations}
with restriction \eqref{XXXeq.05}. 
The above equations transforms to eqs. (A.3) in \cite{Nilsson:2000zg} for the choice $\beta=1$, under the identifications $\theta=y-x, \sigma=y, B^2=K$.
We consider a polytropic equation of state $p=k \mu^\Gamma, \Gamma=1+\frac{1}{n}, n>0$.

\subsubsection{Singularity analysis and integrability}

The FE does not pass the Painlev\`e test, and therefore they are not integrable.

\subsubsection{Equilibrium points in the finite region of the phase space}
\label{SECT_3.5.1}
For the dynamical system's formulation we introduce the variables
\begin{align}
Y=\frac{p}{p+\mu}, \Sigma_1=\frac{y}{\theta}, \Sigma_2=\frac{K}{\theta^2}, P=\frac{p}{\theta^2}, \theta=y-x.
\end{align}
Assuming that the energy density $\mu$ and the pressure $p$ are both non-negative we obtain $0\leq Y \leq 1$, where we have attached the boundaries $Y=0$ and $Y=1$.
On the other hand,  $\Sigma_2\geq 0$ by definition.

Now we define the independent dimensionless variable $\lambda$ by the choice 
$N^2 r^2=Y^2/\theta^2$, that is
\begin{equation}
\frac{d\lambda}{d\tau}=\frac{\theta}{Y},
\end{equation}

and we obtain the dynamical system 
\begin{subequations}
\label{XXXeq:23}
\begin{align}
&\frac{d \Sigma_1}{d\lambda}=-\frac{(\Sigma_2-1) (2 Y+1)}{2 \beta}+2
   \beta \Sigma_1^3 Y+(\Sigma_2-2) \Sigma_1 Y-\frac{1}{2} \Sigma_1^2 (2 Y+1),\\
&\frac{d \Sigma_2}{d\lambda}=	2 \Sigma_2 Y\left(2
   \beta \Sigma_1^2+\Sigma_2-1\right),\\
	&\frac{d Y}{d\lambda}=\frac{(\Gamma -1) \Sigma_1  (Y-1) Y}{\Gamma }.
\end{align}
\end{subequations}
where we used a polytropic equation of state $p=q \mu^\Gamma, \Gamma=1+\frac{1}{n}, n>0$, where $q$ is a nonnegative constant. For $\beta=1$ it is recovered the model of \cite{Nilsson:2000zg}. 

Using the same arguments as in Section \ref{DSSection_2.1.2}, we have that $Y=0, Y=0, \Sigma_2=0$, and $1-\beta\Sigma_1^2-\Sigma_2=0$, are invariant sets for the flow of \eqref{XXXeq:23}, and we can argue that system \eqref{XXXeq:23} defines a flow on the (invariant subset) phase space 
\begin{equation}
\Big\{(\Sigma_1, \Sigma_2,Y)\in\mathbb{R}^3: 0\leq Y\leq 1, \quad \beta\Sigma_1^2+\Sigma_2\leq 1, \quad \Sigma_2\geq 0\Big\}.
\end{equation}

The dynamical system \eqref{XXXeq:23} has the (lines of) equilibrium points at the finite region:
\begin{enumerate}
\item Line of equilibrium points: $L_1:=\left(\Sigma_c, 1-\beta\Sigma_c^2, 0\right)$. Exists for $\beta\leq 0, \Sigma_c\in\mathbb{R}$, or  $\beta> 0, -\sqrt{\frac{1}{\beta}}\leq \Sigma_c \leq \sqrt{\frac{1}{\beta}}$. Eigenvalues: $0, -\frac{\Sigma_c}{1+n}, -\Sigma_c$. Stable 2D manifold for $n>0, \beta\leq 0, \Sigma_c >0$, or  $n>0, \beta> 0, 0<\Sigma_c\leq \sqrt{\frac{1}{\beta}}$. Unstable 2D manifold for $n>0, \beta\leq 0, \Sigma_c <0$, or  $n>0, \beta> 0, -\sqrt{\frac{1}{\beta}}\leq \Sigma_c <0$. This line contains as particular case the line $L_1$ characterized in Table I of \cite{Nilsson:2000zg} when $\beta=1$.

 \item Line of equilibrium points: $L_2:=\left(0, 1, y_c\right)$. Eigenvalues: $0, -y_c, 2 y_c$. Saddle. This line coincides with the line $L_2$ characterized in Table I of \cite{Nilsson:2000zg}.
  
 \item Fixed Point: $P_1:=\left(\frac{2}{3}, \frac{1}{9} (9-8\beta), 1\right)$. Exists for $n>0, 0\leq \beta\leq \frac{9}{8}$. \\ 
 Eigenvalues: $\frac{2}{3 (n+1)}, \frac{1}{6} \left(-3-\sqrt{64\beta-63}\right), \frac{1}{6} \left(-3+\sqrt{64\beta-63}\right)$. Nonhyperbolic for $\beta=\frac{9}{8}$. Saddle (2D stable manifold for $n>0, 0\leq  \beta<\frac{9}{8}$). 
 This point reduces to $P_1$ characterized in Table I of \cite{Nilsson:2000zg} when $\beta=1$.
 
  \item $P_2:= \left(\frac{1}{\sqrt{\beta}}, 0, 1\right)$. Exists for $\beta>0, n>0$. Eigenvalues: $\frac{1}{\sqrt{\beta} (n+1)},2,\frac{4 \sqrt{\beta}-3}{\sqrt{\beta}}$. Nonhyperbolic for $\beta=\frac{9}{16}$. Unstable for $n>0, \beta>\frac{9}{16}$. Saddle, otherwise. 
  This point reduces to $P_2$ characterized in Table I of \cite{Nilsson:2000zg} when $\beta=1$.
  
  \item Fixed Point: $P_3:=\left(-\frac{1}{\sqrt{\beta}}, 0, 1\right)$. Exists for $\beta>0, n>0$. Eigenvalues: 
$-\frac{1}{\sqrt{\beta} (n+1)},2,\frac{4 \sqrt{\beta}+3}{\sqrt{\beta}}$. Saddle.   This point reduces to $P_3$ characterized in Table I of \cite{Nilsson:2000zg} when $\beta=1$.

  \item Fixed Point: $P_4:=\left(\frac{3}{4 \beta}, 0, 1\right)$. Exists for $\beta< 0, n>0$, or  $\beta\geq \frac{9}{16}, n>0$. Eigenvalues: $\frac{3}{4 \beta  (n+1)},-\frac{8 \beta -9}{4 \beta },-\frac{16 \beta -9}{8 \beta }$. Nonhyperbolic for $\beta=\frac{9}{8}$, or  $\beta= \frac{9}{16}$. Stable for $n>0, \beta< 0$. Saddle, otherwise.  This point reduces to $P_4$ characterized in Table I of \cite{Nilsson:2000zg} when $\beta=1$.
 
\end{enumerate}
%%%%%%%%%Figure%%%%%%%%%%%%%%%%%%%%%%%%%%%%%%
\begin{figure*}[!t]
	\subfigure[\label{fig:XXXeq:23a}  $\beta=1$.]{\includegraphics[width=0.4\textwidth]{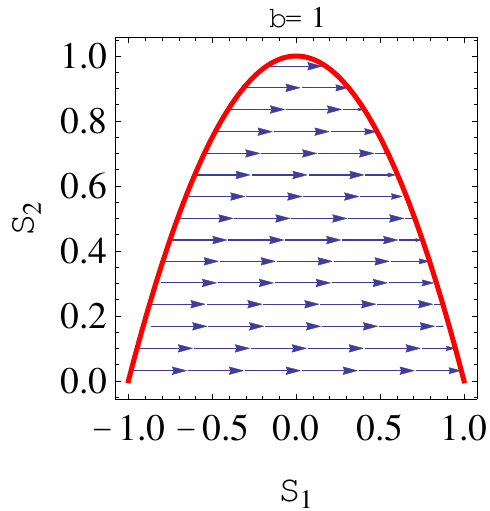}}	\hspace{2cm}
	\subfigure[\label{fig:XXXeq:23b}  $\beta=1.1$.]{\includegraphics[width=0.4\textwidth]{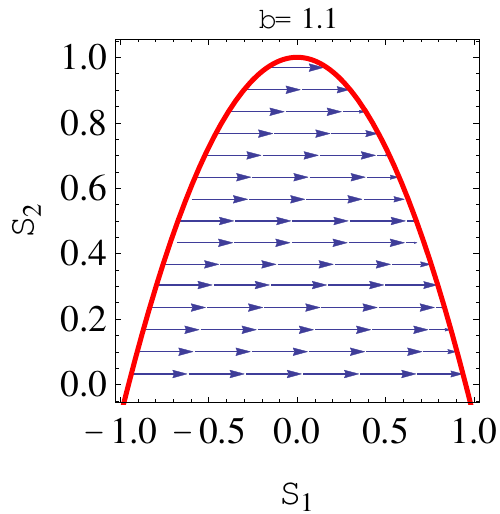}} 
								\caption{\label{fig:XXXeq:23} Streamlines  of the system \eqref{XXXeq:23} in the invariant set $Y=0$. The red (thick) line denotes the singular curve $L_1$. The left branch $\Sigma_1<0$ is unstable and the right branch $\Sigma_1>0$ is stable. }
			\end{figure*}
%%%%%%%%%%%%%%%%%%%%%%%%%%%%%%%
In the Table \ref{Tab3} it is summarized the information of the equilibrium points and the corresponding eigenvalues and stability conditions of the
dynamical system \eqref{XXXeq:23}.

\begin{table}[!t]
\centering
\scalebox{0.9}{
\begin{tabular}{|c|c|c|c|c|}
\hline
Labels & $(\Sigma_1, \Sigma_2,Y)$ &  Eigenvalues & Stability  \\\hline
$L_1$ & $\left(\Sigma_c, 1-\beta\Sigma_c^2, 0\right)$ for $\Sigma_c\neq 1$ &  $0, -\frac{\Sigma_c}{1+n}, -\Sigma_c$ & Nonhyperbolic \\\hline
$L_2$ & $\left(0, 1, y_c\right)$ & $0, -y_c, 2 y_c$ & saddle \\\hline 
$P_1$ & $\left(\frac{2}{3}, \frac{1}{9} (9-8\beta), 1\right)$ & $\begin{array}{c}
\frac{2}{3 (n+1)}, \\ -\frac{1}{6}\left(3-\sqrt{64\beta-63}\right), \\-\frac{1}{6} \left(3+\sqrt{64\beta-63}\right)\end{array}$ & $\begin{array}{c}
\text{Nonhyperbolic for}\;\beta=\frac{9}{8}\\
\text{saddle otherwise} \end{array}$ \\\hline
$P_2$ & $\left(\frac{1}{\sqrt{\beta}}, 0, 1\right)$ & $\frac{1}{\sqrt{\beta} (n+1)},2,\frac{4 \sqrt{\beta}-3}{\sqrt{\beta}}$ & $\begin{array}{c} \text{nonhyperbolic for}\; \beta=\frac{9}{16}\\ \text{unstable for} \; n>0, \beta>\frac{9}{16}\\ \text{saddle, otherwise}\end{array}$\\\hline
$P_3$ & $\left(-\frac{1}{\sqrt{\beta}}, 0, 1\right)$  & 
$-\frac{1}{\sqrt{\beta} (n+1)},2,\frac{4 \sqrt{\beta}+3}{\sqrt{\beta}}$ & saddle \\\hline 
$P_4$ & $\left(\frac{3}{4 \beta}, 0, 1\right)$ & $\frac{3}{4 \beta  (n+1)},-\frac{8 \beta -9}{4 \beta },-\frac{16 \beta -9}{8 \beta }$ & $\begin{array}{c}
\text{Nonhyperbolic for}\; \beta=\frac{9}{8}, \;\text{or}\; \beta= \frac{9}{16}\\ \text{stable for}\; n>0, \beta< 0\\ \text{saddle, otherwise}
\end{array}$ \\\hline
\end{tabular}}
\caption{\label{Tab3} Equilibrium points and the corresponding eigenvalues and stability conditions of the
dynamical system \eqref{XXXeq:23}. }
\end{table}
%%%%%%%%%Figure%%%%%%%%%%%%%%%%%%%%%%%%%%%%%%
\begin{figure*}[!t]
	\includegraphics[width=0.8\textwidth]{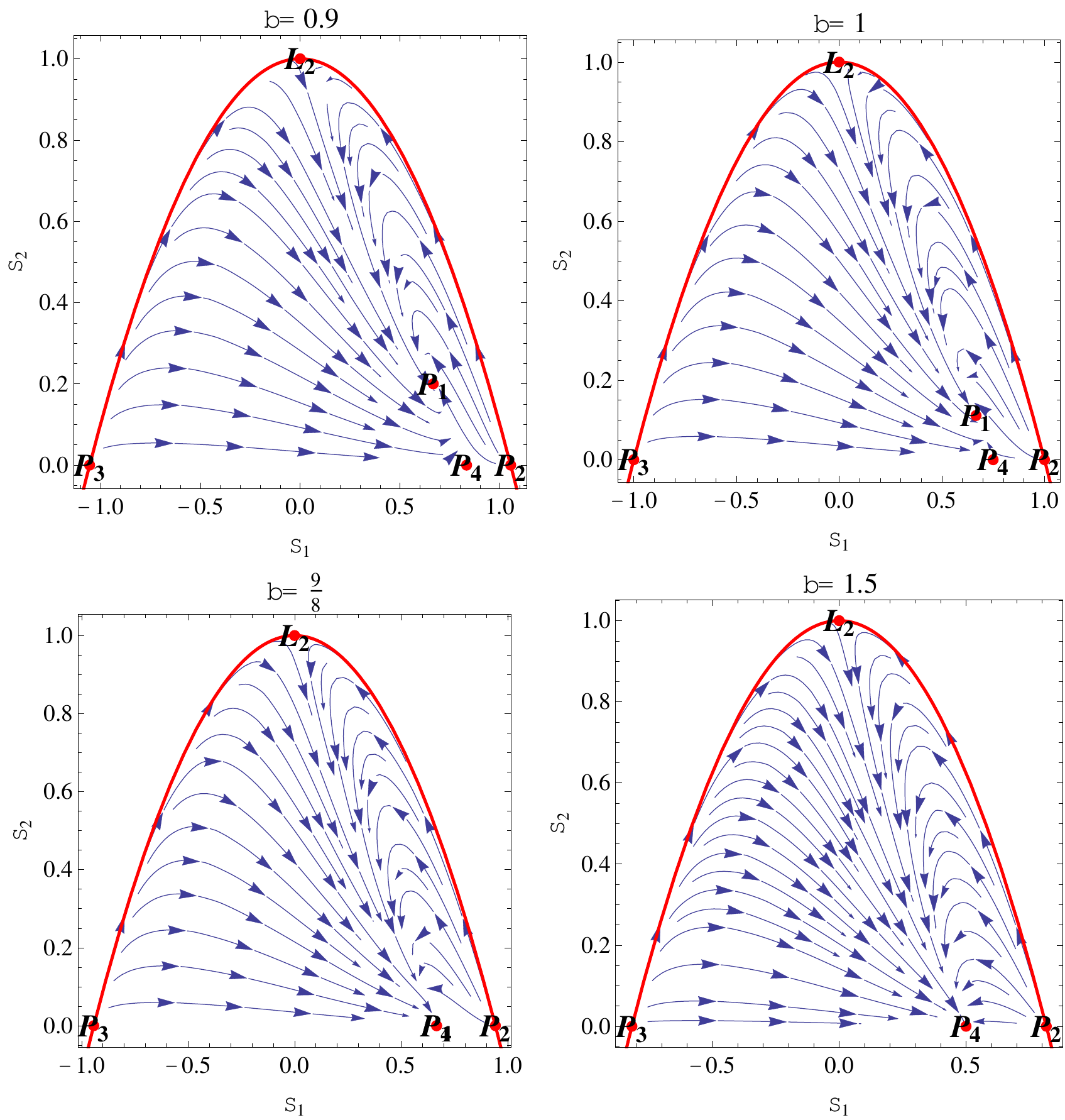} 
								\caption{\label{fig:XXXeq:23c} Streamlines  of the system \eqref{XXXeq:23} in the invariant set $Y=1$. The red (thick) line denotes the singular curve $L_1$. In (a), and (b), the endpoint of line $L_2$ is  a saddle. $P_1$ is the local attractor, $P_2$ and $P_3$ are local sources and $P_4$ is a saddle. In (c) when $\beta=\frac{9}{8}$,  $P_1$ and $P_4$ coincides and they are local attractors. In (d), $P_4$ is a local attractor (it is an hyperbolic saddle in the full phase space and there is a 1-parameter set of orbits that enters the interior of the phase space from it in the direction of the eigenvector associated to the positive eigenvalue for $\beta>0$).}
			\end{figure*} 
%%%%%%%%%%%%%%%%%%%%%%%%%%%%%%%

We have the additional relations 
\begin{align}
& r^2=\frac{P q^n}{\Sigma_2}\left(\frac{1-Y}{Y}\right)^{n+1}, \quad P=1-\beta\Sigma_1^2-\Sigma_2\label{eqra}\\
& \mu =\frac{1}{q^{n}} \left(\frac{Y}{1-Y}\right)^{n}, \quad  p= \frac{1}{q^{n}} \left(\frac{Y}{1-Y}\right)^{n+1},
\mathcal{M}(r):=\frac{m}{r}=\frac{{\Sigma_2}-(1-{\Sigma_1})^2}{2 {\Sigma_2}}. 
\end{align}
From these expressions we obtain the $m-r$ relation 
\begin{equation}
\label{RM2}
r^2=2 {q}^{n} \frac{\mathcal{M}}{D}, \quad m^2=2 {q}^{n} \frac{\mathcal{M}^3}{D}, \quad
D=\frac{\left({\Sigma_2}-(1-{\Sigma_1})^2\right)}{
   \left(1-\beta  {\Sigma_1}^2-{\Sigma_2}\right)} \left(\frac{1-Y}{Y}\right)^{-(n+1)}.
\end{equation}

%%%%%%%%%Figure%%%%%%%%%%%%%%%%%%%%%%%%%%%%%%
\begin{figure*}[!t]
	\includegraphics[width=0.9\textwidth]{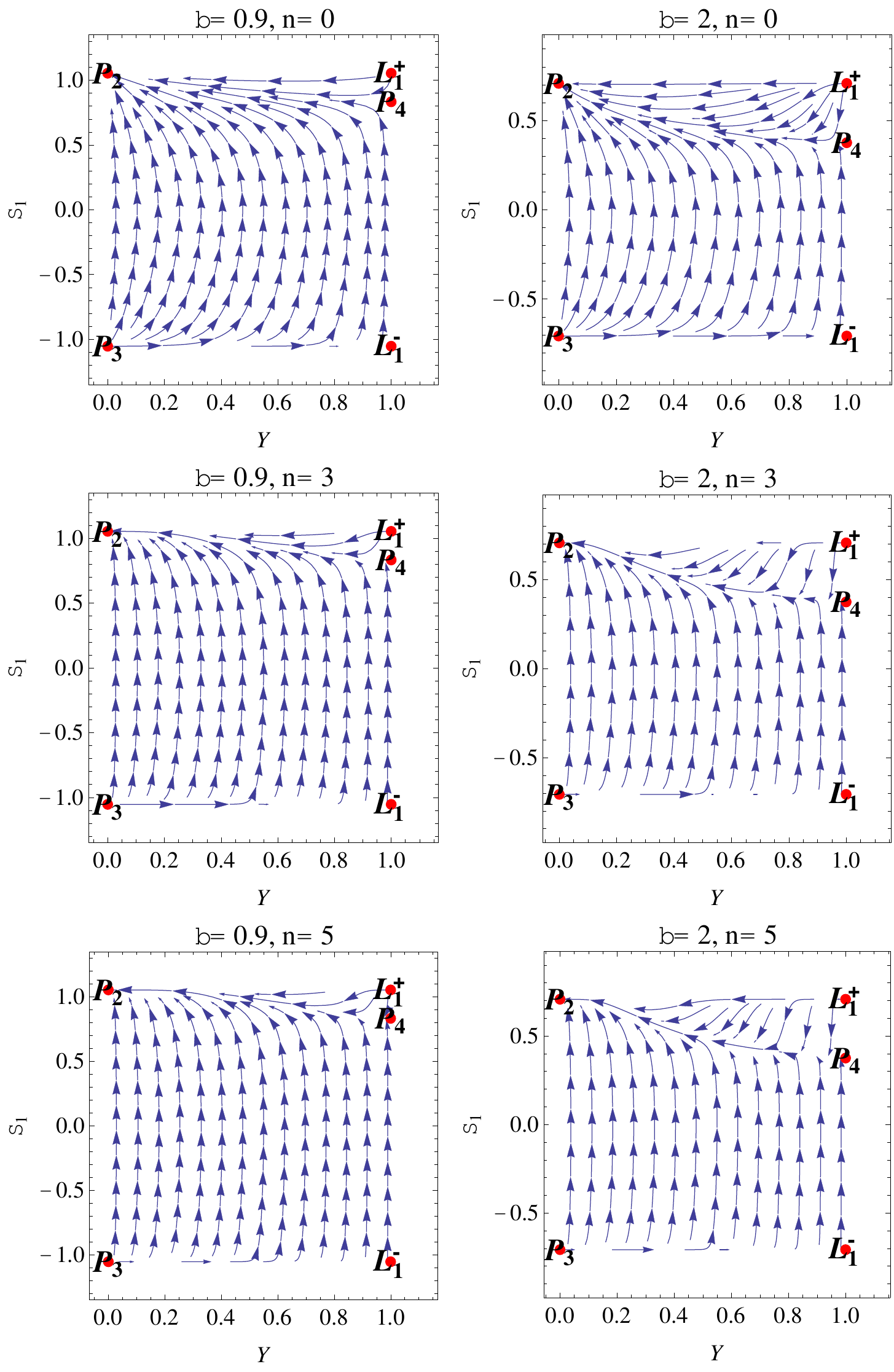} 
								\caption{\label{fig:XXXeq:23d} Streamlines  of the system \eqref{XXXeq:23} in the invariant set $\Sigma_2=0$. }
			\end{figure*}
%%%%%%%%%%%%%%%%%%%%%%%%%%%%%%%
Using the equations \eqref{RM2}, we can define the compact variables 
\begin{equation}
\label{RM2compact}
R_{\text{comp}}=\frac{q^{-\frac{n}{2}}r}{\sqrt{1+ q^{-n} r^2\left(1+{\mathcal{M}}^2\right)}}, \quad M_{\text{comp}}=\frac{q^{-\frac{n}{2}}\mathcal{M}r}{\sqrt{1+ q^{-n} r^2\left(1+{\mathcal{M}}^2\right)}}, 
\end{equation}
which depends only of the phase-space variables $(\Sigma_1,\Sigma_2,Y)$. Evaluating numerically the expressions $R_{\text{comp}}, M_{\text{comp}}$ at the orbits of the system \eqref{XXXeq:23}, we can see whether the resulting model leads to finite radius and finite mass. Futhermore, imposing regularity at the center, that is, as $\lambda\rightarrow -\infty$, and are extrapolated the conditions for relativistic stars as given by the Buchdahl inequalities \cite{Buchdahl:1959zz,hartle1978}, which in units where $8\pi G=1$ are expressed as
\begin{equation}
2 Y (3 \Sigma_1(1-\Sigma_1)-P)\geq P,  \Sigma_2 -(1-\Sigma_1)^2\geq \Sigma_2 P \left[\frac{1-Y}{Y}\right]^{1+n} \left[\frac{Y_c}{1-Y_c}\right]^{1+n},   \Sigma_2 \geq (1-2 \Sigma_1)^2,
\end{equation}
where $Y_c$ is related with the central pressure of the fluid by
$Y_c=
\frac{1}{p_c^{-\frac{1}{n+1}} q^{-\frac{n}{n+1}}+1}, \quad \left[\frac{Y_c}{1-Y_c}\right]^{1+n}= p_c q^n$.

%%%%%%%%%Figure%%%%%%%%%%%%%%%%%%%%%%%%%%%%%%
\begin{figure*}[!t]
	\includegraphics[width=1\textwidth]{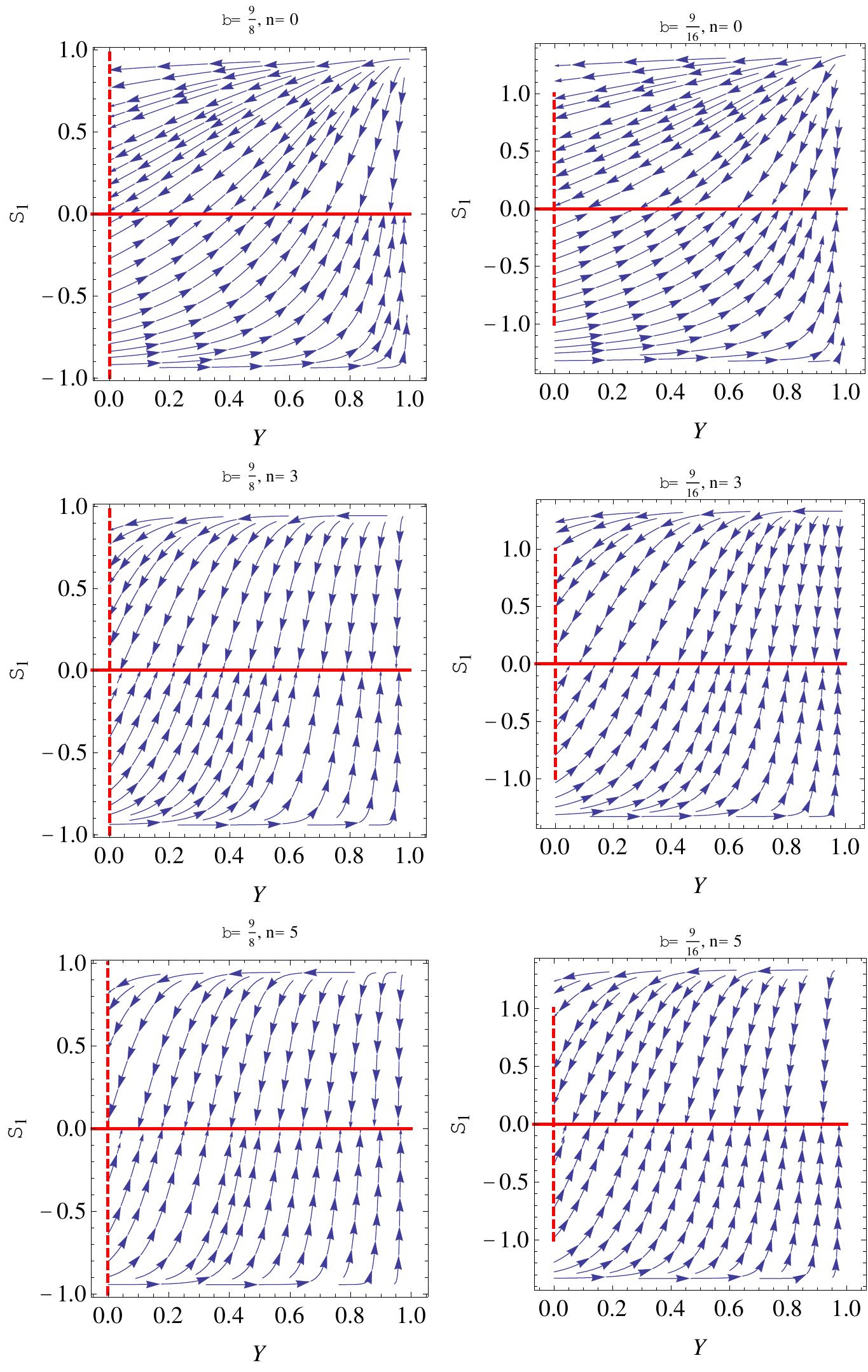} 
								\caption{\label{fig:XXXeq:23e} Streamlines  of the system \eqref{XXXeq:23} in the invariant set $\Sigma_2=1-\beta \Sigma_1^2$ projected on the plane $(Y,\Sigma_1)$. Line $L_1$ is denoted by a vertical dashed red line. Line $L_2$ is denoted by an horizontal red solid line. The orbits on the upper half $\Sigma_1>0$ are attracted by either $L_1$ or $L_2$. $P_2$ is there a local source. In the lower half $\Sigma_1<0$, the line $L_1$ is the local source and $L_2$ acts as the local sink. }
			\end{figure*}
%%%%%%%%%%%%%%%%%%%%%%%%%%%%%%%

\begin{figure*}[!t]
	\subfigure[\label{fig:Syst3104c} $\beta=0.9, n=6$.]{\includegraphics[width=0.4\textwidth]{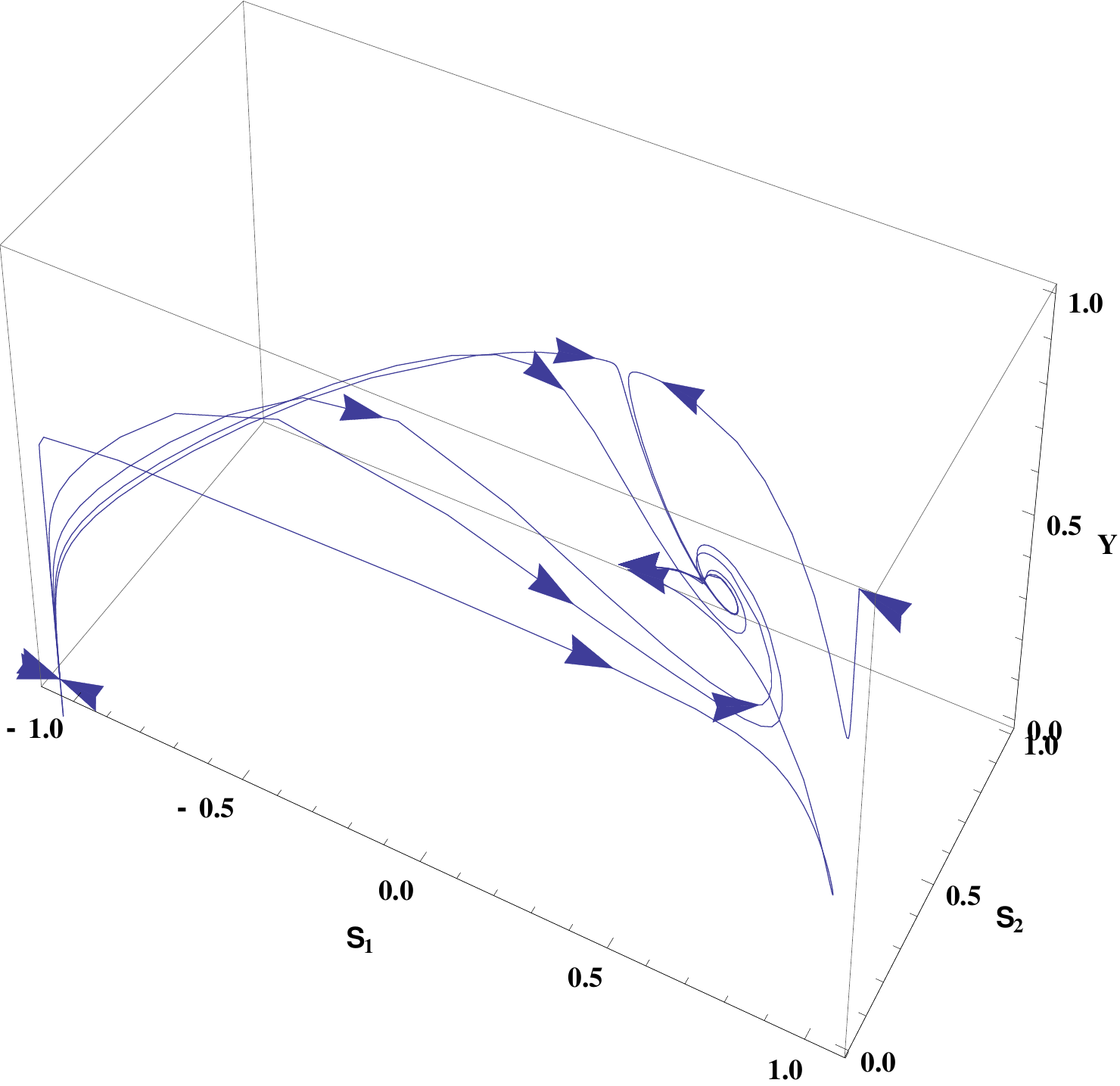}} \hspace{2cm} 	
	\subfigure[\label{fig:Syst3105c} $\beta=1, n=6$.]{\includegraphics[width=0.4\textwidth]{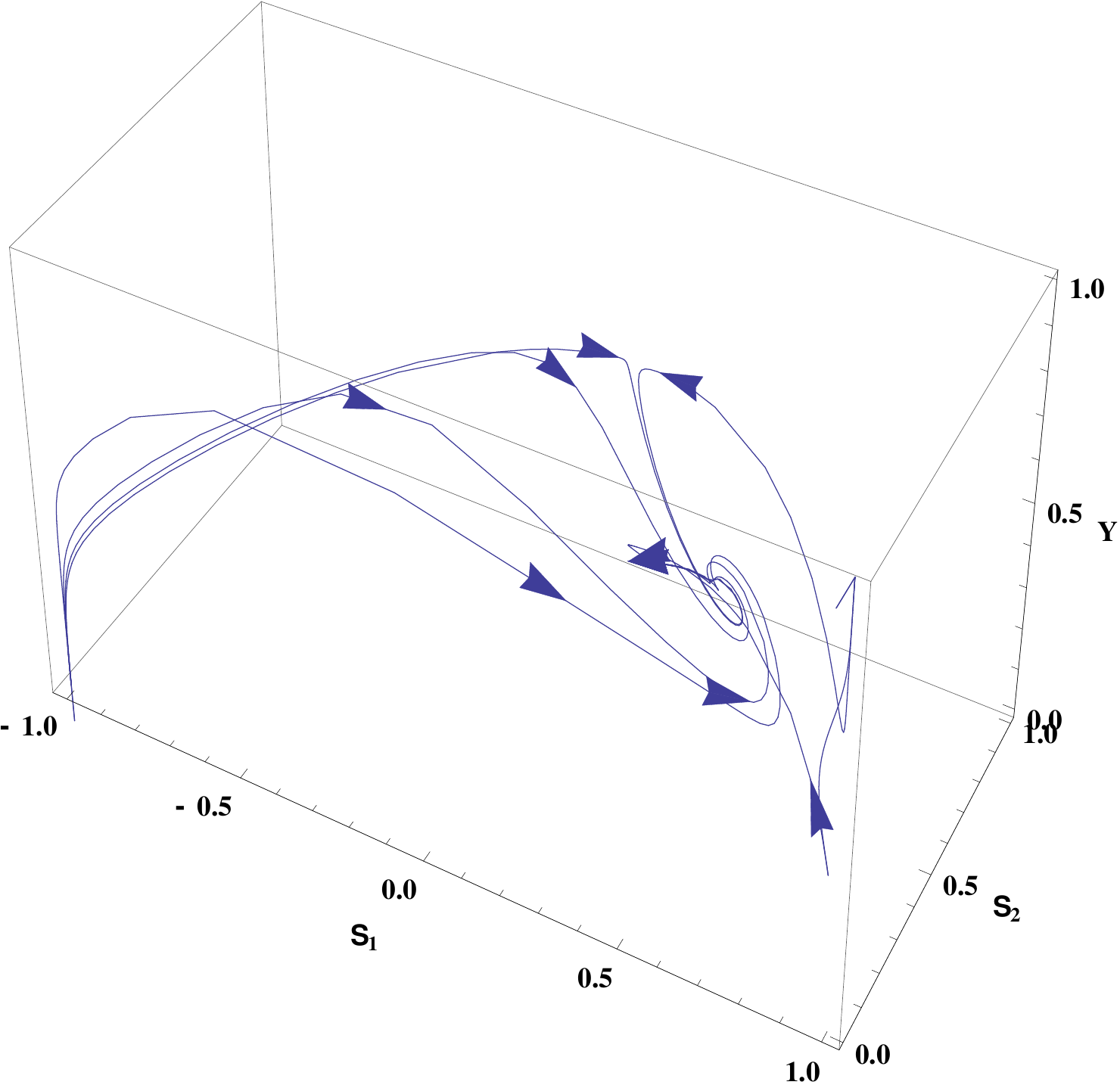}}\\
	\subfigure[\label{fig:Syst3104a} $\beta=0.9, n=6$. ]{\includegraphics[width=0.4\textwidth,height=0.35\textwidth]{FIGURE10C}} \hspace{2cm} 
	\subfigure[\label{fig:Syst3105a} $\beta=1, n=6$. ]{\includegraphics[width=0.4\textwidth, height=0.35\textwidth]{FIGURE10D}}\\
	\subfigure[\label{fig:Syst3104b}  Evolution of the compact variables 
$R_{\text{comp}},  M_{\text{comp}}$ for $\beta=0.9, n=6$.]{\includegraphics[width=0.4\textwidth,height=0.35\textwidth]{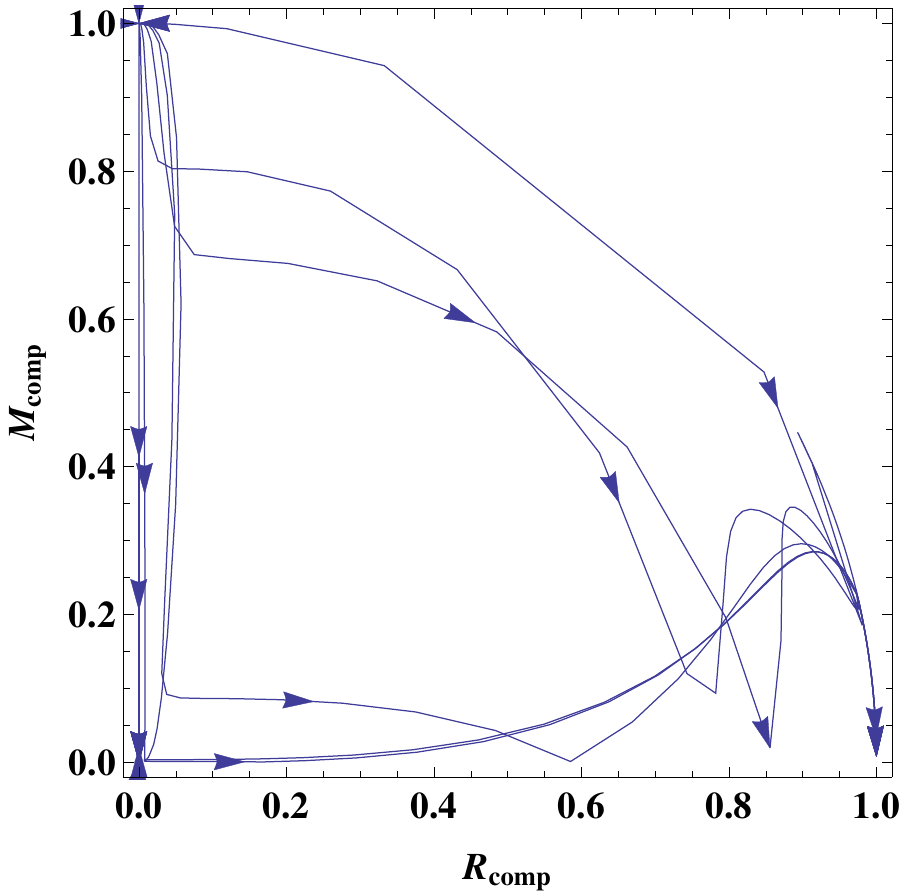}} \hspace{2cm}
	\subfigure[\label{fig:Syst3105b}  Evolution of the compact variables 
$R_{\text{comp}},  M_{\text{comp}}$ for $\beta=1, n=6$.]{\includegraphics[width=0.4\textwidth,height=0.35\textwidth]{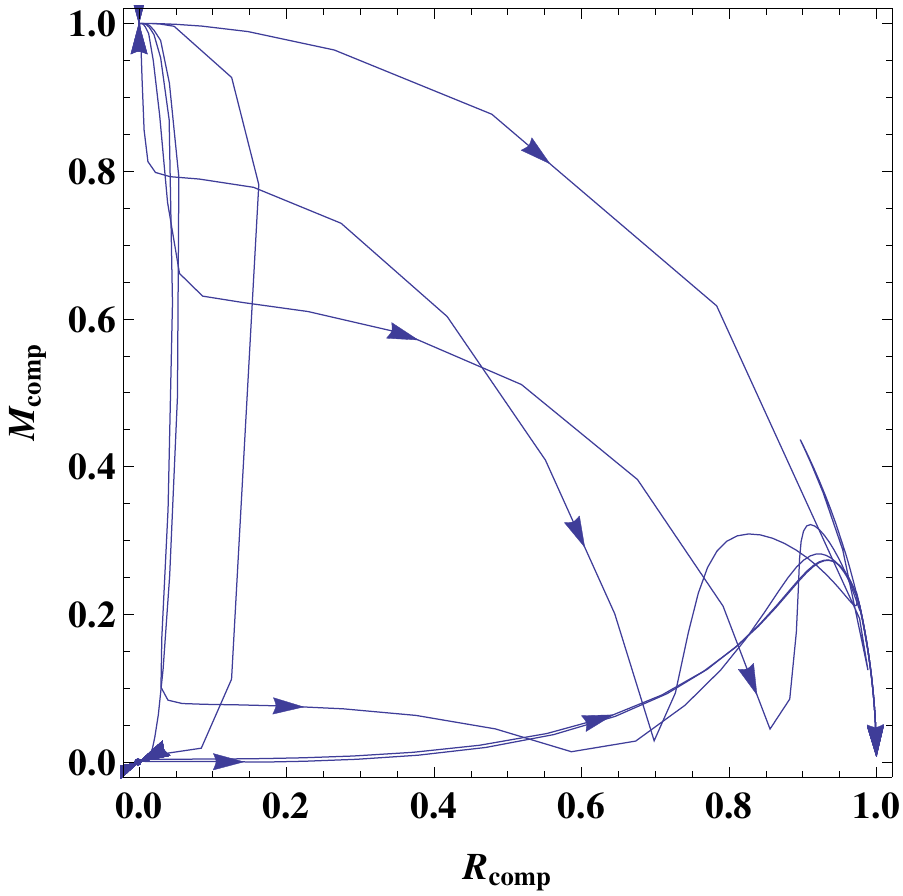}}
\caption{Evolution of system \eqref{XXXeq:23} for some choices of the parameters.}
	\end{figure*}

In the Figure \ref{fig:XXXeq:23} are presented some streamlines  of the system \eqref{XXXeq:23} in the invariant set $Y=0$. The red (thick) line denotes the singular curve $L_1$. The left branch $\Sigma_1<0$ is unstable and the right branch $\Sigma_1>0$ is stable. When an orbit ends on $L_1$, we can determine the total mass and total radius of the corresponding solution. Evaluating at $L_1$, where $\Sigma_2=1-\beta \Sigma_c^2$, we have $M/R=\Sigma_c(1-(1+\beta)\Sigma_c/2)/(1-\beta \Sigma_c^2)$. Thus, the solutions ending  on $L_1$ when $\Sigma_c>0$ has finite radii and masses (say for $0<\Sigma_c(1-(1+\beta)\Sigma_c/2)/(1-\beta \Sigma_c^2)<\text{const.}$).  The orbits ending on $L_1$ when $\Sigma_c=0$ or $\Sigma_c=2/(1+\beta)$, describe solutions with infinite radii and finite masses, or masses that approach infinity slower than $r$.  
In the Figure \eqref{fig:XXXeq:23c} are presented some streamlines  of the system \eqref{XXXeq:23} in the invariant set $Y=1$. The red (thick) line denotes the singular curve $L_1$. In figure \eqref{fig:XXXeq:23c} (a), and (b), the endpoint of line $L_2$ is  a saddle. $P_1$ is the local attractor, $P_2$ and $P_3$ are local sources and $P_4$ is a saddle. In  figure \eqref{fig:XXXeq:23c} (c) with $\beta=\frac{9}{8}$,  $P_1$ and $P_4$ coincides and they are local attractors. In (d), $P_4$ is a local attractor (it is an hyperbolic saddle in the full phase space and there is a 1-parameter set of orbits that enters the interior of the phase space from it in the direction of the eigenvector associated to the positive eigenvalue for $\beta>0$). Recall for $\beta<0$ the attractor is $P_4$. 
In Figure \ref{fig:XXXeq:23d} are displayed some streamlines  of the system \eqref{XXXeq:23} in the invariant set $\Sigma_2=0$. This set corresponds to the plane-symmetric boundary. For $\beta\neq 0$, the dynamics does not differs much of the general relativistic case where one branch of $L_1$ is a local source and the other branch is a local sink. The line $P_2$ is a local source. The self-similar plane symmetric solution $P_4$ is unstable (saddle).  The orbits coming from this point are  associated with a negative mass singularity. 
Finally, in the Figure \ref{fig:XXXeq:23e} are presented some streamlines  of the system \eqref{XXXeq:23} in the invariant set $\Sigma_2=1-\beta \Sigma_1^2$ (surface of pressure zero) projected on the plane $(Y,\Sigma_1)$. The line $L_1$ is denoted by a vertical dashed red line. The line $L_2$ is denoted by an horizontal red solid line. The orbits on the upper half $\Sigma_1>0$ are attracted by either $L_1$ or $L_2$. $P_2$ is there a local source. In the lower half $\Sigma_1<0$, the line $L_1$ is the local source and $L_2$ acts as the local sink.
Each point located at $L_2$ corresponds to the flat Minkowski solution written in spherical symmetric form. An orbit, associated to the eigenvalue $2 y_c, y_c>0$, enters the interior of the phase space from each point of $L_2$  parametrized by $y_c=p_c/(p_c+\mu_c)=q \mu_c^{1/n}/(1+q \mu_c^{1/n})$, where $p_c$ and $\mu_c$ are the central pressure and central density, respectively.

 In GR, the point $(\Sigma_1, \Sigma_2,Y)=\left(\frac{2}{3}, \frac{1}{9}, 1\right)$, corresponds to the so called self-similar Tolman solution discussed in \cite{Nilsson:2000zf}. In the Einstein -aether theory this solution is generalized to a 1-parameter set of solutions given by the equilibrium points $P_1: (\Sigma_1, \Sigma_2,Y)= \left(\frac{2}{3}, \frac{1}{9} (9-8\beta), 1\right)$. The single orbit that enters the interior of the phase space from $P_1$, associated to the positive eigenvalue of $P_1$ is called the Tolman orbit. The solutions having $Y=1$ corresponds to the limiting situation $\frac{p}{\mu}\rightarrow \infty.$ In the case of a linear equation of state $p=(\gamma-1)\mu$, this is consistent to $\gamma\rightarrow \infty$.

In the Figure \eqref{fig:Syst3104c} are shown some orbits of the system \eqref{XXXeq:23} on the plane $\Sigma_2=0$ for $\beta=0.9, n=6$. 
In the Figure \eqref{fig:Syst3105c} are shown some orbits of the system \eqref{XXXeq:23} on the plane $\Sigma_2=0$ for $\beta=0.1, n=6$. In the Figure \eqref{fig:Syst3104a} are shown some projections of the orbits of the system \eqref{XXXeq:23} on the plane $\Sigma_2=0$ for $\beta=0.9, n=6$. 
In the Figure \eqref{fig:Syst3105a} are shown some projections of the orbits of the system \eqref{XXXeq:23} on the plane $\Sigma_2=0$ for $\beta=1, n=6$. 
In the Figure \ref{fig:Syst3104b} it is presented how the compact variables 
$R_{\text{comp}}, M_{\text{comp}}$, defined by \eqref{RM2compact} varies along the orbits of \eqref{XXXeq:23}.
In the Figure \ref{fig:Syst3105b} it is presented how the compact variables 
$R_{\text{comp}}, M_{\text{comp}}$, defined by \eqref{RM2compact} varies along the orbits of \eqref{XXXeq:23}. The last two plots shows that the typical behavior for this choice of parameters is to obtain regions of infinite radius and finite mass, that is, $(R_{\text{comp}},M_{\text{comp}})=(1,0)$ is a local attractor in the $(R_{\text{comp}}, M_{\text{comp}})$ diagram.

\subsubsection{Equilibrium points at infinity.}\label{SECT_3.5.2}

We discussed before the phase space is bounded for $\beta>0$, such that in this section we cover the case $\beta<0$ which gives an unbounded phase space. $Y\in [0,1]$ is bounded but $\Sigma_1, \Sigma_2$ can take infinite values, so that we may define a coordinate system  in which the coordinates consist of $Y$, and the bounded variables \eqref{INFProy} and the radial derivative \eqref{INFProy2} to obtain the dynamical system
\begin{subequations}
\begin{align}
& \frac{d\Pi_1}{d\lambda}=Y \left({\Pi_1} \left(\sqrt{1-{\Pi_1}^2} ({\Pi_2}-2) \sqrt{\left| \beta \right|
   }-{\Pi_1} {\Pi_2}+{\Pi_1}\right)+{\Pi_2}-1\right)-\frac{1}{2} \left({\Pi_1}^2-1\right) ({\Pi_2}-1),\\
&	\frac{d\Pi_2}{d\lambda}=({\Pi_2}-1) {\Pi_2} \left(2 \sqrt{1-{\Pi_1}^2} Y \sqrt{\left| \beta \right| }-{\Pi_1}-2 {\Pi_1} Y\right),\\
& \frac{dY}{d\lambda}=-\frac{(1-Y) Y {\Pi_1}}{n+1},
	\end{align}
	\end{subequations}
defined in the compact phase space
\begin{equation}
\Big\{(Y, \Pi_1, \Pi_2)\in[0,1]\times \left[-1,1\right]\times [0,1]\Big\}.
\end{equation}
Now we discuss about the fixed points at infinity of the system \eqref{XXXeq:23} in the above compact coordinates:
\begin{enumerate}
\item[$I_1$]: $(\Pi_1, \Pi_2 ,Y)=(-1,0,0)$, with eigenvalues $\frac{1}{n+1},-1,-1$. It is a saddle.  

\item[$I_2$]: $(\Pi_1, \Pi_2 ,Y)=(1,0,0)$, with eigenvalues $-\frac{1}{n+1},1,1$. It is a saddle. 

\item[$I_3$]: $(\Pi_1, \Pi_2 ,Y)=\left(-1,0,1\right)$, with eigenvalues $-\frac{1}{n+1},-3,\infty  \sqrt{\left| \beta \right| }$. It is a saddle. 

\item[$I_4$]: $(\Pi_1, \Pi_2 ,Y)=\left(1,0,1\right)$, with eigenvalues $\frac{1}{n+1},3,\infty  \sqrt{\left| \beta \right| }$. It is a source. 

\item[$I_5$]: $(\Pi_1, \Pi_2 ,Y)=\left(-1,1,1\right)$, with eigenvalues $-\frac{1}{n+1},3,\infty  \sqrt{\left| \beta \right| }$. It is a saddle. 

\item[$I_6$]: $(\Pi_1, \Pi_2 ,Y)=\left(1,1,1\right)$, with eigenvalues $\frac{1}{n+1},-3,\infty  \sqrt{\left| \beta \right| }$. It is a saddle.
\end{enumerate}

There are not physical relevant equilibrium points at infinity, apart of the point $I_4$, that has $\Sigma_1\rightarrow \infty$, $\Sigma_2$ finite and $Y=1$, which is a source (it exists only for $\beta<0$). 
This solution corresponds to $y$ finite and $x\rightarrow y$. That is, $\frac{d p}{d\rho}=\frac{1}{\mu+p}, \frac{p}{p+\mu}=1,$ where $\rho$ is the radial coordinate. Then  $\frac{dp}{d\rho}=\frac{1}{p}\implies p(\rho )= \sqrt{ p_{c}^2+2 \rho },$ where $p_c$ is the central pressure (at $\rho=0$).  As $\rho\rightarrow \infty$ the pressure is increasing as the solutions move away this point.

\subsubsection{Dynamical systems analysis based on the Newtonian homology invariants}

As in section \ref{SECTUVY} we can implement the $(U,V,Y)$- formulation introducing the variables 
\begin{align}
U=\frac{\mu  r^2}{2 \mathcal{M}+\mu  r^2}, V=\frac{\mu  \mathcal{M}}{\mu \mathcal{M}+p}, Y=\frac{p}{p+\mu}
\end{align}
where $\mathcal{M}$ is the Misner-Sharp mass defined by $\mathcal{M}(r)=\frac{m}{r}$, $r(\lambda)$ is the usual Schwarzschild radial parameter introduced in \eqref{XXXmet2}, $\mu$ and $p$ are the density and pressure of the scalar field with a polytropic equation of state $p=q \mu^\Gamma, \Gamma=1+\frac{1}{n}, n>0, q\geq 0$. Additionally, we introduce the radial derivative 
\begin{equation}
\frac{d r}{r d{\lambda}}=(1 - U)  (1 - Y)  (1 - V - Y - V Y) .
\end{equation}

The relation between the variables ${U, V}$ and $\Sigma_1, \Sigma_2$ are 
\begin{subequations}
\begin{align}
&U=\frac{(Y-1) \left(\beta  {\Sigma_1}^2+{\Sigma_2}-1\right)}{-\beta  {\Sigma_1}^2-{\Sigma_2}+Y ({\Sigma_1} ((\beta -1) {\Sigma_1}+2)+2 ({\Sigma_2}-1))+1},\\
&V=	\frac{(Y-1) \left(({\Sigma_1}-1)^2-{\Sigma_2}\right)}{({\Sigma_1}-1)^2 (Y-1)+{\Sigma_2} (Y+1)}.
\end{align}
\end{subequations}
We will see that there are several equilibrium points in this formulation that  cannot be studied with the $\{\Sigma_1, \Sigma_2, Y\}$-formulation.

\paragraph{Relativistic equations ($\beta=1$).}

In this example we obtain the dynamical system
\begin{subequations}
\label{EQS:281}
\begin{align}
& \frac{dU}{d\lambda}=\frac{(U-1) U }{n+1}\Big\{n \left((4 U-3) (V+1) Y^2+Y (2 U (V-4)-V+6)-5 U V+4 U+4 V-3\right)\nonumber \\
& +(4 U-3) (Y-1) (V Y+V+Y-1)\Big\},\\
& \frac{dV}{d\lambda}=\frac{(V-1) V}{n+1} \Big\{-n (2 U-1) (Y-1) (V
   Y+V+Y-1) \nonumber \\
& +U \left(V (3-2 Y (Y+1))-2 (Y-1)^2\right)+(Y-1) (V (Y+2)+Y-1)\Big\},\\
& \frac{dY}{d\lambda}=\frac{V (Y-1) Y (U (2 Y-1)-Y+1)}{n+1},
\end{align}
\end{subequations}
defined in the phase space
\begin{equation}
\label{phase-space:2.74}
\Big\{(U,V,Y)\in\mathbb{R}^3: 0\leq U \leq 1, 0\leq Y\leq 1, 0\leq V\leq \frac{1-Y}{1+Y}\Big\}.
\end{equation}

The static surface 
\begin{equation}
\left\{ (V,Y)\in\mathbb{R}^2, 1-Y-V-V Y=0\right\}
\end{equation} is an invariant set, due to the function $\mathcal{S}$ defined by
\begin{equation}
\mathcal{S}(V,Y):= 1-Y-V-V Y,
\end{equation}
satisfies 
\begin{align}
&\frac{d\mathcal{S}}{d\lambda}=\mathcal{S} \left[\frac{V^2 \left(-n (2 U-1) \left(Y^2-1\right)+U (3-2 Y (Y+1))+Y^2+Y-2\right)}{n+1}\right. \nonumber\\
&\left. \;\;\;\;\;\;\;\;\; \;\;\; +\frac{V \left((n+2) (2 U-1) Y^2-2 (n+1) U+n-U
   Y+Y+1\right)}{n+1}\right]. 
\end{align}
Additionally, any of the sets $U=0, U=1, V=0,  V=1, Y=0, Y=1$, defines an invariant set. 
%%%%%%%%%Figure%%%%%%%%%%%%%%%%%%%%%%%%%%%%%%
\begin{figure*}[!t]
	\subfigure[\label{fig:Syst398a}  $n=4$.]{\includegraphics[width=0.3\textwidth]{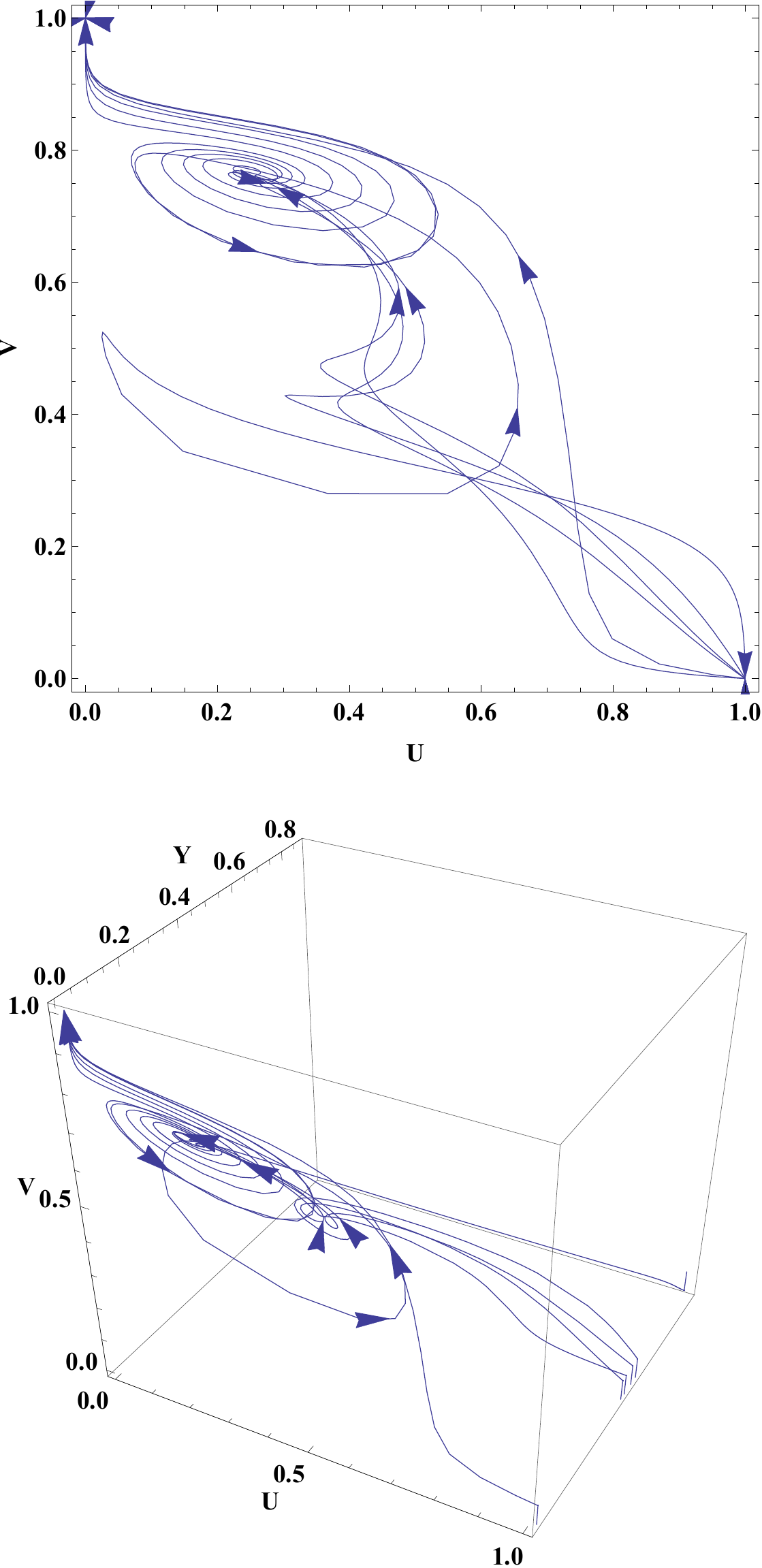}}	
	\subfigure[\label{fig:Syst398b}  $n=5$.]{\includegraphics[width=0.3\textwidth]{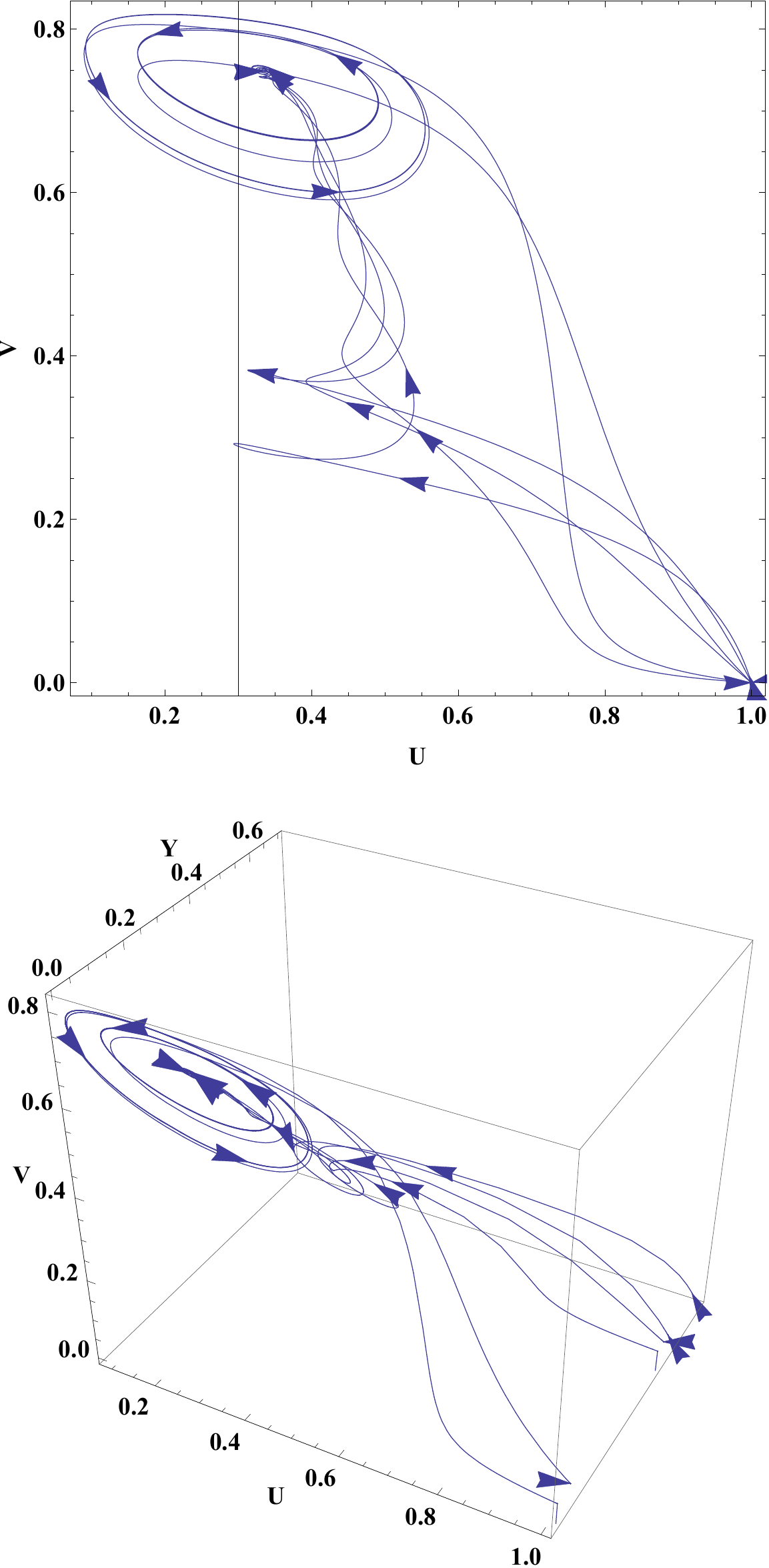}}  
	\subfigure[\label{fig:Syst398c}  $n=6$.]{\includegraphics[width=0.3\textwidth]{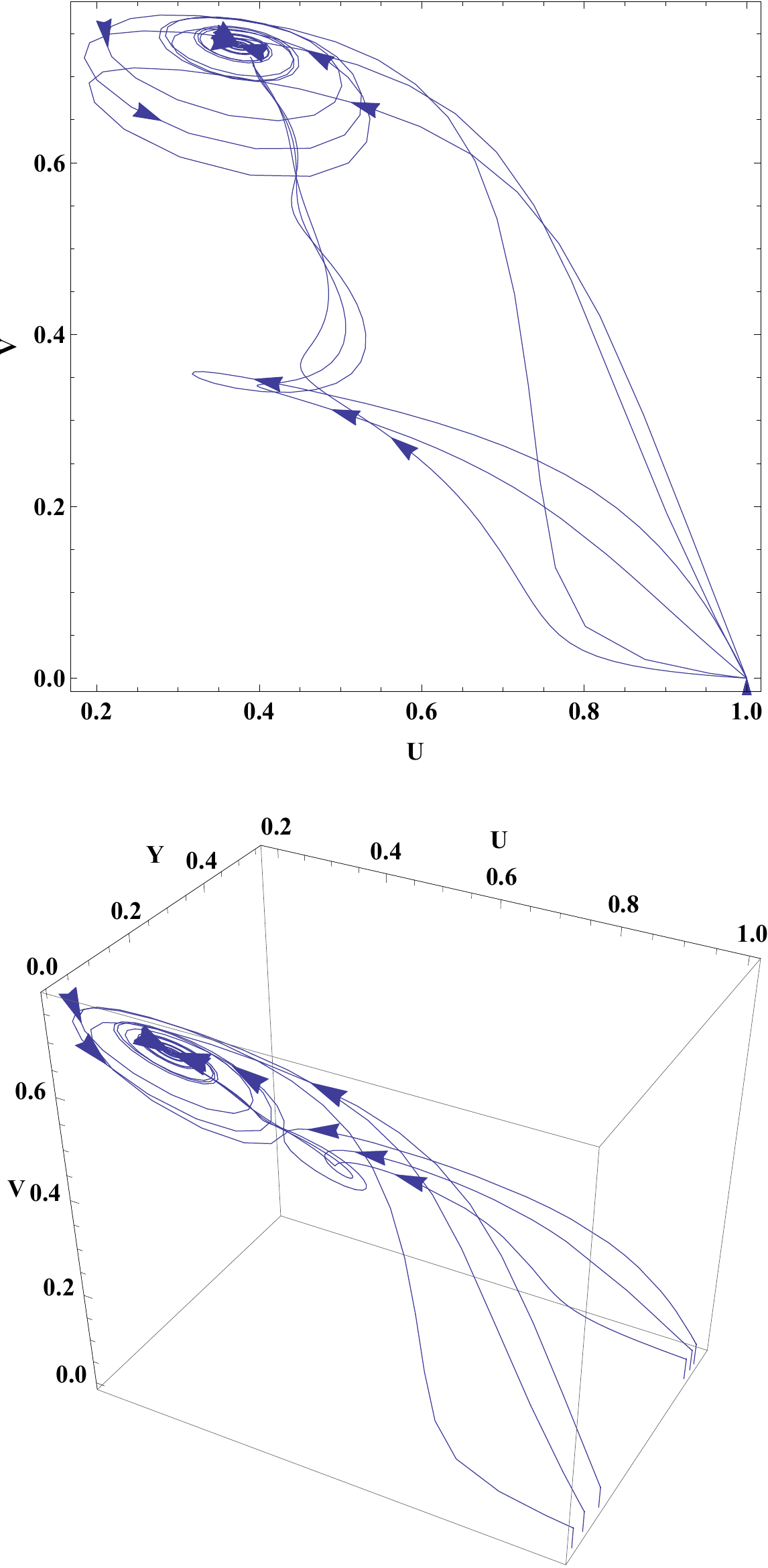}}
								\caption{\label{fig:Syst398} Some orbits of the system \eqref{EQS:281} for different choices of the parameter $n$.}
			\end{figure*}
%%%%%%%%%%%%%%%%%%%%%%%%%%%%%%%	
The equilibrium points of the system \eqref{EQS:281} are  (they were found in the reference \cite{Nilsson:2000zg}, we use here the same notation): 
\begin{enumerate}

\item $L_1: (U,V,Y)=(1, 0, Y_0)$. Exists for $-1\leq Y_0\leq 1$. The eigenvalues are $0,(Y_0-1)^2,(Y_0-1)^2$. It is a source. 

\item $L_2: (U,V,Y)=\left( \frac{3}{4},  0, Y_0\right)$. Exists for $-1\leq Y_0\leq 1$. The eigenvalues are $0,-\frac{3}{4} (Y_0-1)^2,\frac{1}{2} (Y_0-1)^2$. It is nonhyperbolic, behaves as saddle. 

\item $L_3: (U,V,Y)=(U_0,  0, 1)$.  Exists for $0\leq U_0\leq 1$. The eigenvalues are $0,0,0$. It is onhyperbolic. 

\item  $L_4: (U,V,Y)=(0, 0 ,Y_0)$. Exists for $-1\leq Y_0\leq 1$. The eigenvalues are $0,-(Y_0-1)^2,3 (Y_0-1)^2$. It is nonhyperbolic, behaves as saddle. 

\item $L_5: (U,V,Y)=(0, V_0, 1)$. Exists for $-1\leq V_0\leq 1$. The eigenvalues are $0,0,0$. It is nonhyperbolic. 

\item $L_6: (U,V,Y)=(U_0, 1, 0)$. Exists for $n=0$. The eigenvalues are $0,-(1-U_0), -(1-U_0)$. This line is normally hyperbolic and due to the non-zero eigenvalue is negative it is a sink.

\item $P_1: (U,V,Y)=\left(0, \frac{n+1}{n+2},  0 \right)$. Exists for $n\geq 0$. The eigenvalues are $-\frac{1}{n+2},\frac{1}{n+2},-\frac{n-3}{n+2}$.  Saddle with a 2D unstable manifold for $0<n<3$. Saddle with a 2D stable manifold for $n>3$.

\item $P_2: (U,V,Y)=(0,  1,  0)$. Always exists. The eigenvalues are $-\frac{1}{n+1},-\frac{1}{n+1}, -\frac{n}{n+1}$. It is a sink. 

\item $P_3: (U,V,Y)=(1, 1, 1)$. Always exists. The eigenvalues are  $-\frac{1}{n+1},\frac{1}{n+1},\frac{n}{n+1}$. It is a saddle with a 2D unstable manifold. 

\item $P_4: (U,V,Y)=\left(\frac{n-3}{2 (n-2)}, \frac{2 (n+1)}{3 n+1},   0\right)$. Exists for $n>3$. The eigenvalues are\\ $-\frac{n-1}{(n-2) (3 n+1)},-\frac{(n-1) \left(n-5+ \sqrt{1-n (7 n-22)}\right)}{4(n-2) (3 n+1)},-\frac{(n-1) \left(n-5- \sqrt{1-n (7 n-22)}\right)}{4 (n-2) (3 n+1)}$. Stable  for $n>5$ (two complex eigenvalues with negative real parts). 
	
\item $P_5: (U,V,Y)=(1,  1,  0)$. Always exists. The eigenvalues are $0,0,0$. It is nonhyperbolic. 
\end{enumerate}

\paragraph{The Newtonian subset.}
We now consider the ``Newtonian subset'' $Y=0$, which is of great importance for relativistic stars. The stability analysis below is restricted to the dynamics onto the invariant set $Y=0$, and the stability along the $Y$-axis is not taken into account. 
In this subset the dynamical system equations are reduced to 
\begin{subequations}
\label{XXXeq:278}
\begin{align}
& \frac{dU}{d\lambda}=\frac{(U-1) U (V (n (4-5 U)-4 U+3)+(n+1) (4 U-3))}{n+1},\\
& \frac{dV}{d\lambda}=\frac{(V-1) V (n (2 U-1) (V-1)+U (3 V-2)-2 V+1)}{n+1}.
\end{align}
\end{subequations}

%%%%%%%%%Figure%%%%%%%%%%%%%%%%%%%%%%%%%%%%%%
\begin{figure*}[!t]
	\includegraphics[width=1\textwidth]{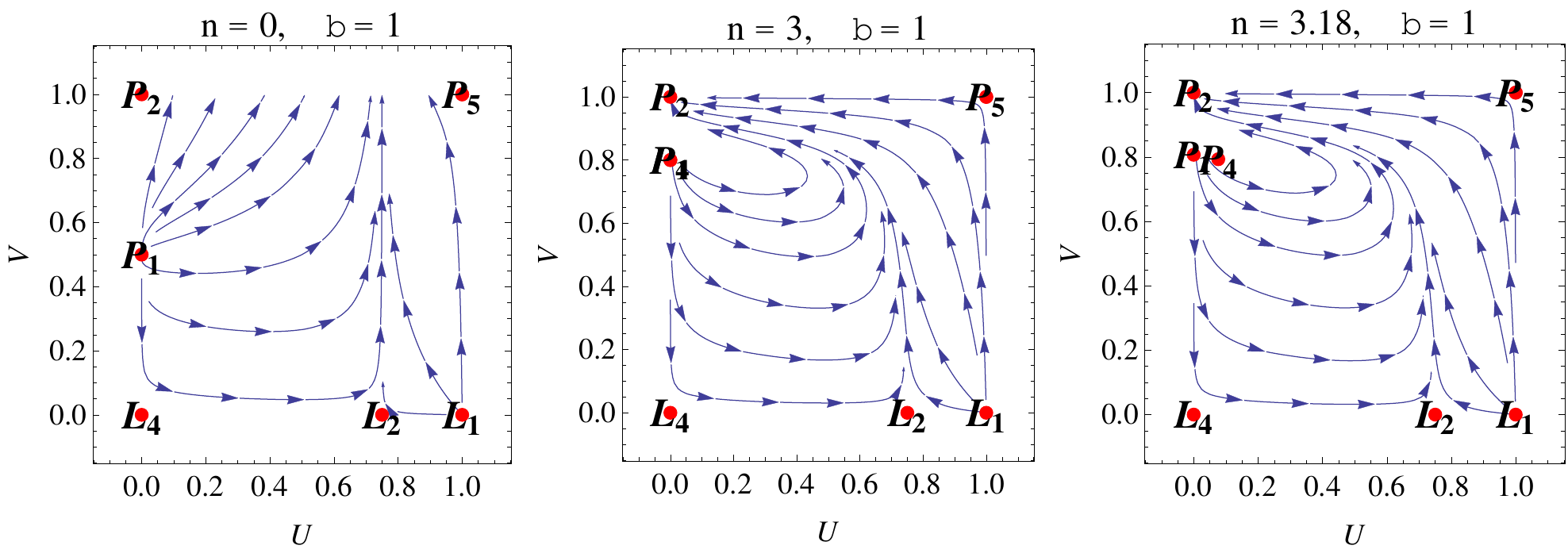} 
		\includegraphics[width=1\textwidth]{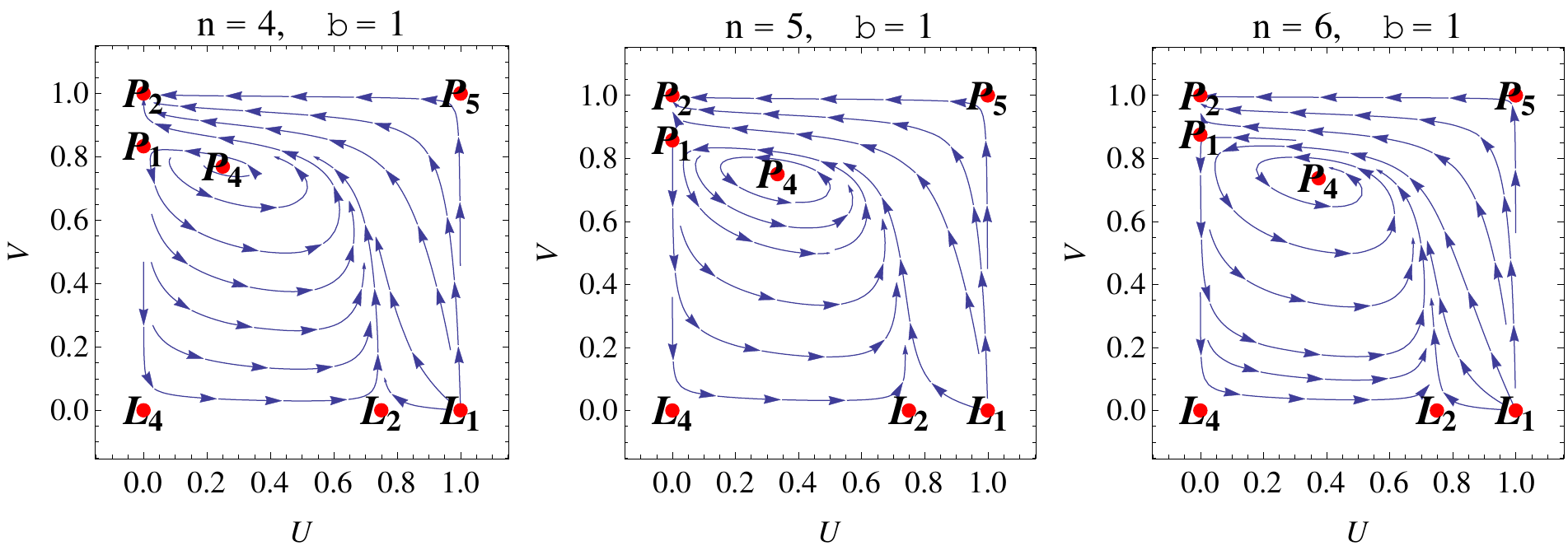} 
								\caption{\label{fig:XXXeq:278} Streamlines  of the system \eqref{XXXeq:278}.}
			\end{figure*}
%%%%%%%%%%%%%%%%%%%%%%%%%%%%%%%

%%%%%%%%%Figure%%%%%%%%%%%%%%%%%%%%%%%%%%%%%%
\begin{figure*}[!t]
	\includegraphics[width=1\textwidth]{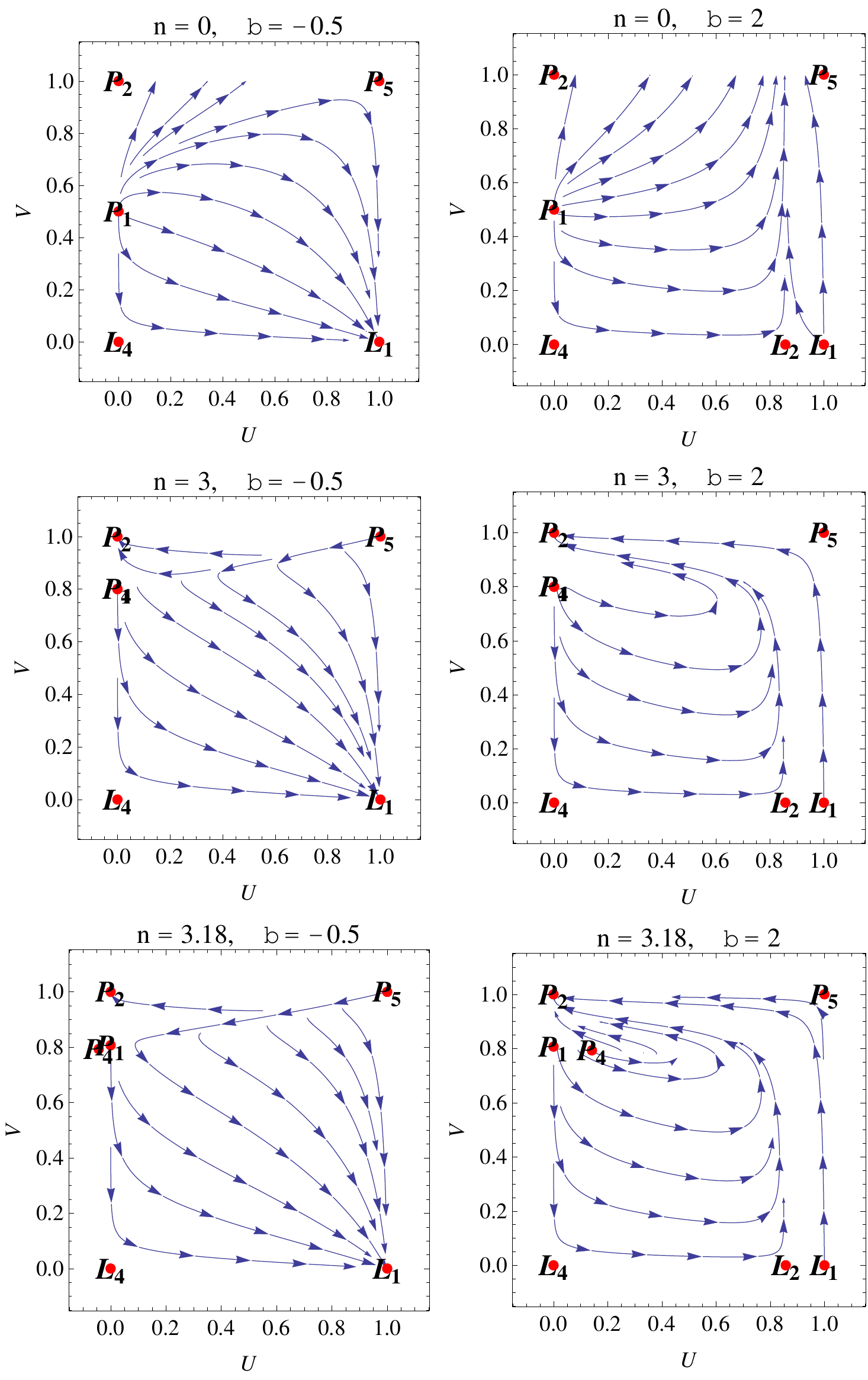} 
								\caption{\label{fig:XXXeq:392A} Streamlines  of the system \eqref{XXXeq:392} for the choices $n=0,  n=3, n=3.18$ and  $\beta=-0.5, 1 , 2$.}
			\end{figure*}
%%%%%%%%%%%%%%%%%%%%%%%%%%%%%%%

%%%%%%%%%Figure%%%%%%%%%%%%%%%%%%%%%%%%%%%%%%
\begin{figure*}[!t]
	\includegraphics[width=1\textwidth]{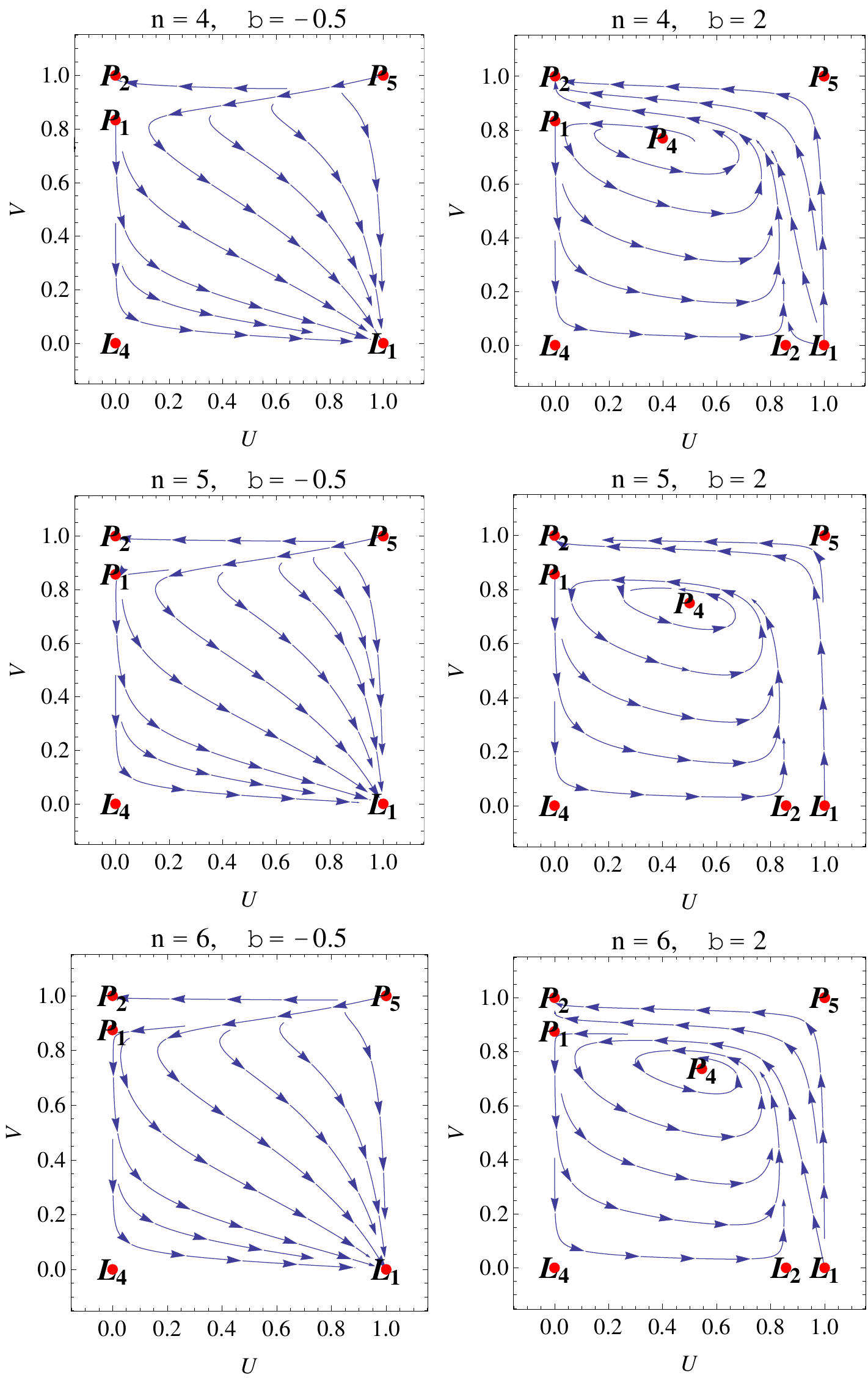} 
								\caption{\label{fig:XXXeq:392B} Streamlines  of the system \eqref{XXXeq:392} for the choices $n=4, n=5, n=6$ and  $\beta=-0.5, 1 , 2$.}
			\end{figure*}
%%%%%%%%%%%%%%%%%%%%%%%%%%%%%%%

The equilibrium points in the Newtonian subset are 
\begin{enumerate}

\item $L_1(0): (U,V)=(1, 0)$, with eigenvalues $1, 1$. Source. 

\item $L_2(0): (U,V)=\left( \frac{3}{4},  0 \right)$, with eigenvalues $-\frac{3}{4},\frac{1}{2}$. Saddle.

\item  $L_4(0): (U,V)=(0, 0)$, with eigenvalues $3, -1$. Saddle.

\item $L_6: (U,V)=(U_0, 1),  U_0\in[0,1]$, with eigenvalues $0, -(1-U_0)$. Exists for $n=0$. This line is normally hyperbolic and due to the non-zero eigenvalue is negative it is a sink.

\item $P_1: (U,V)=\left(0, \frac{n+1}{n+2}\right)$, with eigenvalues $\frac{1}{n+2},-\frac{n-3}{n+2}$. Source for $0\leq n < 3$, nonhyperbolic for $n=3$, saddle when $n>3$.

\item $P_2: (U,V)=(0,  1)$, with eigenvalues $-\frac{1}{n+1},-\frac{n}{n+1}$. Nonhyperbolic for $n=0$, sink for $n>0$. 

\item $P_4: (U,V)=\left(\frac{n-3}{2 (n-2)}, \frac{2 (n+1)}{3 n+1}\right)$, with eigenvalues\\ $-\frac{(n-1) \left(n-5+ \sqrt{1-7 n^2+22 n}\right)}{4 (n-2) (3 n+1)},-\frac{(n-1) \left(n-5- \sqrt{1-7 n^2+22 n-1}\right)}{4 (n-2) (3 n+1)}$. Exists for $n>3$. Unstable node for $3<n\leq \frac{1}{7} \left(11+8 \sqrt{2}\right)\approx 3.18767$. Unstable spiral for $\frac{1}{7} \left(11+8 \sqrt{2}\right)<n<5$. It is a center for $n=5$. Stable spiral for $n>5$.
	
\item $P_5: (U,V)=(1,  1)$ with eigenvalues $0,0$. Nonhyperbolic. 

\end{enumerate}

We deduce the relations
\begin{subequations}
\begin{align}
& \mu =\frac{1}{q^{n}} \left(\frac{Y}{1-Y}\right)^{n}, \quad  p= \frac{1}{q^{n}} \left(\frac{Y}{1-Y}\right)^{n+1},\\
& r^2=\frac{2 q^n  V U}{(1-V)(1-U)}\left(\frac{Y}{1-Y}\right)^{1-n}, \quad \mathcal{M}=\frac{Y V}{(1-V)(1-Y)}.
\end{align}
\end{subequations}
From  these expressions we obtain the $m-r$-relation 
\begin{equation}
\label{RM2A} 
r^2=2 {q}^{n} \frac{\mathcal{M}}{D},\quad 
m^2=2 {q}^{n} \frac{\mathcal{M}^3}{D}, \quad D=\frac{(1-U) }{U}\left(\frac{Y}{1-Y}\right)^n.
\end{equation}
Using the equations \eqref{RM2A} , we can define the compact variables 
\begin{equation}
\label{RM2Acompact}
R_{\text{comp}}=\frac{q^{-\frac{n}{2}}r}{\sqrt{1+ q^{-n} r^2\left(1+{\mathcal{M}}^2\right)}}, \quad M_{\text{comp}}=\frac{q^{-\frac{n}{2}}\mathcal{M}r}{\sqrt{1+ q^{-n} r^2\left(1+{\mathcal{M}}^2\right)}}, 
\end{equation}
which depends only of the phase-space variables $(U,V,Y)$. Evaluating numerically the expressions $R_{\text{comp}}, M_{\text{comp}}$ at the orbits of the system \eqref{EQS:281}, we can see whether the resulting model leads to finite radius and finite mass.

In the Figure \ref{fig:Syst398} are presented some orbits of the system \eqref{EQS:281} for the choices  $n=4, n=5, n=6$ for the relativistic case ($\beta=1$).  

In the Figure \ref{fig:XXXeq:278} it is depicted the typical behavior of the streamlines  of the system \eqref{XXXeq:278}, for the choices
(a) $n=0$, (b) $0< n\leq 3$, (c) $3<n \leq \frac{1}{7}\left(11+8 \sqrt{2}\right)\approx 3.18767$, (d) $\frac{1}{7}\left(11+8 \sqrt{2}\right)<n<5$, (e) $n=5$, and (f) $n>5$ in accordance with results summarized in the list above.	

The system \eqref{EQS:281} was well studied in the reference \cite{Nilsson:2000zg}. According to the results of \cite{Nilsson:2000zg}, for $n\geq 0$ exists regular Newtonian solutions, and when $0\leq n<5$ it is possible to obtain solutions that represents star models with finite radii and mass, while for $n\geq 5$ are obtained infinite regular models (for $n=5$ there are solutions with infinite radius but finite mass).   

The choice $n=0$ corresponds to relativistic incompressible fluid models. The set $U=\frac{3}{4}$ contains all regular solutions. Due to $L_6$ is the attractor line on this case, the solution (Tolman orbit) is the straight line that starts in $L_2$ and ends in the line $L_6$ at $U=\frac{3}{4}, V=1$.

For the choice $0< n\leq 3$, the point $P_2$ is a hyperbolic sink. On the other hand $(V (-1 + Y) Y (1 - Y + U (-1 + 2 Y)))/(1 + n)<0$ for any interior point of the phase space, that means that $Y$ is monotonic decreasing, therefore, any orbit approach the set $Y=0$. This result, combined with the existence of a monotonic function in the subset $Y=0$ implies that $P_2$ is a global sink, which means that the radii and masses of a regular models are finite for $0<n\leq 3$.  

For $3<n<5$, the point $P_2$ is the only sink, but not a global one. There is a 1-parameter set of orbits that ends at $P_1$, acting as a separatrix surface in the state space. The corresponding solutions has finite masses but infinite radii. The boundary of the separatrix surface associated with $P_1$ is described by an orbit in the invariant set $U=0$; an orbit that connects $P_1$ to $P_4$ in the invariant set $Y=0$; and a relativistic orbit that ends at $P_4$ corresponding to a solution with infinite mass and infinite radius. If there is a regular separatrix surface intersecting the separatrix associated to $P_1$ or its boundary orbit at an interior point, the solutions will end at $P_1$ or at $P_4$. That is, the solutions do not end necessarily at $P_2$ (see the numerical elaborations performed in \cite{Nilsson:2000zg}). 

When $n=5$, the separatrix surface that ends at the equilibrium point $P_1$ completely encloses the regular subset of orbits, that prevents any regular orbit to end at $P_2$. The Newtonian regular orbit ends in $P_1$, while all general relativistic regular orbits tends to $P_4$ (which is a center for $n=5$) and to the closed solutions surrounding $P_4$.   

For $n>5$, the separatrix surface associated to $P_1$ encloses all regular solutions (including Newtonian orbit), and all regular solutions end at the hyperbolic sink $P_4$. Therefore, Newtonian and relativistic polytropic models have infinite radii and mass.

\paragraph{Einstein-Aether modification.}

The dynamical system becomes 
\begin{subequations}
\label{GEN-syst}
\begin{align}
& {\frac{dU}{d\lambda}=\frac{1}{(1+n) V Y (-1+\beta ) \beta }(-1+U) U (-1+V+Y+V Y) }\nonumber \\
&
 {((1-Y) \beta  (-1-n+V+Y+G (-1-n+V+(1+n) (1+V) Y)+Y (n-3 V-3 n V+4 (1+n) V \beta ))+}\nonumber \\
&
 {U \left((1+G) (1+n) (-1+Y)^2 \beta +V ((-1-G) \beta +\right.}\nonumber \\
&
 {\left.\left.\left.Y \left(1+n+(3+2 n-G n) \beta -4 (1+n) \beta ^2\right)+(1+n) Y^2 (2+\beta  (-9+G+8 \beta ))\right)\right)\right),}
\\
& {\frac{dV}{d\lambda}=-\frac{1}{(1+n) Y (-1+\beta ) \beta }(-1+V) (-1+V+Y+V Y) }\nonumber \\
&
 {((1-Y) \beta  (-1+Y+G (-1+Y+V Y+n (-1+V+Y+V Y))+V Y (-1+2 \beta )+}\nonumber \\
&
 {n (-1+V+Y-V Y+2 V Y \beta ))+U \left((1+G) (1+n) (-1+Y)^2 \beta +\right.}\nonumber \\
&
 {\left.\left.V \left((-1-G) n \beta +Y \left(1+n+(-G+n) \beta -2 (1+n) \beta ^2\right)+(1+n) Y^2 (2+\beta  (-7+G+6 \beta ))\right)\right)\right),}\\
&{\frac{dY}{d\lambda}=-\frac{(1+G) (-1+U) (-1+Y)^2 (-1+V+Y+V Y)}{(1+n) (-1+\beta )}}. \label{EQQQ277}
\end{align}
\end{subequations}
where \begin{align}
& G:= -\sqrt{\frac{(1-Y) (1-(2 \beta-1)  V Y-V-Y)-U \left(V (Y (-2 \beta +4 \beta  Y-3 Y+2)-1)+(Y-1)^2\right)}{(1-U) (1-Y)(1-V Y-V-Y)}},
\end{align}
defined in the bounded phase space
\begin{align}
&\Big\{(U,V,Y)\in\mathbb{R}^3: 0\leq U \leq 1, 0\leq Y\leq 1, 0\leq V\leq \frac{1-Y}{1+Y}, \nonumber \\
&  \beta \left(2 V Y (U (2
   Y-1)-Y+1)\right) <U \left(V \left(3 Y^2-2 Y+1\right)-(Y-1)^2\right)+(1-V) (Y-1)^2\Big\}.
\end{align}

\begin{table}[!t]
\centering
\scalebox{0.70}{
\begin{tabular}{|c|c|c|c|c|}
\hline
Labels & $(U,V,Y)$&   Eigenvalues & Stability  \\\hline
$L_1$ & $(1, 0, Y_0)$& $0,-\frac{(Y_0-1)^2 ((3 \beta -2) Y_0-1)}{\beta },-\frac{(Y_0-1)^2 ((3 \beta -2) Y_0-1)}{\beta }$ & Attractor for \\
&&& $\eta >1, \beta <0, 0<Y_0<\frac{1}{\eta }$, or \\
&&& $\eta >1, \beta >\frac{\eta +2}{3}, \frac{1}{3 \beta -2}<Y_0<\frac{1}{\eta }$. \\
&&& Source for  \\
&&& $\eta >1, 0<\beta \leq \frac{\eta +2}{3}, 0<Y_0<\frac{1}{\eta }$, or \\
&&& $\eta >1, \beta >\frac{\eta +2}{3}, 0<Y_0<\frac{1}{3 \beta -2}$.\\\hline
$L_2$ & $\left(\frac{3 \beta }{3 \beta +1},  0, Y_0\right)$ & $0,\frac{({Y_0}-1) ((9 \beta -7) {Y_0}-2)}{3 \beta +1},\frac{3 ({Y_0}-1) (3 \beta +(9 (\beta -2) \beta +5) {Y_0}+1)}{(3 \beta +1)^2}$ &  Nonhyperbolic\\
&&& 2D stable manifold for \\
&&& $ \beta <-\frac{1}{3},  \frac{-3 \beta -1}{9 \beta ^2-18 \beta +5}<{Y_0}<1 $, or\\
&&& $ \beta >1,  \frac{2}{9 \beta -7}<{Y_0}<1 $. \\
&&& 2D unstable manifold for\\
&&&$\frac{2}{3}<\beta <1,  \frac{-3 \beta -1}{9 \beta ^2-18 \beta +5}<{Y_0}<1$\\\hline
$L_3$ & $(U_0,  0, 1)$  & $0,0,0$ & Nonhyperbolic. \\\hline
$L_4$ & $(0, 0 ,Y_0)$ & $0,-(Y_0-1)^2,3 (Y_0-1)^2$ & Saddle. \\\hline
$L_5$ & $(0, V_0, 1)$  & $0,0,0$ & Nonhyperbolic. \\\hline
$L_6$ & $(U_0, 1, 0)$, $n=0$ &$0,-\frac{2 \left(\sqrt{\beta }+1\right) (U_0-1)}{\beta -1},-\frac{2 \left(\sqrt{\beta }+1\right) (U_0-1)}{\beta -1}.$ & Nonhyperbolic.\\
&&& 2D stable manifold for \\
&&& $0<U_0<1, 0\leq \beta <1$, or  \\
&&& $\beta <0, 0<U_0<1$. \\
&&& 2D unstable manifold for \\
&&& $0<U_0<1, \beta >1$.\\\hline
$P_1$ & $\left(0, \frac{n+1}{n+2},  0 \right)$& $-\frac{1}{n+2},\frac{1}{n+2},-\frac{n-3}{n+2}$ &  Saddle. \\\hline
$P_2$ & $(0,  1,  0)$ & $-\frac{1}{n+1},-\frac{1}{n+1}, -\frac{n}{n+1}$ & Sink.  \\\hline
$P_3$ & $(1, 1, 1)$ & $-\frac{1}{n+1},\frac{1}{n+1},\frac{n}{n+1}$ & Saddle.  \\\hline
$P_4$ & $\left(\frac{\beta  (n-3)}{\beta  (n-3)+n-1}, \frac{2 (n+1)}{3 n+1},   0\right)$ &$\frac{2-2 n}{(3 n+1) (\beta  (n-3)+n-1)}$, & \\
&& $-\frac{(n-1) \left(n+\sqrt{1-n (7 n-22)}-5\right)}{2 (3 n+1) (\beta  (n-3)+n-1)}$, & \\
&& $\frac{(n-1) \left(-n+\sqrt{1-n (7 n-22)}+5\right)}{2 (3 n+1) (\beta  (n-3)+n-1)}$& Stable spiral  for $\beta \geq 0, n>5$.  \\\hline
$P_5$ & $(1,  1,  0)$  & $0,0,0$ & Nonhyperbolic.  \\\hline
$P_6$ & $\left(0, \frac{1}{2 \Delta +1},\frac{\Delta }{\Delta +1}\right)$  & $0,0,\frac{4 \Delta }{(\Delta +1)^2 (2 \Delta +1)}$ & Nonhyperbolic.\\
&&& 1D unstable manifold.  \\\hline
$P_7$ & $\left(1, \frac{1}{2 \Delta +1},\frac{\Delta }{\Delta +1}\right)$ & $0,0,\frac{2 \Delta  (4 \beta  \Delta -3 \Delta -1)}{\beta  (\Delta +1)^2 (2
   \Delta +1)}$. & Nonhyperbolic. \\
&&& 1D unstable manifold for \\
&&& $\beta <0, n>0$, or \\
&&& $\frac{3}{4}<\beta <\frac{1}{8} \left(7+\sqrt{5}\right), 0<n<n_0$, or \\
&&& $\beta \geq \frac{1}{8} \left(7+\sqrt{5}\right),  n>0$\\
&&& 1D stable manifold for \\
&&& $0<\beta \leq \frac{3}{4}, n>0$, or \\
&&& $\frac{3}{4}<\beta <\frac{1}{8} \left(7+\sqrt{5}\right), n>n_0$.\\\hline
$P_8$ & $\left(\frac{1}{2},\frac{2}{(\Delta +1)^2},
   1-\frac{4}{\Delta  (\Delta +2)+3}\right)$ & $0,0,0$ & Nonhyperbolic. \\\hline
$P_9$ & $\left(1, 1, Y_0\right)$, $(0,1)$ & $0,\frac{2Y_0 ((2-4 \beta )Y_0+1)}{\beta },-\frac{2Y_0 ((2-4 \beta )Y_0+1)}{\beta}$& Saddle.	\\\hline 
$P_{10}(0)$ &  $\left(0, V_0, \frac{1-V_0}{1+V_0}\right)$, $\left(0,0\right)$ & $0,0,-\frac{8 ({V_0}-1) {V_0}^2}{({V_0}+1)^2}$ & Nonhyperbolic. \\
&&& 1D unstable manifold.	\\\hline 
$P_{10}(1)$ &  $\left(1, V_0, \frac{1-V_0}{1+V_0}\right)$ & $0,0,\frac{2 ({V_0}-1) {V_0} (4 \beta  ({V_0}-1)-{V_0}+3)}{\beta  ({V_0}+1)^2}$ & Nonhyperbolic. 	\\\hline 
$P_{11}$& $\left(1, V_0, \frac{1}{2 (2 \beta -1)}\right)$, $\beta>\frac{3}{4}$ & $0,0,0$ & Nonhyperbolic. 	\\\hline
$P_{12}$& $\left(U_0, \frac{1}{2 \Delta +1},\frac{\Delta }{\Delta +1}\right)$ & $0,0,\lambda_c$ & Nonhyperbolic.	\\\hline 
\end{tabular}}
\caption{\label{Tab2BXXXN}  Equilibrium points and the corresponding eigenvalues of the dynamical system \eqref{GEN-syst}. We use the notations $\Delta=\frac{\sqrt{n (5 n+4)}-n}{2 n}$, $n_0=-\frac{(3-4 \beta )^2}{16 \beta ^2-28 \beta +11}$. $\lambda_c$ is a function of $n$ and $\beta$. }
\end{table}

The details of the stability of the relevant equilibrium point of the dynamical system \eqref {GEN-syst} are presented  in Table \ref{Tab2BXXXN}. 
Evaluating numerically the expressions $R_{\text{comp}}, M_{\text{comp}}$ defined by \eqref{RM2Acompact} at the orbits of the system \eqref{GEN-syst}, we can see whether the resulting model leads to finite radius and finite mass.

The system \eqref{GEN-syst} admits the relevant invariant sets $Y=0$ (the low-pressure Newtonian subset) and the invariant set $Y=1$ (high-pressure subset). Now we discuss about the dynamics on these invariant sets. 

\paragraph{Low pressure regime (The Newtonian subset).}

In the invariant set $Y=0$ the dynamics is governed by the equations 
\begin{subequations}
\label{XXXeq:392}
\begin{align}
& \frac{dU}{d\lambda}=(U-1) U \left((U-1) \left(3-\frac{(4 n+3) V}{n+1}\right)+\frac{U(1-V)}{\beta }\right),\\
& \frac{dV}{d\lambda}=(V-1) V \left(\frac{(n+2) (U-1) V}{n+1}+\frac{U(V-1)}{\beta }-U+1\right).
\end{align}
\end{subequations}
In the especial case $\beta=1$ are recovered the equations \eqref{XXXeq:278}, and therefore we have the know results for stellar and relativistic dynamics investigate in the reference \cite{Nilsson:2000zg}. 
The equilibrium points that lies on the invariant set $Y=0$ are:
\begin{enumerate}
\item $L_1(0)$: $(U,V)=(1, 0)$. The eigenvalues are $\frac{1}{\beta },\frac{1}{\beta }$. It is a source for $\beta\geq 0$ and a sink for $\beta<0$. Therefore, the dynamics is richer as compared with the previous case where it is always a source. 

\item $L_2(0)$: $(U,V)=\left(\frac{3 \beta }{3 \beta +1}, 0\right)$. The eigenvalues are $-\frac{3}{3 \beta +1},\frac{2}{3 \beta +1}$. It is a saddle. 
\item $L_4(0)$: $(U,V)=(0, 0)$. The eigenvalues are $-1,3$. It is a saddle. 
\item $L_6$: $(U,V)=(U_0, 1),  U_0\in[0,1]$, with eigenvalues $0, -(1-U_0)$. Exists for $n=0$. This line is normally hyperbolic and due to the non-zero eigenvalue is negative it is a sink.
\item $P_1$: $(U,V)=\left(0, \frac{n+1}{n+2}\right)$. The eigenvalues are $\frac{1}{n+2},-\frac{n-3}{n+2}$. Source for $0\leq n < 3$, nonhyperbolic for $n=3$, saddle when $n>3$. This equilibrium point has the same properties as for the Newtonian subset. 
\item $P_2$: $(U,V))=(0, 1)$. The eigenvalues are $-\frac{n}{n+1},-\frac{1}{n+1}$.  Nonhyperbolic for $n=0$, sink for $n>0$.
\item $P_4$: $(U,V)=\left(\frac{\beta  (n-3)}{\beta  (n-3)+n-1}, \frac{2 (n+1)}{3 n+1}\right)$. Exists for $\beta <0,  1\leq n\leq 3$, or $\beta =0,  n>1$, or $\beta >0,  n=1$, or $\beta >0,  n\geq 3$, with  eigenvalues $-\frac{(n-1) \left(n+\sqrt{1-n (7 n-22)}-5\right)}{2 (3 n+1) (\beta  (n-3)+n-1)},\frac{(n-1) \left(-n+\sqrt{1-n (7 n-22)}+5\right)}{2 (3
   n+1) (\beta  (n-3)+n-1)}$. As compared with the Newtonian case, the dynamical features of this point are more intricate (starting that it not only exists when $n>3$, but also that it is a two-parametric solution). It is an unstable node for $\beta \geq 0, 3<n\leq \frac{1}{7} \left(11+8 \sqrt{2}\right)\approx 3.18767$. Unstable spiral for $\beta \geq 0, \frac{1}{7} \left(11+8 \sqrt{2}\right)<n<5$. Stable spiral for $\beta \geq 0, n>5$. Center for $\beta \neq 0, n=1$, or $n=3$. Saddle for $\beta \leq 0, 1<n<3$.

\item $P_5$: $(U,V)=(1, 1)$,  with eigenvalues $0,0$. Nonhyperbolic. 
\end{enumerate}

In the Figure \ref{fig:XXXeq:392A} are presented some streamlines  of the system \eqref{XXXeq:392} for the choices $n=0, n=3, n=3.18$ and for comparison with the relativistic case ($\beta=1$), that it is presented in the middle row, we presented two cases $\beta<0$ and $\beta>0$.  
In the Figure \ref{fig:XXXeq:392B} are presented some streamlines  of the system \eqref{XXXeq:392} for the choices  $n=4, n=5, n=6$ and for comparison with the relativistic case ($\beta=1$), that it is presented in the middle row, we presented two cases $\beta<0$ and $\beta>0$. 
%%%%%%%%%Figure%%%%%%%%%%%%%%%%%%%%%%%%%%%%%%
\begin{figure*}[!t]
	\subfigure[\label{fig:XXXeq:393a}  $\beta=-1.1$.]{\includegraphics[width=0.4\textwidth]{FIGURE15A}}	\hspace{2cm}
	\subfigure[\label{fig:XXXeq:393b}  $\beta=1.1$.]{\includegraphics[width=0.4\textwidth]{FIGURE15B}} 
								\caption{\label{fig:XXXeq:393} Streamlines  of the system \eqref{XXXeq:393} in the invariant set $Y=1$. }
			\end{figure*}
%%%%%%%%%%%%%%%%%%%%%%%%%%%%%%%
Some streamlines of the system \eqref{XXXeq:393} in the invariant set $Y=1$ are presented in Figure \ref{fig:XXXeq:393}.

\paragraph{High pressure regime.} 
In the invariant set $Y=1$ the dynamics is given by the equations 
 \begin{equation}
\label{XXXeq:393}
 \frac{dU}{d\lambda}=\frac{2 (4 \beta -3) (U-1) U^2 V}{\beta },\quad 
 \frac{dV}{d\lambda}=-\frac{2 (4 \beta -3) U (V-1) V^2}{\beta }.
\end{equation}
 
The equilibrium points in the invariant set $Y=1$ are
\begin{enumerate}
\item $L_3$: $(U,V)=(U_0, 0), \quad 0\leq U_0\leq 1$, with eigenvalues $0,0$. Nonhyperbolic. 
\item $L_5$: $(U,V)=(0, V_0), \quad 0\leq V_0\leq 1$, with eigenvalues $0,0$. Nonhyperbolic. 
\item $P_3$: $(U,V)=(1,1)$, with eigenvalues $-\frac{2 (4 \beta -3)}{\beta },\frac{2 (4 \beta -3)}{\beta }$. Nonhyperbolic for $\beta=\frac{3}{4}$. It is a saddle otherwise. 
\end{enumerate}
In this example the system is integrable. The orbit passing through $(U_0, V_0)$ is 
\begin{subequations}
\begin{align}
& V(U)= \frac{(U-1) U_{0} V_{0}}{U  (U_{0}V_{0}-1)-U_{0} V_{0}}, \;\text{with}\\
& {\lambda}=c +\frac{\beta  \left(\frac{({U_0}-1) ({V_0}-1)}{U-1}+\frac{{U_0} {V_0}}{U}+2 \tanh ^{-1}(1-2 U) ({U_0} (2
   {V_0}-1)-{V_0}+1)\right)}{2 (4 \beta -3) {U_0} {V_0}}.
\end{align}
\end{subequations}

\section{Stationary comoving aether with perfect fluid and scalar field in
static metric}
\label{sf}

We use now the metric
\begin{equation}
ds^{2}=-N^{2}(r)dt^{2}+e_{1}{}^{1}(r)^{-2}dr^{2}+e_{2}{}^{2}(r)^{-2}(d%
\vartheta ^{2}+\sin ^{2}\vartheta d\phi ^{2}),  \label{met2-2}
\end{equation}%
where we defined $x=\mathbf{e}_{1}\ln e_{2}{}^{2},y=\mathbf{e}_{1}\ln N$, and the
differential operator $\mathbf{e}_{1}=e_{1}{}^{1}\partial _{r}$. The
equations for the variables $x,y,p,\phi ,K$ are:
\begin{subequations}
\label{static}
\begin{align}
& \mathbf{e}_{1}\left( x\right) =\frac{\mu +3p}{2\beta }+%
\mathbf{e}_{1}(\phi ){}^{2}-\frac{W(\phi)}{\beta}+2(\beta-1)y^2+3xy+K,
\label{statica} \\
& \mathbf{e}_{1}\left( y\right) =\frac{\mu +3p}{2\beta }-%
\frac{W(\phi)}{\beta}+2xy-y^{2},  \label{staticb} \\
& {\mathbf{e}_{1}}\left( p\right) =-y(\mu +p)  \label{staticc1} \\
& \mathbf{e}_{1}(\mathbf{e}_{1}(\phi ))=-\left( y-2x\right) \mathbf{e}%
_{1}(\phi )+W'(\phi),  \label{staticd1} \\
& \mathbf{e}_{1}(K)=2xK,  \label{eqKstatic}
\end{align}%
where $W(\phi)$ is the scalar field self-interacting potential. The system satisfies the
restriction
\end{subequations}
\begin{equation}
\label{static_sf}
-(x-y)^{2} +\beta y^2+p+\frac{1}{2}\mathbf{e}_{1}(\phi ){}^{2}-W(\phi)+K=0.
\end{equation}%
Taking the differential operator $\mathbf{e}_{1}(...)$ in both sides of %
\eqref{static_sf}, using the equations \eqref{static} to substitute the
spatial derivatives, and using again the restriction \eqref{static_sf}
solved for $K,$ we obtain an identity. Thus, \eqref{static} (Gauss
constraint) is a first integral of the system. The aether constraint is
identically satisfied.

In order to apply Tolman-Oppenheimer-Volkoff (TOV) approach 
(we use units where $8  \pi G=1$)
we use the metric
\begin{equation}
ds^{2}=-N(\rho)^2 dt^{2}+\frac{d{\rho}^{2}}{1-\frac{2 m(\rho)}{\rho}}+\rho^2 (d\vartheta ^{2}+\sin
^{2}\vartheta d\varphi ^{2}),  
\end{equation}%
where $m(\rho)$ denotes the mass up to the radius $\rho$, with range $0\leq \rho <\infty$.
Using the expressions \eqref{EQ_58} and \eqref{EQ_59}, and the equations \eqref{statica}, \eqref{staticd1} and \eqref{static_sf}, we obtain the equations
\begin{small}
\begin{subequations}
\label{modified-TOV-eqsSf}
\begin{align}
& m'(\rho )=\frac{2 (\beta -1) \rho  (\rho -2 m) p'(\rho )^2}{(\mu+p)^2}+\frac{3 (\rho -2 m) p'(\rho )}{\mu +p}+\frac{3 m}{\rho }+\frac{\rho ^2 (\mu +3 p)}{2 \beta } +\rho  (\rho -2 m) \phi '(\rho )^2-\frac{\rho ^2 W(\phi
  )}{\beta },\\
	& \frac{(\beta -1) (\rho -2 m) p'(\rho )^2}{\rho  (\mu +p)^2}+\frac{2 (\rho -2 m) p'(\rho )}{\rho ^2 (\mu
   +p)}+\frac{2 m}{\rho ^3}+\frac{(\rho -2 m) \phi '(\rho )^2}{2 \rho }+p-W(\phi)=0,\\
	&\left(\frac{m'(\rho )}{\rho
   }-\frac{2 \rho -3 m}{\rho ^2}\right) \phi '(\rho )+\frac{(\rho -2 m) p'(\rho ) \phi '(\rho )}{\rho  (\mu +p)}-\frac{(\rho
   -2 m) \phi ''(\rho )}{\rho }+W'(\phi)=0.
\end{align}
\end{subequations} 
\end{small}
Equations \eqref{modified-TOV-eqsSf} have two solutions for $p'(\rho), m'(\rho), \phi''(\rho)$. We consider the branch 
\begin{small}
\begin{subequations}
\label{modified-TOV-eqs_Sf_2}
\begin{align}
& p'(\rho )= \frac{(\mu +p) \left(\sqrt{2} \sqrt{\rho -2 m} \Xi+4 m-2 \rho \right)}{2 (\beta -1) \rho  (\rho -2 m)},\\
	& m'(\rho )= -\frac{-2 \beta  \rho +2 \beta  (\beta +1) m+\sqrt{2} \beta  \sqrt{\rho -2 m} \Xi+(\beta -1) \rho ^3 (-\mu+(4 \beta
   -3) p+(2-4 \beta ) W(\phi))}{2 (\beta -1) \beta  \rho },\\
	& \phi ''(\rho )= \frac{\phi '(\rho ) \left(-4 \beta  \rho +4 \beta  m+\rho ^3 (\mu +(3-4 \beta ) p+(4 \beta -2) W(\phi))\right)+2 \beta  \rho ^2 W'(\phi)}{2 \beta  \rho  (\rho -2 m)},\\
	& \Omega'(\rho)=-\frac{\sqrt{2} \sqrt{\rho -2 m} \Xi+4 m-2 \rho }{2 (\beta -1) \rho  (\rho -2 m)},
\end{align}
\end{subequations}
where $\Xi\equiv\sqrt{2 \left(-2 \beta  m-(\beta -1) \rho ^3 (p-W(\phi))+\rho \right)-(\beta -1) \rho ^2 (\rho -2 m) \phi '(\rho )^2}$.\\

Taking the limit $\beta\rightarrow 1$ in Eqs. \eqref{modified-TOV-eqs_Sf_2}, we obtain the proper relativistic equations:
\begin{subequations}
\begin{align}
& p'(\rho )=-\frac{\left(p \rho ^3+2 m \right) (p +\mu  )}{2 (\rho  (\rho -2 m ))}-\frac{1}{4} \rho  (\mu +p) \phi '(\rho )^2+\frac{\rho ^2 (\mu+p) W(\phi)}{2 (\rho -2 m)}\label{eq-130a},\\
& m'(\rho )=\frac{\mu  \rho ^2}{2}+\frac{1}{4} \rho  (\rho -2 m) \phi '(\rho )^2+\frac{1}{2} \rho ^2 W(\phi), \label{eq-130b}\\
& \phi''(\rho)=\frac{\rho ^2 (\mu -p) \phi '(\rho )}{2 (\rho -2 m)}+\frac{2 (m-\rho ) \phi '(\rho )}{\rho  (\rho -2 m)}+\frac{\rho 
   W'(\phi)}{\rho -2 m}+\frac{\rho ^2 \phi '(\rho ) W(\phi)}{\rho -2 m}, \label{eq-130c}\\
& \Omega'(\rho )=\frac{\left(p \rho ^3+2 m \right)}{2 \rho  (\rho -2 m )}+\frac{1}{4} \rho \phi '(\rho )^2-\frac{\rho ^2 W(\phi)}{2 (\rho -2 m)},
	\end{align}
	\end{subequations}
	\end{small} 
\\where $\phi$ is the scalar field and $\Omega$ is the gravitational potential. The above TOV formulation is general. For the analysis we have to specify the equation of state and the potential of the scalar field. 

\subsection{Model 3: Perfect fluid with linear equation of state and a scalar field with an exponential potential}

Let us now show how \eqref{static} can be used to obtain exact solutions and in
addition, be used in the dynamical systems approach for investigating the
structure of the whole solution space. We use the linear equation of state
\begin{equation}
\mu =\mu _{0}+(\eta -1)p,  \label{ss1}
\end{equation}%
where the constants $\mu _{0}$ and $\eta $ satisfy $\mu _{0}\geq 0,\eta \geq
1.$ The case $\eta =1$ corresponds to an incompressible fluid with constant
energy density, while the case $\mu _{0}=0$ describes a scale-invariant
equation of state.

\subsubsection{Singularity analysis and integrability}

Without loss of generality we assume  the differential operator in the form  $\mathbf{e}%
_{1}=\partial _{{\ell}}$, where we have introduced the radial rescaling 
$r=e^{\ell}$, such that $\ell\rightarrow -\infty$ as $r\rightarrow 0$ and $\ell\rightarrow \infty$ as $r\rightarrow \infty$. Furthermore with the use of (\ref{staticc1}) and (\ref%
{eqKstatic}) we rewrite the system (\ref{statica})-(\ref{eqKstatic}) as that
of three second-order ordinary differential equations in terms of the
variables,~$p\left( \ell\right) $, $K\left( \ell\right) $ and $\phi \left(
\ell\right) $, where for the perfect fluid we assume the EoS (\ref{ss1}) and assume that the potential for the scalar field is $%
W\left( \phi \right) =W_{0}e^{-k\phi }$. Recall that equations  (\ref%
{staticc1}) and (\ref{eqKstatic}) provide%
\begin{equation}
x=\frac{\mathbf{e}_{1}(K)}{2K}~,~y=-\frac{\mathbf{e}_{1}\left( p\right) }{\left( \mu +p\right)
}.
\end{equation}

Consider first that $\mu _{0}=0$, and we substitute $p\left( {\ell}\right)
=p_0 {\ell}^{p_{1}},K\left( {\ell}\right) =K_{0}{\ell}^{p_{2}},~\phi =\phi _{0}{\ell}^{p_{3}}$
in the system of second order equations. We search for the dominant
behavior and we find that $p_{1}=p_{2}=p_{3}=-2$, while when the
constants $p_0,K_{0}$ and $\phi _{0}$ are
\begin{equation}
p_0=\frac{4\beta k^{2}-4\eta }{\eta ^{2}k^{2}}~,~\phi
_{0}=-\frac{2\left( 2+\eta \right) }{\eta k^{2}W_{0}},
\end{equation}%
and%
\begin{equation}
K_{0}=-\frac{\left( 4\left(2\beta-1\right) -5\eta ^{2}\right) +4\eta ^{2}}{%
\eta ^{2}k^{2}}.
\end{equation}%
The dominant behavior is also a solution of the FE.

We follow the ARS algorithm which we described above and we find the six
resonances to be
\begin{equation}
s_{1}=-1~,~s_{2}=\frac{2\left( 2+\eta \right) }{\eta },
\end{equation}%
\begin{equation}
s_{\pm }^{1}=-\frac{1}{2}-\frac{1}{\eta }\pm \frac{\sqrt{32\eta
^{2}+k^{2}\left( 36\left(2\beta-1\right) -7\eta \left( 4+\eta \right)
\right) }}{2\eta ^{2}k^{2}},
\end{equation}

\begin{equation}
s_{\pm }^{2}=-\frac{2+\eta }{2\eta }\pm \frac{\sqrt{\beta
k^{2}\left( 2+\eta \right) \left( \beta \left( 2-7\eta
\right) +8\eta ^{2}\right) }}{2k^{2}\eta \beta }.
\end{equation}%
We conclude that the
FE pass the singularity test.

On the other hand, for $\mu _{0}\neq 0$, the dominant terms are not a
specific solution of the FE. Furthermore, the ARS algorithm
fails and the system does not pass the singularity test. Now that we have 
studied the integrability of the field FE, we continue with the
phase-space analysis.

\subsubsection{Equilibrium points in the finite region of the phase space}
\label{Section:3.1.2}
In this section  we introduce a new dynamical system formulation best suited to the case when a  scalar field is included in the matter content. We investigate here both the general relativistic case ($\beta=1$) and the Einstein-aether modification using the equations \eqref{static}.  The advantage of this formulation is that the resulting system is polynomial, and we do not have rational functions as in the TOV formulation. Therefore, we define the normalized variables 
\begin{align}
\label{Vars-Model_3}
& X_\phi= \frac{\e_1(\phi)}{\sqrt{2}\sqrt{\theta^2+W}}, Y_\phi= \frac{\theta}{\sqrt{\theta^2+W}}, Y=\frac{p}{p+\mu}, S_1=\frac{y}{\sqrt{\theta^2+W}}, S_2=\frac{K}{\theta^2+W},\nonumber\\
& \Pi=\frac{p}{\theta^2+ W}, \theta=y-x,
\end{align}
and the time derivative 
 \begin{equation}
\frac{d f}{d\lambda}=\frac{Y}{\sqrt{\theta^2+W}} \e_1(f),
\end{equation}
to obtain the dynamical system
\begin{subequations}
\label{EQS-3-13}
\begin{align}
& \frac{dX_\phi}{d\lambda}=\frac{Y \left(\sqrt{2} {X_\phi} \left(2 \beta  {S_1}^2 {Y_\phi}-{S_1} {Y_\phi}^2+{S_1}+{Y_\phi} \left({S_2}+2 {X_\phi}^2-2\right)\right)-k \left({X_\phi}^2-1\right) \left({Y_\phi}^2-1\right)\right)}{\sqrt{2}},\\
& \frac{dY_\phi}{d\lambda}=	-\frac{1}{2} Y \left({Y_\phi}^2-1\right) \left(\sqrt{2} k {X_\phi} {Y_\phi}+2 {S_1} ({Y_\phi}-2 \beta  {S_1})-2
   {S_2}-4 {X_\phi}^2\right),\\
& \frac{dS_1}{d\lambda}=\frac{\left(\beta  {S_1}^2-1\right) (-2 Y {Y_\phi} ({Y_\phi}-2 \beta  {S_1})-1)+{S_2} (2 Y (\beta  {S_1} {Y_\phi}-1)-1)+{X_\phi}^2 (4 \beta  {S_1} Y {Y_\phi}-2 Y-1)}{2
   \beta }\nonumber \\
	& -\frac{k {S_1} {X_\phi} Y \left({Y_\phi}^2-1\right)}{\sqrt{2}},\end{align}
	\begin{align}
& \frac{dS_2}{d\lambda}=	{S_2} Y \left(-\sqrt{2} k {X_\phi} \left({Y_\phi}^2-1\right)+4 \beta  {S_1}^2 {Y_\phi}-2 {S_1} \left({Y_\phi}^2-1\right)+2 {Y_\phi} \left({S_2}+2 {X_\phi}^2-1\right)\right),\\
& \frac{dY}{d\lambda}=	{S_1} Y (\eta  Y-1). 
	\end{align}
\end{subequations}
Additionally, 
\begin{equation}
\Pi=1-X_\phi^2-\beta S_1^2-S_2, \quad W+Y_\phi^2=1. 
\end{equation}\\
	
\begin{table}[!t]
\centering
\scalebox{0.9}{
\begin{tabular}{|c|c|}
\hline
Labels  & Stability \\\hline
$L_1$ & Nonhyperbolic. 
  2D unstable  (resp. stable) manifold for $S_1^*<0$ (resp. $S_1^*>0$). \\
\hline
$L_2^{\varepsilon}$ &   $L_2^+$ (resp. $L_2^-$ ) nonhyperbolic. 3D unstable (resp. stable) manifold for  $Y_0>0, k<2\sqrt{2}$. \\
\hline
$L_3^{\varepsilon}$& $L_3^+$ (resp. $L_3^-$) is nonhyperbolic. 3D unstable (resp. stable) manifold for 
$Y_0>0, k>-2\sqrt{2}$. \\
\hline
$L_4^\varepsilon$ & saddle  \\
\hline
$L_5^\varepsilon$ &   $L_5^+$ (resp. $L_5^-$) is nonhyperbolic with a 4D unstable (resp. stable) manifold for \\
& $ k\in \mathbb{R},  {X_\phi^*}=0,  \eta >1,  \beta >\frac{1}{16} (\eta +2)^2 $, or  \\
& $ -1<{X_\phi^*}<0,  k\geq
   \frac{2 \sqrt{2}}{3 {X_\phi^*}},  \eta >1,  \beta >-\frac{1}{16} (\eta +2)^2 ({X_\phi^*}^2-1) $, or \\
& $ 0<{X_\phi^*}<1,  k\leq \frac{2 \sqrt{2}}{3 {X_\phi^*}},  \eta >1,  \beta >-\frac{1}{16} (\eta +2)^2 ({X_\phi^*}^2-1)$, or \\
& $ -1<{X_\phi^*}<0,  \frac{2 \sqrt{2}}{{X_\phi^*}}<k<\frac{2 \sqrt{2}}{3 {X_\phi^*}},  \eta >-\frac{2 \sqrt{2} k {X_\phi^*}}{\sqrt{2} k {X_\phi^*}-4},  
	\beta >-\frac{1}{16} (\eta +2)^2 ({X_\phi^*}^2-1) $, or \\ 
&	$ 0<{X_\phi^*}<1,  \frac{2 \sqrt{2}}{3 {X_\phi^*}}<k<\frac{2 \sqrt{2}}{{X_\phi^*}}, 
   \eta \geq -\frac{2 \sqrt{2} k {X_\phi^*}}{\sqrt{2} k {X_\phi^*}-4},  
	\beta >-\frac{1}{16} (\eta +2)^2 ({X_\phi^*}^2-1) $, or \\ 
& $ -1<{X_\phi^*}<0,  \frac{2 \sqrt{2}}{{X_\phi^*}}<k<\frac{2 \sqrt{2}}{3 {X_\phi^*}}, 
   1<\eta \leq -\frac{2 \sqrt{2} k {X_\phi^*}}{\sqrt{2} k {X_\phi^*}-4},  
	\beta >\frac{2 ({X_\phi^*}^2-1)}{-k^2 {X_\phi^*}^2+4 \sqrt{2} k {X_\phi^*}-8} $, or \\ 
& $0<{X_\phi^*}<1,  \frac{2 \sqrt{2}}{3 {X_\phi^*}}<k<\frac{2 \sqrt{2}}{{X_\phi^*}},  1<\eta <-\frac{2 \sqrt{2} k {X_\phi^*}}{\sqrt{2} k {X_\phi^*}-4},  \beta >\frac{2
   ({X_\phi^*}^2-1)}{-k^2 {X_\phi^*}^2+4 \sqrt{2} k {X_\phi^*}-8}$.  \\
\hline
$L_6^\varepsilon$ &  saddle.\\ \hline
$P_1^\varepsilon$ & nonhyperbolic. \\
\hline
$P_2^\varepsilon$ & saddle. \\
\hline
$P_3^\varepsilon$ &  $P_3^+$ (resp. $P_3^-$) is nonhyperbolic with a 4D unstable (resp. stable) manifold for $\eta >1,  \beta >\frac{1}{16} (\eta +2)^2$  \\
\hline
$P_4^\varepsilon$ & saddle. \\
\hline 
$P_5^\varepsilon$&   $P_5^+$ (resp. $P_5^-$) is a sink (resp. a source) for $\beta <0,  \eta >1$. They are saddle otherwise.  \\
\hline
$P_6^\varepsilon$ & saddle.  \\
\hline
$P_7^\varepsilon$ & $P_7^+$ (resp. $P_7^-$) is a source (resp. sink) for \\
& $ k\leq -2 \sqrt{2},  \beta >\frac{1}{k^2},  1<\eta <\beta  k^2 $, or\\ 
& $ -2 \sqrt{2}<k<-2 \sqrt{\frac{2}{3}}, 
   \frac{1}{k^2}<\beta <-\frac{2}{k^2-8},  1<\eta <\beta  k^2 $, or\\ 
& $ 2 \sqrt{\frac{2}{3}}<k<2 \sqrt{2},  \frac{1}{k^2}<\beta
   <-\frac{2}{k^2-8},  1<\eta <\beta  k^2 $, or \\ 
&$ k\geq 2 \sqrt{2},  \beta >\frac{1}{k^2},  1<\eta <\beta  k^2 $.  \\ 
	 \hline
$P_8^\varepsilon$  &  saddle. \\\hline 
$P_{9}^\varepsilon$ & For some choices of parameters, say  $\beta=1, \eta=1$, or $\beta=1, \eta=2$, they are saddle.\\ 
\hline 
\end{tabular}}
\caption{\label{XXX}  Stability of the equilibrium points of the dynamical system \eqref{EQS-3-13}.}
\end{table}
%%%%%%%%%%%%%%%%%%%%%%%
Therefore, the system defines a flow on the invariant set 
\begin{equation}
\Big\{(X_\phi, Y_\phi, S_1, S_2,Y)\in \mathbb{R}^5: X_\phi^2+ \beta S_1^2+S_2\leq 1, -1\leq Y_\phi \leq 1, S_2\geq 0, 0\leq Y\leq 1\Big\}.
\end{equation}

The system is invariant under the change of coordinates\\$(X_\phi, Y_\phi, S_1) \rightarrow (-X_\phi, -Y_\phi, -S_1)$ with the simultaneous reversal in the independent variable ${\lambda}\rightarrow -{\lambda}$.   In relation to the phase space
dynamics this implies that  for two points related by this symmetry, say $P^+$ and $P^-$, each has the
opposite dynamical behavior to the other; that is, if the equilibrium point $P^+$  is an attractor for a given
set of parameters, then $P^-$  is a source under the same choice. 

The details
of the stability of the equilibrium points (curves/surfaces of equilibrium points) of the dynamical system \eqref{EQS-3-13} are presented in
the Appendix \ref{App_A}. These results are summarized in Tables \ref{XXX} and \ref{YYY}.

\begin{table}[!t]
\centering
\scalebox{0.8}{
\begin{tabular}{|c|c|}
\hline
Labels  & Stability \\\hline
$P_{10}^{\varepsilon}$ & $P_{10}^+$ (resp. $P_{10}^-$) is nonhyperbolic with a 3D unstable (resp. stable) manifold for \\
& $ -2 \sqrt{2}<k<0 ,  \beta <-\frac{2}{k^2} ,  1<\eta <4 \sqrt{\frac{\beta  k^4-8 \beta  k^2+2 k^2}{\left(k^2-8\right)^2}}-\frac{2 k^2}{k^2-8}$, or \\ 
& $2 \sqrt{\frac{2}{3}}<k<2 \sqrt{2} ,  \frac{9 k^2-8}{16 k^2}<\beta \leq -\frac{2}{k^2-8} ,  1<\eta <-\frac{2
   k^2}{k^2-8}-4 \sqrt{\frac{\beta  k^4-8 \beta  k^2+2 k^2}{\left(k^2-8\right)^2}}$, or \\
& $k=2 \sqrt{2} ,  \beta >\frac{1}{2} , 
   1<\eta <4 \beta -1$, or \\
&$k>2 \sqrt{2} ,  \beta >\frac{9 k^2-8}{16 k^2} ,  1<\eta <4 \sqrt{\frac{\beta  k^4-8 \beta  k^2+2
   k^2}{\left(k^2-8\right)^2}}-\frac{2 k^2}{k^2-8}$. \\
& $P_{10}^+$ (resp. $P_{10}^-$) nonhyperbolic with a 3D stable (resp. unstable) manifold for \\
& $ k=-2 \sqrt{\frac{2}{5}} ,  -\frac{5}{4}<\beta <\frac{1}{4} ,  1<\eta <\frac{1}{2} \sqrt{5-16 \beta }+\frac{1}{2} $,  or  \\ 
& $k=-2
   \sqrt{\frac{2}{3}} ,  \frac{1}{4}<\beta <\frac{3}{8} ,  1<\eta <\sqrt{3-8 \beta }+1 $,  or  \\ 
& $ k=-2 \sqrt{\frac{2}{3}} , 
   -\frac{3}{4}<\beta <\frac{1}{4} ,  1<\eta <\sqrt{3-8 \beta }+1 $,  or  \\  
& $ -2 \sqrt{\frac{2}{3}}<k<-2 \sqrt{\frac{2}{5}} , 
   \frac{1}{4}<\beta <\frac{9 k^2-8}{16 k^2} ,  1<\eta <4 \sqrt{\frac{\beta  k^4-8 \beta  k^2+2 k^2}{\left(k^2-8\right)^2}}-\frac{2
   k^2}{k^2-8} $,  or  \\  
& $ -2 \sqrt{2}<k<-2 \sqrt{\frac{2}{3}} ,  \beta =-\frac{2}{k^2-8} ,  1<\eta <8 \beta -2 $,  or  \\  
& $ -2 \sqrt{\frac{2}{5}}<k<0 ,  -\frac{2}{k^2}<\beta <\frac{9 k^2-8}{16 k^2} ,  1<\eta <4 \sqrt{\frac{\beta  k^4-8 \beta  k^2+2
   k^2}{\left(k^2-8\right)^2}}-\frac{2 k^2}{k^2-8} $,  or  \\  
& $-2 \sqrt{2}<k<-2 \sqrt{\frac{2}{3}} ,  \frac{1}{4}<\beta
   <-\frac{2}{k^2-8} ,  1<\eta <4 \sqrt{\frac{\beta  k^4-8 \beta  k^2+2 k^2}{\left(k^2-8\right)^2}}-\frac{2 k^2}{k^2-8} $,  or \\ 
& $ -2
   \sqrt{2}<k<-2 \sqrt{\frac{2}{3}} ,  -\frac{2}{k^2}<\beta <\frac{1}{4} ,  1<\eta <4 \sqrt{\frac{\beta  k^4-8 \beta  k^2+2
   k^2}{\left(k^2-8\right)^2}}-\frac{2 k^2}{k^2-8} $,  or  \\  
& $ -2 \sqrt{\frac{2}{3}}<k<-2 \sqrt{\frac{2}{5}} ,  -\frac{2}{k^2}<\beta
   <\frac{1}{4} ,  1<\eta <4 \sqrt{\frac{\beta  k^4-8 \beta  k^2+2 k^2}{\left(k^2-8\right)^2}}-\frac{2 k^2}{k^2-8} $. \\
	\hline
$P_{11}^\varepsilon$&  $P_{11}^+$ (resp. $P_{11}^-$) is nonhyperbolic with a 3D stable (resp. unstable) manifold for \\
& $ \beta >\frac{1}{2} ,  k=-2 \sqrt{2} ,  1<\eta <4 \beta -1 $,  or \\ 
& $ \beta \geq \frac{9}{16} ,  k<-2 \sqrt{2} ,  1<\eta <4
   \sqrt{\frac{\beta  k^4-8 \beta  k^2+2 k^2}{\left(k^2-8\right)^2}}-\frac{2 k^2}{k^2-8} $,  or  \\ 
& $ \beta <-\frac{1}{4} ,  \sqrt{2}
   \sqrt{-\frac{1}{\beta }}<k<2 \sqrt{2} ,  1<\eta <4 \sqrt{\frac{\beta  k^4-8 \beta  k^2+2 k^2}{\left(k^2-8\right)^2}}-\frac{2
   k^2}{k^2-8} $,  or  \\ 
& $ \frac{1}{2}<\beta <\frac{9}{16} ,  -2 \sqrt{2} \sqrt{-\frac{1}{16 \beta -9}}<k<-2 \sqrt{2} ,  1<\eta <4
   \sqrt{\frac{\beta  k^4-8 \beta  k^2+2 k^2}{\left(k^2-8\right)^2}}-\frac{2 k^2}{k^2-8} $,  or  \\  
& $ \beta >\frac{1}{2} ,  -2
   \sqrt{2}<k\leq -\sqrt{2} \sqrt{\frac{4 \beta -1}{\beta }} ,  1<\eta <-\frac{2 k^2}{k^2-8}-4 \sqrt{\frac{\beta  k^4-8 \beta  k^2+2
   k^2}{\left(k^2-8\right)^2}} $,  or \\  
& $ \frac{3}{8}<\beta \leq  \frac{1}{2} ,  -2 \sqrt{2} \sqrt{-\frac{1}{16 \beta -9}}<k\leq
   -\sqrt{2} \sqrt{\frac{4 \beta -1}{\beta }} ,  1<\eta <-\frac{2 k^2}{k^2-8}-4 \sqrt{\frac{\beta  k^4-8 \beta  k^2+2
   k^2}{\left(k^2-8\right)^2}} $. \\
& $P_{11}^+$ (resp. $P_{11}^-$)  is nonhyperbolic with a 3D unstable (resp. stable) manifold for  \\
& $\beta >\frac{3}{8} ,  k=\sqrt{2} \sqrt{\frac{4 \beta -1}{\beta }} ,  1<\eta <8 \beta -2$,  or  \\ 
& $\beta >\frac{3}{8} , 
   \sqrt{2} \sqrt{\frac{4 \beta -1}{\beta }}<k<2 \sqrt{2} ,  1<\eta <4 \sqrt{\frac{\beta  k^4-8 \beta  k^2+2
   k^2}{\left(k^2-8\right)^2}}-\frac{2 k^2}{k^2-8}$,  or \\  
& $-\frac{1}{4}<\beta <\frac{1}{4} ,  2 \sqrt{2} \sqrt{-\frac{1}{16 \beta
   -9}}<k<2 \sqrt{2} ,  1<\eta <4 \sqrt{\frac{\beta  k^4-8 \beta  k^2+2 k^2}{\left(k^2-8\right)^2}}-\frac{2 k^2}{k^2-8}$,  or \\ 
& $\frac{1}{4}<\beta \leq \frac{3}{8} ,  2 \sqrt{2} \sqrt{-\frac{1}{16 \beta -9}}<k<2 \sqrt{2} ,  1<\eta <4 \sqrt{\frac{\beta  k^4-8
   \beta  k^2+2 k^2}{\left(k^2-8\right)^2}}-\frac{2 k^2}{k^2-8}$,  or  \\  
& $\beta \leq -\frac{1}{4} ,  2 \sqrt{2} \sqrt{-\frac{1}{16
   \beta -9}}<k<\sqrt{2} \sqrt{-\frac{1}{\beta }} ,  1<\eta <4 \sqrt{\frac{\beta  k^4-8 \beta  k^2+2 k^2}{\left(k^2-8\right)^2}}-\frac{2
   k^2}{k^2-8}$.
\\\hline 
\end{tabular}}
\caption{\label{YYY}  Stability of the equilibrium points of the dynamical system \eqref{EQS-3-13} (cont.).}
\end{table}
%%%%%%%%%Figure%%%%%%%%%%%%%%%%%%%%%%%%%%%%%%
\begin{figure*}[!t]
\includegraphics[width=1\textwidth]{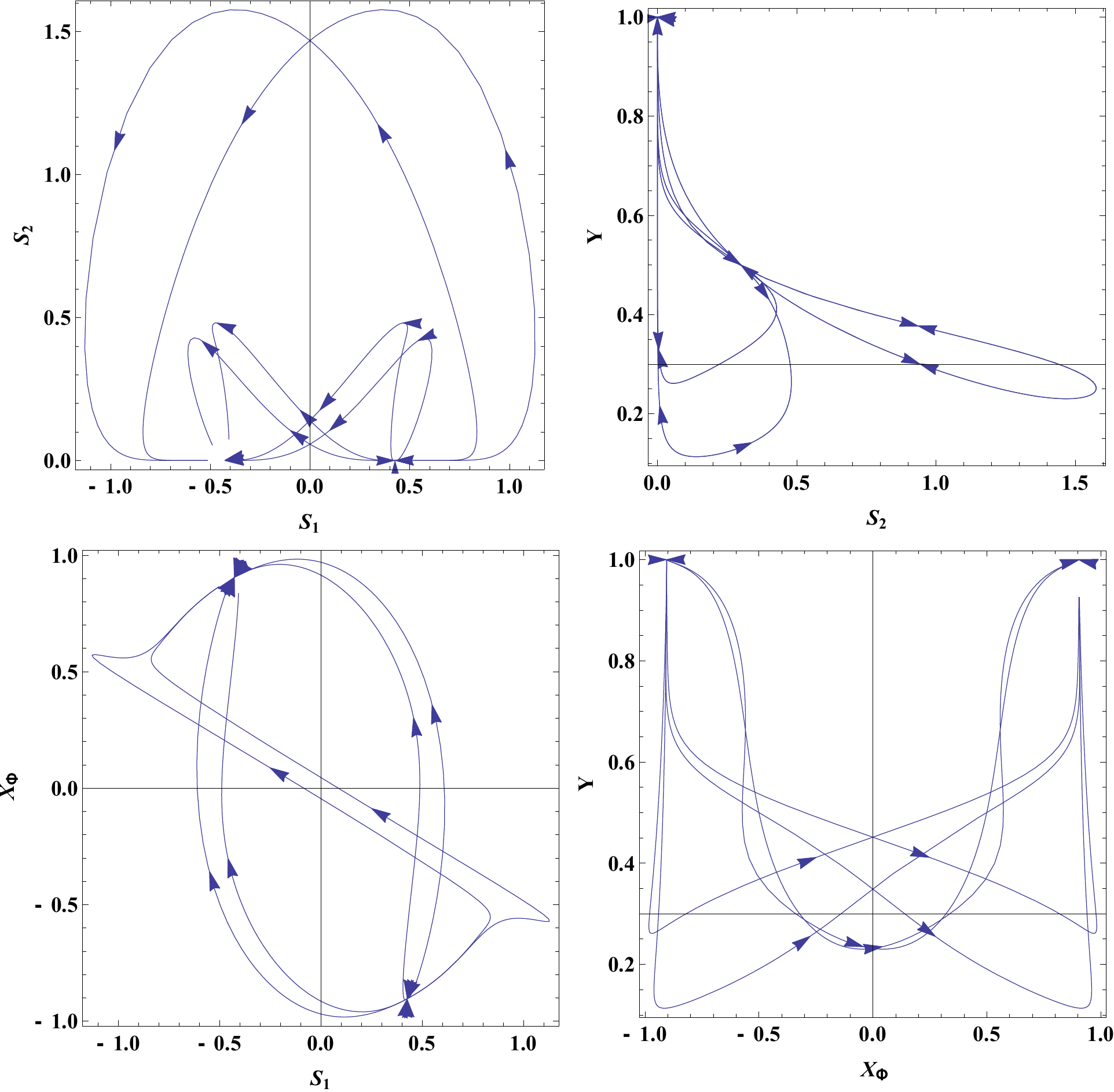}
								\caption{\label{fig:EQS-3-13} Some 2D projections of the orbits  of the system \eqref{XXXeq:393} for $\beta=1, \eta=1, k=-3$.}
			\end{figure*}
%%%%%%%%%%%%%%%%%%%%%%%%%%%%%%%
%%%%%%%%%Figure%%%%%%%%%%%%%%%%%%%%%%%%%%%%%%
\begin{figure*}[!t]
\includegraphics[width=1\textwidth]{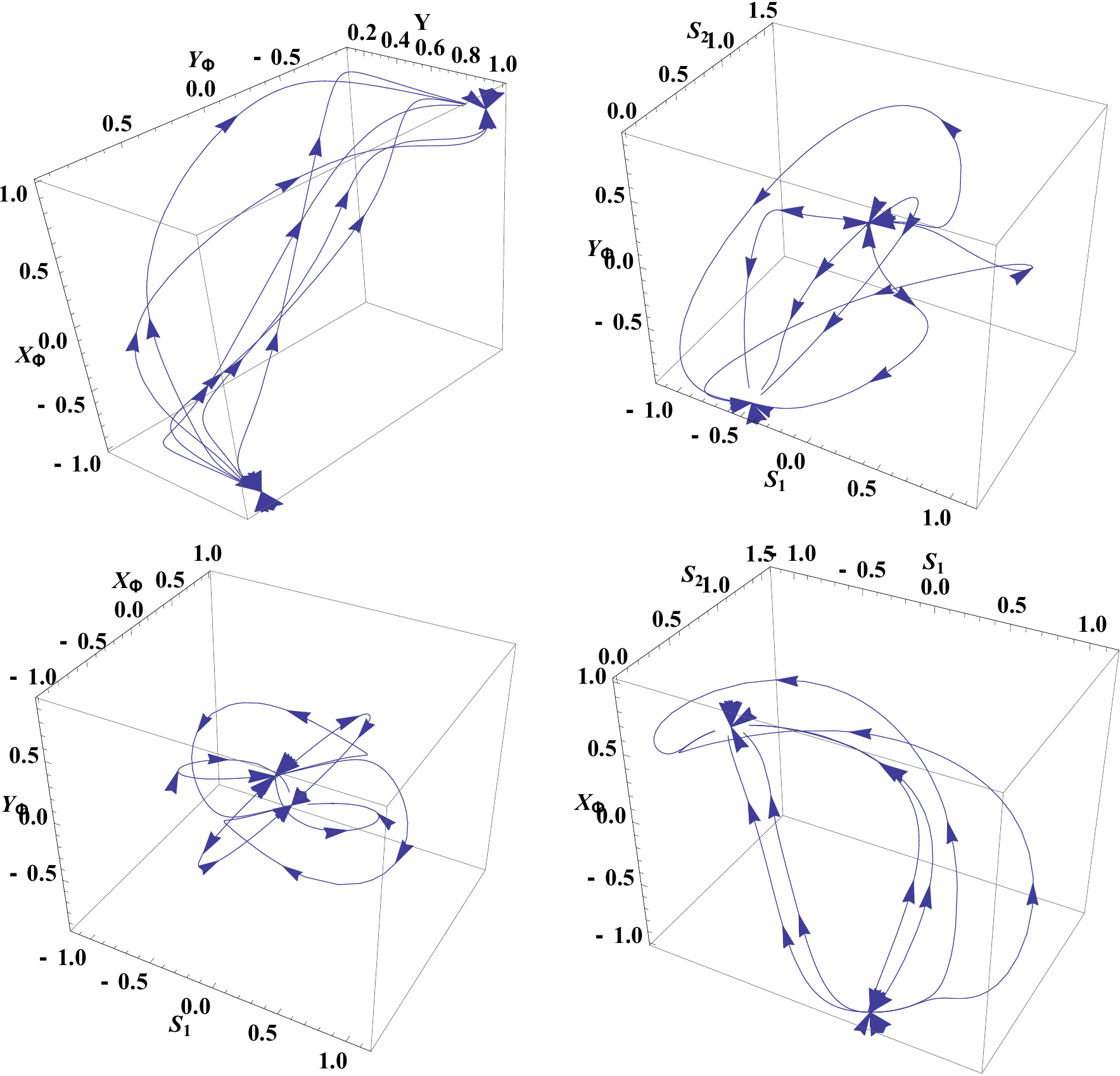}
								\caption{\label{fig:EQS-3-13B} Some 3D projections of the orbits  of the system \eqref{XXXeq:393} for $\beta=1, \eta=1, k=-3$.}
			\end{figure*}
%%%%%%%%%%%%%%%%%%%%%%%%%%%%%%%

We have the useful equations
\begin{subequations}
\begin{align}
& r^2=\frac{\left({S_2}-({Y_\phi}-{S_1})^2\right)
   (1-\eta  Y) \left(1-X_\phi^2-\beta S_1^2-S_2\right)}{\mu_{0}
   {S_2} \left(Y \left({S_2}-({Y_\phi}-{S_1})^2\right)+Y_\phi^2\right)}, \quad 
\mu =\frac{\mu_0 (1-Y)}{1-\eta  Y}, \quad  p= \frac{\mu_{0} Y}{1-\eta  Y},
\end{align}
\begin{align}
&\mathcal{M}(r):=\frac{m}{r}=\frac{{S_2}-({Y_\phi}-{S_1})^2}{2 {S_2}}, 
\end{align}
\end{subequations}
where $\mathcal{M}(r)$ denotes the Misner-Sharp mass \cite{Misner:1964je}. From  these expressions we obtain the $m-r$-relation 
\begin{equation}
\label{RM3}
r^2=\frac{2}{\mu_0}\frac{\mathcal{M}}{D}, \quad  m^2=\frac{2}{\mu_0} \frac{\mathcal{M}^3}{D}, \quad D=\frac{\left(Y \left({S_2}-({Y_\phi}-{S_1})^2\right)+Y_\phi^2\right)}{(1-\eta  Y) \left(1-X_\phi^2-\beta S_1^2-S_2\right)}.
\end{equation}

Using the equations \eqref{RM3}, we can define the compact variables 
\begin{equation}
\label{RM3compact}
R_{\text{comp}}=\frac{\sqrt{\mu_0}r}{\sqrt{1+ \mu_0 r^2\left(1+{\mathcal{M}}^2\right)}}, \quad M_{\text{comp}}=\frac{\sqrt{\mu_0}\mathcal{M}r}{\sqrt{1+ \mu_0 r^2\left(1+{\mathcal{M}}^2\right)}}, 
\end{equation}
which depends only of the phase-space variables $(X_\phi, Y_\phi, S_1,S_2,Y)$. Evaluating numerically the expressions for $R_{\text{comp}}, M_{\text{comp}}$ on the orbits of the system\eqref{EQS-3-13}, we can determine whether the resulting model leads to finite radius and finite mass.

According to our analytical results, the equilibrium points with  5D unstable or stable manifolds are $P_5^\varepsilon$: $(X_\phi, Y_\phi, S_1, S_2,Y)= \left(0,\varepsilon,\frac{\eta +2}{4 \beta }\varepsilon,0,\frac{1}{\eta }\right)$,   and $P_7^\varepsilon$: $(X_\phi, Y_\phi, S_1, S_2,Y)= \left(\frac{k \sqrt{\beta }}{\sqrt{\beta  k^2+2}}\varepsilon, \frac{2 \sqrt{2} \sqrt{\beta }}{\sqrt{\beta  k^2+2}}\varepsilon, \frac{\sqrt{2}}{\sqrt{\beta } \sqrt{\beta  k^2+2}}\varepsilon, 0, \frac{1}{\eta }\right)$.  $P_5^+$ (resp. $P_5^-$) is a sink (resp. a source) for $\beta <0,  \eta >1$. They are saddle otherwise.  $P_7^+$ (resp. $P_7^-$) is a source (resp. sink) for  $ k\leq -2 \sqrt{2},  \beta >\frac{1}{k^2},  1<\eta <\beta  k^2 $, or  $ -2 \sqrt{2}<k<-2 \sqrt{\frac{2}{3}}, 
   \frac{1}{k^2}<\beta <-\frac{2}{k^2-8},  1<\eta <\beta  k^2 $, or  $ 2 \sqrt{\frac{2}{3}}<k<2 \sqrt{2},  \frac{1}{k^2}<\beta
   <-\frac{2}{k^2-8},  1<\eta <\beta  k^2 $, or  $ k\geq 2 \sqrt{2},  \beta >\frac{1}{k^2},  1<\eta <\beta  k^2$. Following the discussion in  \cite{Coley:2019tyx}, 
$P_5^{\pm}$ satisfies $(Q,S,C,A_\phi,A_W)= \left(\pm 1,\pm\frac{\eta +2}{4 \beta},0,0,0\right)$. The physical solutions has the asymptotic metric  \\$ds^{2}=-\bar{N}_0^2 e^{\pm\frac{(\eta+2)\bar{\lambda}}{2\beta}}dt^{2}+\frac{(\eta +2)^2 e^{\pm\frac{\eta  (\eta +2) \bar{\lambda}}{4 \beta}}}{16 c_1^2 \beta^2}d\bar{\lambda}^2+{\bar{K}_0}^{-1} e^{\pm 2 \left(1-\frac{\eta+2}{4\beta}\right)\bar{\lambda}}(d\vartheta ^{2}+\sin
^{2}\vartheta d\varphi ^{2})$. On the introduction of $\rho_\pm=e^{\pm\left(1-\frac{\eta+2}{4\beta}\right)\bar{\lambda}}$, the line element becomes $-\bar{N}_0^2\rho_\pm ^{\frac{2 (\eta +2)}{4 \beta -\eta -2}}dt^2+ \frac{(\eta+2)^2}{2 c_1^2 (4\beta-\eta-2)^2}\rho_\pm ^{-\frac{16\beta-\eta ^2-6 \eta +24}{2 (4\beta-\eta-2)}}d\rho_\pm^2+\frac{\rho_\pm^2}{\bar{K}_0}(d\vartheta ^{2}+\sin^{2}\vartheta d\varphi ^{2})$. For $\beta=1$, these solutions are the analogues of $P_5^{\pm}$ investigated in \cite{Nilsson:2000zf}, which correspond to self-similar plane-symmetric perfect fluid models. For $\eta \geq 1, 16 \beta \geq (\eta +2) (\eta +4)$ the exponent of the component $tt$ is positive and the exponent of the $\rho_\pm\rho_\pm$ component is negative.  Thus the singularity has a horizon at  $\rho_\pm= 0^+$ .

The points $P_{7}^\varepsilon$ has $(Q,S,C,A_\phi,A_W)= \left(1,\frac{\varepsilon}{2 \beta },0,\frac{k \varepsilon}{2 \sqrt{2}},\sqrt{\frac{-8 \beta +\beta  k^2+2}{8\beta }}\right)$. That is, $P_7^\pm$ is exactly the equilibrium point $P_{18}^{\pm}$ studied in \cite{Coley:2019tyx}. According to the results in  \cite{Coley:2019tyx}, the asymptotic metric is given by 
$N=\bar{N}_0 e^{\pm\frac{\bar{\lambda}}{2 \beta}}, K=\bar{K}_0 e^{\mp\frac{(2 \beta -1) \bar{\lambda}}{\beta}}, y=\bar{y}_0 e^{\mp \frac{1}{4} k^2\bar{\lambda}}$. That is,  the metric can be written as 
$ds^2=-{\bar{N}_0}^2 e^{\pm\frac{\bar{\lambda}}{\beta}}dt^2+\frac{1}{4\bar{y}_0^2 \beta^2} e^{\pm \frac{k^2 \bar{\lambda} }{2}}d\bar{\lambda}^2+{\bar{K}_0}^{-1} e^{\pm \frac{(2\beta -1)\bar{\lambda}}{\beta}} \left(d\vartheta ^{2}+\sin ^{2}\vartheta d\varphi ^{2}\right)$. The sink is $P_7^{-}$ with asymptotic line element $ds^2=-{\bar{N}_0}^2 e^{-\frac{\bar{\lambda}}{\beta}}dt^2+\frac{1}{4\bar{y}_0^2 \beta^2} e^{-\frac{k^2 \bar{\lambda} }{2}}d\bar{\lambda}^2+{\bar{K}_0}^{-1} e^{-\frac{(2\beta -1)\bar{\lambda}}{\beta}} \left(d\vartheta ^{2}+\sin ^{2}\vartheta d\varphi ^{2}\right)$, for some conditions of the parameters. Imposing the additional condition $\beta>\frac{1}{2}$, i.e., when $k\leq -2 \sqrt{2},  \beta >\frac{1}{2},  1<\eta <\beta  k^2$, or $-2 \sqrt{2}<k<-2,  \frac{1}{2}<\beta <-\frac{2}{k^2-8},  1<\eta <\beta  k^2$, or $2<k<2 \sqrt{2}, 
   \frac{1}{2}<\beta <-\frac{2}{k^2-8},  1<\eta <\beta  k^2$, or $k\geq 2 \sqrt{2},  \beta >\frac{1}{2},  1<\eta <\beta  k^2$, the solution becomes the Minkowski solution as $\lambda\rightarrow +\infty$. 
However, in any of the cases $k\leq -2, \frac{1}{k^2}<\beta <\frac{1}{2}, 1<\eta <\beta  k^2$, or $-2<k<-2 \sqrt{\frac{2}{3}}, \frac{1}{k^2}<\beta <-\frac{2}{k^2-8}, 1<\eta <\beta  k^2$, or $2
   \sqrt{\frac{2}{3}}<k<2, \frac{1}{k^2}<\beta <-\frac{2}{k^2-8},1<\eta <\beta  k^2$, or $k\geq 2, \frac{1}{k^2}<\beta <\frac{1}{2}, 1<\eta <\beta  k^2$ (all have $\beta<\frac{1}{2}$), the metric component  $e^{-\frac{(2\beta -1)\bar{\lambda}}{\beta}}\rightarrow \infty$ as $\lambda\rightarrow \infty$. Therefore, the solution becomes singular (in two directions the length is stretched, and in one direction it becomes large; cigar singularity). This case does not appears in GR where $\beta=1$. 

To finish this section we present some numerical integrations of the system \eqref{EQS-3-13}. 
In figure \ref{fig:EQS-3-13} we present some 2D projections of the orbits of the dynamical system \eqref{EQS-3-13} for $\beta=1, \eta=1, k=-3$. In figure \ref{fig:EQS-3-13B} we present some 3D projections of the orbits  of the system \eqref{XXXeq:393} for the same choice of parameters and the same selection of the initial conditions as in figure \ref{fig:EQS-3-13}.   
In this figures it is illustrated that the attractor as $\lambda\rightarrow \infty$, is $P_7^-$, that for this choice of parameters is is asymptotically the Minkowski solution (as discussed before). 
The past attractor as $\lambda\rightarrow -\infty$, is $P_7^+$.

\subsection{Model 4: Perfect fluid with polytropic of state and a scalar field with an exponential potential}

Using the parametrization of section \eqref{model2}, that is, introducing the line element 
\begin{equation}
ds^{2}=-e^{2\Omega(\lambda)}dt^{2}+r(\lambda)^2 \left[N(\lambda)^2d\lambda^{2}+d\vartheta ^{2}+\sin^{2} \vartheta d\phi ^{2}\right],  \label{XXXmet2_2}
\end{equation}%
the new radial coordinate $\tau$ such that
\begin{equation}
\frac{d\lambda}{d\tau}=\frac{1}{N(\lambda) r(\lambda)},
\end{equation}
and the differential operator
$\bf{e}_{1}=\partial _{\tau}$, the
equations for the variables $x,y,p,\phi ,K$ become:
\begin{subequations}
\label{static_2}
\begin{align}
&x'(\tau)=\frac{\mu +3p}{2\beta }+{\phi'(\tau)}^{2}-\frac{W(\phi)}{\beta}+2(\beta-1)y^2+3xy+K,
\label{statica_2} \\
& y'(\tau) =\frac{\mu +3p}{2\beta }-\frac{W(\phi)}{\beta}+2xy-y^{2},  \label{staticb_2} \\
& p'(\tau) =-y(\mu +p)  \label{staticc1_2} \\
& \phi''(\tau)=-\left( y-2x\right) \phi'(\tau)+W'(\phi),  \label{staticd1_2} \\
& K'(\tau)=2xK,  \label{eqKstatic_2}
\end{align}%
with the restriction
\begin{equation}
\label{static_sf_2}
-(x-y)^{2} +\beta y^2+p+\frac{1}{2}{\phi'(\tau)}^{2}-W(\phi)+K=0.
\end{equation}%
\end{subequations}
We consider the polytropic equation of state $p=q \mu^\Gamma, \Gamma=1+\frac{1}{n}, n>0$, and assume the scalar field potential to be $%
W\left( \phi \right) =W_{0}e^{-k\phi }$. 

\subsubsection{Singularity analysis and integrability}

The FE do not pass the Painlev\`e test, and therefore they are not integrable. 

\subsubsection{Equilibrium points in the finite region of the phase space}
\label{Section:3.2.2}
Introducing the variables
\begin{align}
& X_\phi= \frac{\partial_\tau \phi}{\sqrt{2}\sqrt{\theta^2+W}}, Y_\phi= \frac{\theta}{\sqrt{\theta^2+W}}, Y=\frac{p}{p+\mu}, S_1=\frac{y}{\sqrt{\theta^2+W}}, S_2=\frac{K}{\theta^2+W},\nonumber\\
& \Pi=\frac{p}{\theta^2+ W}, \theta=y-x,
\end{align}
\\and using the independent dimensionless variable $\lambda$ defined by the choice 
$N^2 r^2=Y^2/(\theta^2+W)$, that is
\begin{equation}
\frac{d\lambda}{d\tau}=\frac{\sqrt{\theta^2+W}}{Y},
\end{equation}
we obtain the dynamical system 
\begin{subequations}
\label{EQS-EQS}
\begin{align}
& \frac{dX_\phi}{d\lambda}=\frac{Y \left(\sqrt{2} X_\phi \left(2 \beta  S_1^2 Y_\phi-S_1 Y_\phi^2+S_1+Y_\phi \left(S_2+2 X_\phi^2-2\right)\right)-k \left(X_\phi^2-1\right) \left(Y_\phi^2-1\right)\right)}{\sqrt{2}},\\
& \frac{dY_\phi}{d\lambda}=	-\frac{1}{2} Y \left(Y_\phi^2-1\right) \left(\sqrt{2} k X_\phi Y_\phi+2
   S_1 (Y_\phi-2 \beta  S_1)-2 S_2-4 X_\phi^2\right),\\
& \frac{dS_1}{d\lambda}=\frac{Y \left(\left(\beta  S_1^2-1\right) (-2 Y Y_\phi (Y_\phi-2 \beta  S_1)-1)+S_2 (2 Y (\beta  S_1
   Y_\phi-1)-1)+X_\phi^2 (4 \beta  S_1 Y Y_\phi-2 Y-1)\right)}{2 \beta }\nonumber \\
	& -\frac{k S_1 X_\phi Y
   \left(Y_\phi^2-1\right)}{\sqrt{2}},\\
& \frac{dS_2}{d\lambda}=	S_2 Y \left(-\sqrt{2} k X_\phi \left(Y_\phi^2-1\right)+4 \beta 
   S_1^2 Y_\phi-2 S_1 \left(Y_\phi^2-1\right)+2 Y_\phi \left(S_2+2 X_\phi^2-1\right)\right),\\
	& \frac{dY}{d\lambda}=\frac{S_1 (Y-1) Y}{n+1}.
\end{align}
\end{subequations}
Additionally, 
\begin{equation}
\Pi=1-X_\phi^2-\beta S_1^2-S_2, \quad W+Y_\phi^2=1. 
\end{equation}

\begin{table}[!t]
\centering
\scalebox{0.9}{
\begin{tabular}{|c|c|}
\hline
Labels & Stability   \\\hline
$L_1^\varepsilon$ & Nonhyperbolic with a 2D unstable manifold (resp. stable) for $S_1^*<0$ (resp. $S_1^*>0$).\\\hline
$L_2^\varepsilon$ & 
 $L_2^+$ (resp. $L_2^-$) is nonhyperbolic with a 3D unstable (resp. stable) manifold for $Y_0>0, k<2\sqrt{2}$. \\\hline
$L_3^\varepsilon$ & $L_3^+$ (resp. $L_3^-$) is nonhyperbolic with a 3D unstable (resp. stable) manifold for $Y_0>0, k>-2\sqrt{2}$.\\\hline
$L_4^\varepsilon$ & saddle. \\\hline
$L_5^\varepsilon$ &$L_5^+$ (resp. $L_5^-$) is nonhyperbolic, with a 4D unstable (resp. stable) manifold for\\
& $ k\in \mathbb{R},  {X_\phi^*}=0,  \beta >\frac{9}{16},  n> 0 $, or \\
& $ -1<{X_\phi^*}<0,  k>\frac{2
   \sqrt{2}}{3 {X_\phi^*}},  \beta >\frac{9}{16} \left(1- {X_\phi^*}^2\right),  n> 0$, or \\
& $0<{X_\phi^*}<1,  k\leq \frac{2 \sqrt{2}}{3 {X_\phi^*}},  \beta >\frac{9}{16} \left(1- {X_\phi^*}^2\right),  n> 0$, or \\
& $-1<{X_\phi^*}<0,  \frac{2 \sqrt{2}}{{X_\phi^*}}<k\leq \frac{2 \sqrt{2}}{3 {X_\phi^*}},  \beta >\frac{2-2
   {X_\phi^*}^2}{k^2 {X_\phi^*}^2-4 \sqrt{2} k {X_\phi^*}+8},  n> 0$, or \\
&	$0<{X_\phi^*}<1,  \frac{2
   \sqrt{2}}{3 {X_\phi^*}}<k<\frac{2 \sqrt{2}}{{X_\phi^*}},  \beta >\frac{2-2 {X_\phi^*}^2}{k^2 {X_\phi^*}^2-4 \sqrt{2}
   k {X_\phi^*}+8},  n> 0$. \\\hline
$L_6^\varepsilon$ & saddle. \\\hline
$P_1^\varepsilon$ & Nonhyperbolic. \\\hline 
$P_2^\varepsilon$ & saddle. \\\hline
$P_3^\varepsilon$ & saddle. \\\hline
$P_4^\varepsilon$ & $P_4^+$ (resp. $P_4^-$) is nonhyperbolic with a 4D unstable (resp. stable) manifold for $n> 0, \beta >\frac{9}{16}$. \\\hline 
$P_5^\varepsilon$ & $P_5^+$ (resp. $P_5^-$) is a sink (resp. a source) for $\beta <0, n> 0$ or a saddle otherwise. \\\hline 
$P_6^\varepsilon$ & saddle. \\\hline
$P_7^\varepsilon$ & 
	$P_7^+$ (resp. $P_7^-$) is a source (resp. a sink) for \\
& $k\leq -2 \sqrt{2},  \beta >\frac{1}{k^2},  n>0$, or \\
& $-2 \sqrt{2}<k<-2 \sqrt{\frac{2}{3}},  \frac{1}{k^2}<\beta
   <-\frac{2}{k^2-8},  n>0$, or \\
& $2 \sqrt{\frac{2}{3}}<k<2 \sqrt{2},  \frac{1}{k^2}<\beta <-\frac{2}{k^2-8}, 
   n>0$, or \\
& $k\geq 2 \sqrt{2},  \beta >\frac{1}{k^2},  n>0$.
	\\
&It is a saddle otherwise. \\\hline 
$P_8^\varepsilon$& saddle. \\\hline
$P_9^\varepsilon$ &	For $\beta=1$ they are saddle. 
\\\hline 
\end{tabular}}
\caption{\label{XXXX} Stability of the equilibrium points of the dynamical system \eqref{EQS-EQS}.}
\end{table}

%%%%%%%%%Figure%%%%%%%%%%%%%%%%%%%%%%%%%%%%%%
\begin{figure*}[!t]
\includegraphics[width=1\textwidth]{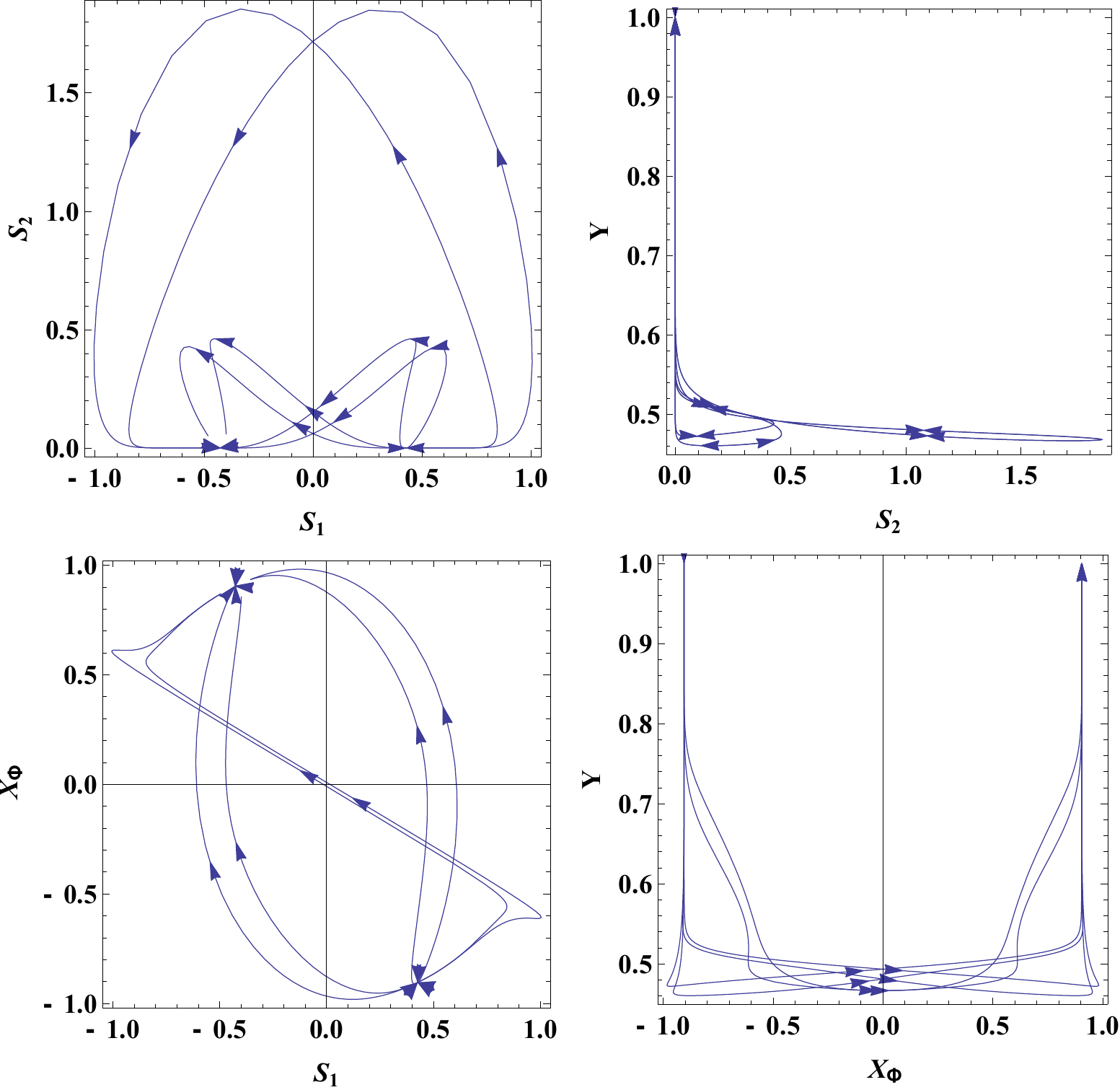}
								\caption{\label{fig:EQS-EQSA} Some 2D projections of the orbits  of the system \eqref{EQS-EQS} for $\beta=1, n=6, k=-3$.}
			\end{figure*}
%%%%%%%%%%%%%%%%%%%%%%%%%%%%%%%
%%%%%%%%%Figure%%%%%%%%%%%%%%%%%%%%%%%%%%%%%%
\begin{figure*}[!t]
\includegraphics[width=1\textwidth]{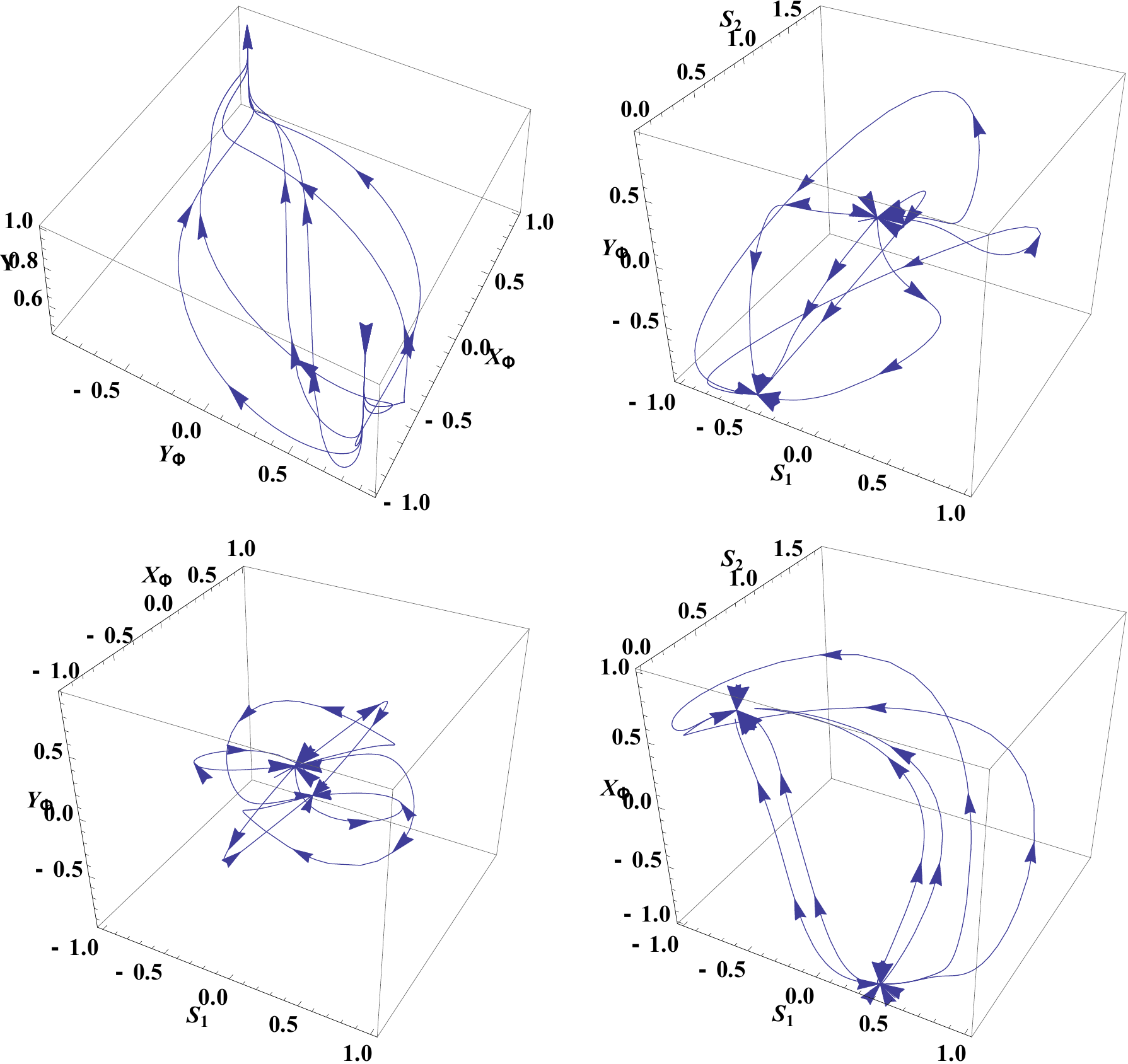}
								\caption{\label{fig:EQS-EQSB} Some 3D projections of the orbits  of the system \eqref{EQS-EQS} for $\beta=1, n=6, k=-3$.}
			\end{figure*}
%%%%%%%%%%%%%%%%%%%%%%%%%%%%%%%

\begin{table}[!t]
\centering
\scalebox{1}{
\begin{tabular}{|c|c|}
\hline
Labels & Stability   \\\hline
 $P_{10}^\varepsilon$& $P_{10}^+$ (resp. $P_{10}^-$) has a 3D stable (resp. unstable) manifold for \\
& $-2 \sqrt{2}<k<-2 \sqrt{\frac{2}{3}},  n\geq 0,  \frac{1}{4}<\beta \leq -\frac{2}{k^2-8}$, or \\
& $-2 \sqrt{2}<k<-2 \sqrt{\frac{2}{3}},  n\geq 0,  -\frac{2}{k^2}<\beta <\frac{1}{4}$, or \\
& $k=-2 \sqrt{\frac{2}{3}},  n\geq 0,  -\frac{3}{4}<\beta <\frac{1}{4}$, or \\
& $k=-2 \sqrt{\frac{2}{3}},  n\geq 0,  \frac{1}{4}<\beta <\frac{3}{8}$, or \\
& $-2 \sqrt{\frac{2}{3}}<k<-2
   \sqrt{\frac{2}{5}},  n\geq 0,  -\frac{2}{k^2}<\beta <\frac{1}{4}$, or \\
& $-2 \sqrt{\frac{2}{3}}<k<-2 \sqrt{\frac{2}{5}},  n\geq 0,  \frac{1}{4}<\beta <\frac{9 k^2-8}{16 k^2}$, or \\
& $k=-2 \sqrt{\frac{2}{5}},  n\geq 0,  -\frac{5}{4}<\beta <\frac{1}{4}$, or \\
& $-2 \sqrt{\frac{2}{5}}<k<-\frac{2 \sqrt{2}}{3},  n\geq 0,  -\frac{2}{k^2}<\beta <\frac{9 k^2-8}{16 k^2}$, or \\
& $k=-\frac{2 \sqrt{2}}{3},  n\geq 0,  -\frac{9}{4}<\beta <0$, or \\
&$-\frac{2 \sqrt{2}}{3}<k<0,  n\geq 0,  -\frac{2}{k^2}<\beta <\frac{9 k^2-8}{16 k^2}$. \\
& $P_{10}^+$ (resp. $P_{10}^-$) has a 3D unstable (resp. stable) manifold for \\
& $-2 \sqrt{2}<k<0,  n\geq 0,  \beta <-\frac{2}{k^2}$, or \\
& $2 \sqrt{\frac{2}{3}}<k<2 \sqrt{2},  n\geq 0,  \frac{9 k^2-8}{16 k^2}<\beta \leq -\frac{2}{k^2-8}$, or \\
& $k\geq 2 \sqrt{2}, n\geq 0,  \beta >\frac{9 k^2-8}{16 k^2}$.\\\hline
$P_{11}^\varepsilon$ & $P_{11}^+$ (resp. $P_{11}^-$) has a 3D stable (resp. unstable) manifold for \\ 
& $k\leq -2 \sqrt{2},  n\geq 0,  \beta >\frac{9 k^2-8}{16 k^2}$, or \\
& $-2 \sqrt{2}<k<-2 \sqrt{\frac{2}{3}},  n\geq 0,  \frac{9 k^2-8}{16 k^2}<\beta \leq -\frac{2}{k^2-8}$, or \\
& $0<k<2\sqrt{2},  n\geq 0,  \beta <-\frac{2}{k^2}$. \\
&		$P_{11}^+$ (resp. $P_{11}^-$) has a 3D unstable (resp. stable) manifold for \\
& $k=\frac{2 \sqrt{2}}{3},  n\geq 0,  -\frac{9}{4}<\beta <0$, or \\
& $2 \sqrt{\frac{2}{5}}<k<2 \sqrt{2},  n\geq 0,  -\frac{2}{k^2}<\beta <\frac{1}{4}$, or \\
& $0<k<\frac{2 \sqrt{2}}{3},  n\geq   0,  -\frac{2}{k^2}<\beta <\frac{9 k^2-8}{16 k^2}$, or \\ 
& $\frac{2 \sqrt{2}}{3}<k\leq 2 \sqrt{\frac{2}{5}},  n\geq 0,  -\frac{2}{k^2}<\beta <\frac{9 k^2-8}{16 k^2}$, or \\
& $2  \sqrt{\frac{2}{5}}<k\leq 2 \sqrt{\frac{2}{3}},  n\geq 0,  \frac{1}{4}<\beta <\frac{9 k^2-8}{16 k^2}$, or \\
& $2 \sqrt{\frac{2}{3}}<k<2 \sqrt{2},  n\geq 0,  \frac{1}{4}<\beta \leq -\frac{2}{k^2-8}$.
\\\hline 
\end{tabular}}
\caption{\label{YYYY} Stability of the equilibrium points of the dynamical system \eqref{EQS-EQS} (cont.).}
\end{table}
Therefore, the system defines a flow on the invariant set 
\begin{equation}
\Big\{(X_\phi, Y_\phi, S_1, S_2,Y)\in \mathbb{R}^5: X_\phi^2+ \beta S_1^2+S_2\leq 1, -1\leq Y_\phi \leq 1, S_2\geq 0, 0\leq Y\leq 1\Big\}.
\end{equation}
The system is invariant under the change of coordinates $(X_\phi, Y_\phi, S_1) \rightarrow (-X_\phi, -Y_\phi, -S_1)$ with the simultaneous reversal in the independent variable ${\lambda}\rightarrow -{\lambda}$.  In relation to the phase space
dynamics this implies that for two points related by this symmetry, say $P^+$ and $P^-$, each has the
opposite dynamical behavior to the other; that is, if the equilibrium point $P^+$  is an attractor for a
choice of parameters, then $P^-$  is a source under the same choice. The details
of the stability of the equilibrium points (curves/surfaces of equilibrium points) of the dynamical system \eqref{EQS-EQS} are presented in
Appendix \ref{App_A2}. These results are summarized in Tables \ref{XXXX} and  \ref{YYYY}. 

We have the additional relations 
\begin{align}
& r^2=\frac{\left(1-X_\phi^2-\beta S_1^2-S_2\right) q^n}{S_2}\left(\frac{1-Y}{Y}\right)^{n+1}, \quad 
\mu =\frac{1}{q^{n}} \left(\frac{Y}{1-Y}\right)^{n}, \quad  p= \frac{1}{q^{n}} \left(\frac{Y}{1-Y}\right)^{n+1},\\
&\mathcal{M}(r):=\frac{m}{r}=\frac{{S_2}-({Y_\phi}-{S_1})^2}{2 {S_2}}. 
\end{align}
From these expressions we obtain the $m-r$ relation: 
\begin{equation}
\label{RM4}
r^2=2 {q}^{n} \frac{\mathcal{M}}{D},\quad 
m^2=2 {q}^{n} \frac{\mathcal{M}^3}{D}, \quad 
D= \frac{\left({S_2}-({S_1}-Y_{\phi})^2\right)}{
   \left(1-\beta  {S_1}^2-{S_2}-X_{\phi}^2\right)} \left(\frac{1-Y}{Y}\right)^{-(n+1)}.
\end{equation}
Using these equation we define the compact variables 
\begin{equation}
\label{RM4compact}
R_{\text{comp}}=\frac{q^{-\frac{n}{2}}r}{\sqrt{1+ q^{-n} r^2\left(1+{\mathcal{M}}^2\right)}}, \quad M_{\text{comp}}=\frac{q^{-\frac{n}{2}}\mathcal{M}r}{\sqrt{1+ q^{-n} r^2\left(1+{\mathcal{M}}^2\right)}}, 
\end{equation}
which depend only of the phase-space variables $(X_\phi, Y_\phi, S_1,S_2,Y)$. Evaluating numerically the expressions $R_{\text{comp}}, M_{\text{comp}}$ on the orbits of the system \eqref{EQS-EQS}, we can determine whether the resulting model leads to finite radius and finite mass. 
To complete this section we make some numerical integrations of the system \eqref{EQS-EQS}. In the figure \ref{fig:EQS-EQSA} are presented some 2D projections of the orbits  of the system \eqref{EQS-EQS} for $\beta=1, n=6, k=-3$. In the figure \eqref{fig:EQS-EQSB} are presented some 3D projections of the orbits  of the system \ref{EQS-EQS} for the same choice of parameters. 
According to our analytical results, the points with the 5D unstable or stable manifolds are
$P_5^\varepsilon$: $(X_\phi, Y_\phi, S_1, S_2,Y)=\Big(0,\varepsilon, \frac{3\varepsilon}{4 \beta },0,1\Big)$, exists for  $\beta <0$, or $\beta \geq \frac{9}{16}$,  and  $P_7^\varepsilon$: $(X_\phi, Y_\phi, S_1, S_2,Y)=\Big(\frac{k \sqrt{\beta }}{\sqrt{\beta  k^2+2}}\varepsilon, \frac{2 \sqrt{2} \sqrt{\beta }}{\sqrt{\beta  k^2+2}}\varepsilon, \frac{\sqrt{2}}{\sqrt{\beta } \sqrt{\beta  k^2+2}}\varepsilon,0,1\Big)$.
$P_5^+$ (resp. $P_5^-$) is a sink (resp. a source) for $\beta <0, n> 0$ or a saddle otherwise. $P_7^+$ (resp. $P_7^-$) is a source (resp. a sink) for $k\leq -2 \sqrt{2},  \beta >\frac{1}{k^2},  n>0$, or $-2 \sqrt{2}<k<-2 \sqrt{\frac{2}{3}},  \frac{1}{k^2}<\beta
   <-\frac{2}{k^2-8},  n>0$, or  $2 \sqrt{\frac{2}{3}}<k<2 \sqrt{2},  \frac{1}{k^2}<\beta <-\frac{2}{k^2-8}, 
   n>0$, or $k\geq 2 \sqrt{2},  \beta >\frac{1}{k^2},  n>0$.
In the simulations presented in Figs. \ref{fig:EQS-EQSA}, \ref{fig:EQS-EQSB} the attractor as $\lambda \rightarrow +\infty$ is thereof the equilibrium point $P_7^{-}$, and as $\lambda \rightarrow -\infty$ the past attractor is $P_7^+$.
Comparing the models 3 and 4 it can be argued that the dynamics of the scalar field is very similar irrespectively the equation of state. 
\section{Conclusions}
\label{conclusions}

In this paper we have investigated the FE in the Einstein-aether
model in a static spherically symmetric spacetime. We have studied if the gravitational FE posses the Painlev\`e property, and consequently whether an  analytic explicit integration can be performed for the FE. We have applied the classical treatment for the singularity analysis as summarized in the ARS algorithm. As far as  the dynamical system with only the perfect fluid is concerned, we showed that the FE are integrable and we wrote the analytic solution in mixed Laurent expansions. That is, in the case of
the static spherically symmetric spacetime with perfect fluid there always exist  a negative
resonance which means that the Laurent expansion is expressed in a Left and
Right Painlev\'{e} Series, because we have integrated
over annulus around the singularity which has two borders \cite{Andriopoulos}.
On the other hand, in the presence of the scalar field, the FE do not generally pass the Painlev\`e test. 
We have shown that if the scalar field is not present, and assuming the perfect fluid has the EoS $\mu =\mu _{0}+\left( \eta -1\right) p,~\eta >1$, the system passes 
the singularity test which means that the FE are integrable (in a similar way to what we have done for Kantowski-Sachs Einstein-aether theories in \cite{Latta:2016jix}).

We have also studied the case of a polytropic equation of state $p=k \mu^\Gamma, \Gamma=1+\frac{1}{n}, n>0$ in the static Eistein-aether theory, by means of the $(\Sigma, \Sigma_2,Y)$-formulation given by the the model \eqref{XXXeq:23}.  For $\beta=1$ the model of \cite{Nilsson:2000zg} is recovered. We have found  that in GR the point $(\Sigma, \Sigma_2,Y)=\left(\frac{2}{3}, \frac{1}{9}, 1\right)$, corresponds to the so called self-similar Tolman solution discussed in \cite{Nilsson:2000zf}. In the AE theory this solution is generalized to a 1-parameter set of solutions given by the equilibrium points $P_1: (\Sigma, \Sigma_2,Y)= \left(\frac{2}{3}, \frac{1}{9} (9-8\beta), 1\right)$. The solutions having $Y=1$ correspond to the limiting situation $\frac{p}{\mu}\rightarrow \infty.$ In the case of a linear equation of state $p=(\gamma-1)\mu$, this is consistent with $\gamma\rightarrow \infty$. 

We have also taken a natural extension of the previous analysis, as in the general relativistic static case \cite{Nilsson:2000zg}, by studying a stationary comoving aether with a perfect fluid and a scalar field with exponential potential.

Furthermore, we have applied the Tolman-Oppenheimer-Volkoff (TOV) approach \cite{Tolman:1939jz,Oppenheimer:1939ne,Misner:1964zz}
(we use units where $8  \pi G=1$)
 and found that the usual relativistic TOV equations
are then drastically modified in Einstein-aether theory for a perfect fluid to \eqref{modified-TOV-eqs},
where $\beta=c_a+1$, and $c_a$ is the aether parameter. This modification has several physical implication that we have discussed.  Then we have constructed a 3D dynamical system in compact variables and obtained a picture of the entire solution space for different choices of the EoS  that can visualized in a geometrical way. For higher dimensional systems we still can obtain information by numerical integrations and the use of projections. The results obtained have been discussed physically, and we have obtained an appropriate description of the universe both on local and larger scales.

Finally, we have studied the model with an additional scalar field with an exponential potential. First we show that the singularity test is not passed, except in the special case of a linear EoS with parameter $\mu_0=0$. We then performed a phase space analysis and presented all of the equilibrium points in the cases of both a linear EoS and a  polytropic EoS, and described the corresponding exact solutions and discussed their physical properties in the appendix. We note that the sources and sinks in the linear EoS are the pairs $P_5^{\varepsilon}$ and $P_7^{\varepsilon}$ for appropriate values of the model parameters (and the corresponding pairs for the polytropic EoS case studied subsequently), which all have $S_1=\frac{K}{\theta^2+W}=0$. In particular in the case of $P_5^{\varepsilon}$ (where $C=0$ and $Q=\pm 1$) the asymptotic metric, corresponding a a self-similar plane symmetric model containing a singularity with a non-trivial horizon, was explicitly displayed.  

\section*{Acknowledgments}

G. L. was funded by Comisi\'on Nacional de Investigaci\'on Cient\'{\i}fica y Tecnol\'ogica (CONICYT) through FONDECYT Iniciaci\'on grant no.
11180126 and by
Vicerrector\'ia de Investigaci\'on y Desarrollo Tecnol\'ogico at Universidad
Cat\'olica del Norte. A. C. was supported by NSERC of Canada and partially supported by FONDECYT Iniciaci\'on grant no.
11180126. The authors would like to thank Prof. PGL Leach and to J. Latta for useful comments and suggestions.

\appendix

\section{Equilibrium points for Model 3}
\label{App_A}

The relation between the new variables \eqref{Vars-Model_3} and the set of variables used in \cite{Coley:2019tyx}, say \\$e_{1}{}^{1}\frac{d \bar{\lambda}}{d r}=\sqrt{\frac{\mu_0}{\eta }+\theta^2}, 
Q=\frac{\theta}{\sqrt{\frac{\mu_0}{\eta }+\theta^2}},   S=\frac{\sigma}{\sqrt{\frac{\mu_0}{\eta }+\theta^2}},  C=\frac{\eta  K}{\mu_0+\eta \theta^2},
A_{\phi}=\frac{\Phi}{\sqrt{2}\sqrt{\frac{\mu_0}{\eta }+\theta^2}},  A_{W}=\frac{\sqrt{W_0} e^{-\frac{k}{2} \phi }}{\sqrt{\frac{\mu_0}{\eta }+\theta^2}}$, 
is
\begin{subequations} 
\begin{align}
& \frac{d\bar{\lambda}}{d\lambda}=\sqrt{Y \left(Y \left(\beta  {S_1}^2+{S_2}+{X_\phi}^2+{Y_\phi}^2-1\right)-\frac{\beta  {S_1}^2+{S_2}+{X_\phi}^2-1}{\eta }\right)}, 
\\
& Q=\frac{{Y_\phi} \sqrt{\eta  Y}}{\sqrt{(1-\eta  Y) \left(-\beta  {S_1}^2-{S_2}-{X_\phi}^2+1\right)+\eta  Y {Y_\phi}^2}},\\
& S= \frac{{S_1}
   \sqrt{\eta  Y}}{\sqrt{(1-\eta  Y) \left(-\beta  {S_1}^2-{S_2}-{X_\phi}^2+1\right)+\eta  Y {Y_\phi}^2}},\\
& C=	\frac{\eta  {S_2} Y}{-\beta  {S_1}^2-\eta  Y \left(-\beta 
   {S_1}^2-{S_2}-{X_\phi}^2-{Y_\phi}^2+1\right)-{S_2}-{X_\phi}^2+1},\\
& A_\phi=	\frac{\eta  {X_\phi} Y}{\sqrt{(1-\eta  Y) \left(-\beta  {S_1}^2-{S_2}-{X_\phi}^2+1\right)+\eta  Y
   {Y_\phi}^2}},\\
& A_W=\frac{\sqrt{\eta  Y \left(1-{Y_\phi}^2\right)}}{\sqrt{-\beta  {S_1}^2-\eta  Y \left(-\beta  {S_1}^2-{S_2}-{X_\phi}^2-{Y_\phi}^2+1\right)-{S_2}-{X_\phi}^2+1}}.
\end{align} 
\end{subequations}

The details
of the stability of the relevant equilibrium point (curves/surfaces of equilibrium point) of the dynamical system \eqref{EQS-3-13} are as follows.

\begin{enumerate}
\item $L_1$: $(X_\phi, Y_\phi, S_1, S_2,Y)=\left(X_{\phi}^*, Y_{\phi}^*,S_1^*,1-\beta  {S_1^{*}}^2-{X_{\phi}^*}^2,0\right)$.  Always exists. The eigenvalues are $0,0,0,-S_1^*,-S_1^*$. It is nonhyperbolic with a 2D unstable manifold (resp. stable) for $S_1^*<0$ (resp. $S_1^*>0$).

\item $L_2^{\varepsilon}$: $(X_\phi, Y_\phi, S_1, S_2,Y)= (\varepsilon,\varepsilon,0,0,Y_0)$,  where $\varepsilon=\pm 1$ is the sign of $\theta$. Exists for $-1\leq Y_0\leq 1$. The eigenvalues are $0,0,2 Y_0 \varepsilon, 4 Y_0 \varepsilon,\left(4-\sqrt{2} k\right) Y_0 \varepsilon$.  $L_2^+$ (resp. $L_2^-$ ) is nonhyperbolic with a 3D unstable (resp. stable) manifold for $Y_0>0, k<2\sqrt{2}$. Otherwise the unstable (resp. stable) manifold  is 2D.

\item $L_3^{\varepsilon}$: $(X_\phi, Y_\phi, S_1, S_2,Y)=(-\varepsilon,\varepsilon,0,0,Y_0)$. Always exists. The eigenvalues are \\$0,0,4 Y_0 \varepsilon,2 Y_0 \varepsilon,\left(\sqrt{2} k+4\right) Y_0 \varepsilon$. $L_3^+$ (resp. $L_3^-$) is nonhyperbolic with a 3D unstable (resp. stable) manifold for $Y_0>0, k>-2\sqrt{2}$. Otherwise the unstable (resp. stable) manifold is 2D.

\item $L_4^\varepsilon$: $(X_\phi, Y_\phi, S_1, S_2,Y)=(0,\varepsilon,0,1,Y_0)$. Exists for $-1\leq Y_0\leq 1$. The eigenvalues are\\ $0,-Y_0 \varepsilon,-Y_0 \varepsilon,2 Y_0 \varepsilon,2 Y_0 \varepsilon$.  They are nonhyperbolic and behaves as saddle. 

\item $L_5^\varepsilon$: $(X_\phi, Y_\phi, S_1, S_2,Y)=\left(\varepsilon X_{\phi}^*,\varepsilon ,\varepsilon \frac{\sqrt{1- {X_{\phi}^*}^2}}{\sqrt{\beta }},0,\frac{1}{\eta }\right)$. Exists for $\beta<0, {X_{\phi}^*}^2\geq 1$ or $\beta>0, {X_{\phi}^*}^2\leq 1$. The eigenvalues are 
$0,\frac{2}{\eta }\varepsilon,\frac{-\sqrt{2} k { X_{\phi}^*}-\frac{2 \sqrt{1-{ X_{\phi}^*}^2}}{\sqrt{\beta }}+4}{\eta } \varepsilon,\frac{\sqrt{1-{ X_{\phi}^*}^2}}{\sqrt{\beta }} \varepsilon,\frac{4 \sqrt{\beta }-(\eta +2) \sqrt{1-{ X_{\phi}^*}^2}}{\sqrt{\beta } \eta }\varepsilon$.   $L_5^+$ (resp. $L_5^-$) is nonhyperbolic with a 4D unstable (resp. stable) manifold for 
\begin{enumerate}
\item $ k\in \mathbb{R},  {X_\phi^*}=0,  \eta >1,  \beta >\frac{1}{16} (\eta +2)^2 $, or  
\item $ -1<{X_\phi^*}<0,  k\geq
   \frac{2 \sqrt{2}}{3 {X_\phi^*}},  \eta >1,  \beta >-\frac{1}{16} (\eta +2)^2 ({X_\phi^*}^2-1) $, or 
\item  $ 0<{X_\phi^*}<1,  k\leq \frac{2 \sqrt{2}}{3 {X_\phi^*}},  \eta >1,  \beta >-\frac{1}{16} (\eta +2)^2 ({X_\phi^*}^2-1)$, or  
\item	$ -1<{X_\phi^*}<0,  \frac{2 \sqrt{2}}{{X_\phi^*}}<k<\frac{2 \sqrt{2}}{3 {X_\phi^*}},  \eta >-\frac{2 \sqrt{2} k {X_\phi^*}}{\sqrt{2} k {X_\phi^*}-4},  
	\beta >-\frac{1}{16} (\eta +2)^2 ({X_\phi^*}^2-1) $, or  
\item	$ 0<{X_\phi^*}<1,  \frac{2 \sqrt{2}}{3 {X_\phi^*}}<k<\frac{2 \sqrt{2}}{{X_\phi^*}}, 
   \eta \geq -\frac{2 \sqrt{2} k {X_\phi^*}}{\sqrt{2} k {X_\phi^*}-4},  
	\beta >-\frac{1}{16} (\eta +2)^2 ({X_\phi^*}^2-1) $, or  
\item $ -1<{X_\phi^*}<0,  \frac{2 \sqrt{2}}{{X_\phi^*}}<k<\frac{2 \sqrt{2}}{3 {X_\phi^*}}, 
   1<\eta \leq -\frac{2 \sqrt{2} k {X_\phi^*}}{\sqrt{2} k {X_\phi^*}-4},  
	\beta >\frac{2 ({X_\phi^*}^2-1)}{-k^2 {X_\phi^*}^2+4 \sqrt{2} k {X_\phi^*}-8} $, or  
\item	$0<{X_\phi^*}<1,  \frac{2 \sqrt{2}}{3 {X_\phi^*}}<k<\frac{2 \sqrt{2}}{{X_\phi^*}},  1<\eta <-\frac{2 \sqrt{2} k {X_\phi^*}}{\sqrt{2} k {X_\phi^*}-4},  \beta >\frac{2
   ({X_\phi^*}^2-1)}{-k^2 {X_\phi^*}^2+4 \sqrt{2} k {X_\phi^*}-8}$. 
\end{enumerate}

\item $L_6^\varepsilon$: $(X_\phi, Y_\phi, S_1, S_2,Y)=\left( \varepsilon X_{\phi}^*,\varepsilon,- \varepsilon\frac{\sqrt{1- {X_{\phi}^*}^2}}{\sqrt{\beta }},0,\frac{1}{\eta }\right)$. Exists for $\beta<0, {X_{\phi}^*}^2\geq 1$ or $\beta>0, {X_{\phi}^*}^2\leq 1$. The eigenvalues are 
$0,\frac{2}{\eta }  \varepsilon,\frac{-\sqrt{2} k { X_{\phi}^*}+\frac{2 \sqrt{1-{ X_{\phi}^*}^2}}{\sqrt{\beta }}+4}{\eta } \varepsilon,-\frac{\sqrt{1-{ X_{\phi}^*}^2}}{\sqrt{\beta }} \varepsilon,\frac{4 \sqrt{\beta }+(\eta +2) \sqrt{1-{ X_{\phi}^*}^2}}{\sqrt{\beta } \eta }  \varepsilon$.   Nonhyperbolic, behaves as saddle.  

\item $P_1^\varepsilon$: $(X_\phi, Y_\phi, S_1, S_2,Y)= \left(\varepsilon,\frac{2 \sqrt{2}}{k}\varepsilon,0,0,0\right)$.  Exists for $k\leq -2 \sqrt{2}$ or $k\geq 2 \sqrt{2}$. The eigenvalues are $0,0,0,0,0$. They are nonhyperbolic.

\item $P_2^\varepsilon$: $(X_\phi, Y_\phi, S_1, S_2,Y)= \left(0,\varepsilon,0,1,\frac{1}{\eta }\right)$. Always exists. The eigenvalues are $0,-\frac{1}{\eta }\varepsilon,-\frac{1}{\eta }\varepsilon,\frac{2}{\eta }\varepsilon,\frac{2}{\eta }\varepsilon$.  They are nonhyperbolic. Behaves as a saddle. 

\item $P_3^\varepsilon$: $(X_\phi, Y_\phi, S_1, S_2,Y)= \left(0,\varepsilon,\frac{1}{\sqrt{\beta }}\varepsilon,0,\frac{1}{\eta }\right)$. Exists for $\beta>0$. The  eigenvalues are\\  $0,\frac{4-\frac{2}{\sqrt{\beta }}}{\eta }\varepsilon, \frac{2}{\eta }\varepsilon, \frac{4 \sqrt{\beta }-\eta -2}{\sqrt{\beta } \eta }\varepsilon,\frac{1}{\sqrt{\beta }}\varepsilon$.  $P_3^+$ (resp. $P_3^-$) is nonhyperbolic with a 4D unstable (resp. stable) manifold for $\eta >1,  \beta >\frac{1}{16} (\eta +2)^2$.

\item $P_4^\varepsilon$: $(X_\phi, Y_\phi, S_1, S_2,Y)= \left(0,\varepsilon,-\frac{1}{\sqrt{\beta }}\varepsilon,0,\frac{1}{\eta }\right)$.  Exists for $\beta>0$. The  eigenvalues are\\ $0,\frac{2 \left(\frac{1}{\sqrt{\beta }}+2\right)}{\eta }\varepsilon, \frac{2}{\eta }\varepsilon, -\frac{1}{\sqrt{\beta }}\varepsilon, \frac{4 \sqrt{\beta }+\eta +2}{\sqrt{\beta } \eta }\varepsilon$.
They are nonhyperbolic and behaves as saddle. 

\item $P_5^\varepsilon$: $(X_\phi, Y_\phi, S_1, S_2,Y)= \left(0,\varepsilon,\frac{\eta +2}{4 \beta }\varepsilon,0,\frac{1}{\eta }\right)$. Exists for $\beta <0,  \eta \geq 1$ or $\eta \geq 1,  \beta \geq \frac{1}{16} (\eta +2)^2$. The eigenvalues are $\frac{\eta +2}{4 \beta }\varepsilon, \frac{\eta +2}{4 \beta }\varepsilon, \frac{(\eta +2)^2-8 \beta }{4 \beta  \eta }\varepsilon, \frac{(\eta +2)^2-16 \beta }{8 \beta  \eta }\varepsilon, \frac{(\eta +2)^2-16 \beta }{8 \beta  \eta }\varepsilon$.   $P_5^+$ (resp. $P_5^-$) is a sink (resp. a source) for $\beta <0,  \eta >1$. They are saddle otherwise. 

\item $P_6^\varepsilon$: $(X_\phi, Y_\phi, S_1, S_2,Y)= \left(0,\varepsilon,\frac{2}{\eta +2} \varepsilon,1-\frac{8 \beta }{(\eta +2)^2},\frac{1}{\eta }\right)$.  Exists for $\eta \geq 1, 0\leq \beta \leq \frac{1}{8} (\eta +2)^2$. The eigenvalues are
$\frac{2}{\eta +2}\varepsilon, \frac{2}{\eta +2}\varepsilon, -\frac{1}{\eta }\varepsilon,-\frac{\sqrt{64 \beta -7 (\eta +2)^2}+\eta +2}{2 \eta  (\eta +2)}\varepsilon, \frac{\sqrt{64 \beta -7 (\eta +2)^2}-\eta -2}{2 \eta  (\eta +2)}\varepsilon$. 
		  They are saddle. 
		
\item $P_7^\varepsilon$: $(X_\phi, Y_\phi, S_1, S_2,Y)= \left(\frac{k \sqrt{\beta }}{\sqrt{\beta  k^2+2}}\varepsilon, \frac{2 \sqrt{2} \sqrt{\beta }}{\sqrt{\beta  k^2+2}}\varepsilon, \frac{\sqrt{2}}{\sqrt{\beta } \sqrt{\beta  k^2+2}}\varepsilon, 0, \frac{1}{\eta }\right)$.  Exists for
\begin{enumerate}
\item $k\leq -2 \sqrt{2},  \beta \geq 0$, or 
\item $-2 \sqrt{2}<k<2 \sqrt{2},  0\leq \beta \leq -\frac{4}{2 k^2-16}$, or
\item $k\geq 2 \sqrt{2},  \beta \geq 0$. 
\end{enumerate}
 The eigenvalues are
$\frac{\sqrt{2} \left(\beta  \left(k^2-4\right)+2\right)}{\sqrt{\beta } \eta  \sqrt{\beta  k^2+2}}\varepsilon, \frac{\beta 
   \left(k^2-8\right)+2}{\sqrt{2} \sqrt{\beta } \eta  \sqrt{\beta  k^2+2}}\varepsilon, \frac{\beta  \left(k^2-8\right)+2}{\sqrt{2} \sqrt{\beta } \eta 
   \sqrt{\beta  k^2+2}}\varepsilon, \frac{\sqrt{2} \left(\beta  k^2-\eta \right)}{\sqrt{\beta } \eta  \sqrt{\beta  k^2+2}}\varepsilon, \frac{\sqrt{2}}{\sqrt{\beta }
   \sqrt{\beta  k^2+2}}\varepsilon$. \\ $P_7^+$ (resp. $P_7^-$) is a source (resp. sink) for
	 \begin{enumerate}
	 \item $ k\leq -2 \sqrt{2},  \beta >\frac{1}{k^2},  1<\eta <\beta  k^2 $, or  
	 \item $ -2 \sqrt{2}<k<-2 \sqrt{\frac{2}{3}}, 
   \frac{1}{k^2}<\beta <-\frac{2}{k^2-8},  1<\eta <\beta  k^2 $, or 
	 \item $ 2 \sqrt{\frac{2}{3}}<k<2 \sqrt{2},  \frac{1}{k^2}<\beta
   <-\frac{2}{k^2-8},  1<\eta <\beta  k^2 $, or 
	 \item $ k\geq 2 \sqrt{2},  \beta >\frac{1}{k^2},  1<\eta <\beta  k^2 $.
	 \end{enumerate}

\item $P_8^\varepsilon$:  $(X_\phi, Y_\phi, S_1, S_2,Y)=\left(\frac{k \beta \varepsilon}{\sqrt{\frac{\beta  k^2}{2}+1} \sqrt{\left(k^2-2\right) \beta +2}},\frac{\varepsilon\sqrt{\beta  k^2+2}}{\sqrt{\left(k^2-2\right) \beta +2}},\frac{2 \varepsilon}{\sqrt{\beta  k^2+2} \sqrt{\left(k^2-2\right) \beta
   +2}},\frac{\left(k^2-4\right) \beta +2}{\left(k^2-2\right) \beta +2},\frac{1}{\eta }\right)$. 	Exists for
	\begin{enumerate}
	\item $k<0,  -\frac{2}{k^2}\leq \beta \leq 0$, or 
	\item $k=0,  \beta \leq 0$, or 
	\item $k>0,  -\frac{2}{k^2}\leq \beta \leq 0$.
	\end{enumerate}
		The eigenvalues are 
	$-\frac{\sqrt{\beta  k^2+2}}{\eta  \sqrt{\beta  \left(k^2-2\right)+2}}\varepsilon, \frac{-\sqrt{\beta  k^2+2}-\sqrt{\beta  \left(-\left(7
   k^2-32\right)\right)-14}}{2 \eta  \sqrt{\beta  \left(k^2-2\right)+2}}\varepsilon$,\\$\frac{\sqrt{\beta  \left(-\left(7 k^2-32\right)\right)-14}-\sqrt{\beta 
   k^2+2}}{2 \eta  \sqrt{\beta  \left(k^2-2\right)+2}}\varepsilon, \frac{2 \beta  k^2-2 \eta }{\eta  \sqrt{\beta  k^2+2} \sqrt{\beta 
   \left(k^2-2\right)+2}}\varepsilon, \frac{2}{\sqrt{\beta  k^2+2} \sqrt{\beta  \left(k^2-2\right)+2}}\varepsilon$.  They are saddle. 

\item $P_{9}^\varepsilon$:  $(X_\phi, Y_\phi, S_1, S_2,Y)=$\\\noindent\( {\left(\varepsilon\frac{k \eta  (2+\eta )}{\sqrt{32 k^4 \beta ^2-16 \eta ^3+k^2 \eta  \left(16 \beta  (-2+\eta )+(2+\eta )^3\right)}},\right.}
 {\varepsilon\frac{2 \sqrt{2} \left(2 k^2 \beta +\eta ^2\right)}{\sqrt{32 k^4 \beta ^2-16 \eta ^3+k^2 \eta  \left(16 \beta  (-2+\eta )+(2+\eta )^3\right)}},}\\
 {\left. \varepsilon\frac{\sqrt{2} k^2 (2+\eta )}{\sqrt{32 k^4 \beta ^2-16 \eta ^3+k^2 \eta  \left(16 \beta  (-2+\eta )+(2+\eta )^3\right)}},0,\frac{1}{\eta
}\right)}\).  Exists for
\begin{enumerate}
\item $\eta \geq 1,  k>2 \sqrt{2},  \beta >\frac{1}{4}$, or 
\item $\eta \geq 1,  k=-2 \sqrt{2},  \beta >\frac{1}{4}$, or
\item $\eta \geq 1,  k=2 \sqrt{2},  \beta >\frac{1}{4}$, or 
\item $\eta \geq 1,  k<-2 \sqrt{2},  \beta >\frac{1}{4}$, or 
\item $\eta \geq 1,  k=0,  \beta <\frac{1}{4}$, or 
\item $\eta \geq 1,  k=-2 \sqrt{2},  \beta <-\frac{1}{4}$, or 
\item $\eta \geq 1,  k=2 \sqrt{2}, \beta <-\frac{1}{4}$, or 
\item $\eta \geq 1,  k>2 \sqrt{2},  -\frac{2}{k^2}<\beta <\frac{1}{4}$, or
\item $\eta \geq 1,  k<-2 \sqrt{2},  -\frac{2}{k^2}<\beta <\frac{1}{4}$, or
\item $\eta \geq 1,  k>2 \sqrt{2},  -\frac{2}{k^2-8}\leq \beta <-\frac{2}{k^2}$, or 
\item $\eta \geq 1,  k<-2 \sqrt{2}, -\frac{2}{k^2-8}\leq \beta <-\frac{2}{k^2}$, or 
\item $\eta \geq 1,  k=-2 \sqrt{2},  -\frac{1}{4}<\beta <\frac{1}{4}$, or
 \item $\eta \geq 1,  k=2 \sqrt{2},  -\frac{1}{4}<\beta <\frac{1}{4}$, or 
\item $\eta \geq 1,  0<k<2 \sqrt{2},  \beta  <-\frac{2}{k^2}$, or 
\item $\eta \geq 1,  -2 \sqrt{2}<k<0,  \beta <-\frac{2}{k^2}$, or 
\item $\eta \geq 1,  0<k<2 \sqrt{2},  \frac{1}{4}<\beta \leq -\frac{2}{k^2-8}$, or 
\item $\eta \geq 1,  -2 \sqrt{2}<k<0,  \frac{1}{4}<\beta \leq -\frac{2}{k^2-8}$, or 
\item $\eta \geq 1,  0<k<2 \sqrt{2},  -\frac{2}{k^2}<\beta <\frac{1}{4}$, or 
\item $\eta \geq 1,  -2 \sqrt{2}<k<0,  -\frac{2}{k^2}<\beta <\frac{1}{4}$.
\end{enumerate}

The eigenvalues are\\\noindent\( {0,0,\varepsilon\frac{2 \left(2 k \beta +\sqrt{2+\left(-8+k^2\right) \beta }\right) \sqrt{2+\beta  \left(-8+k \left(k+4 k \beta -4 \sqrt{2+\left(-8+k^2\right)
\beta }\right)\right)}}{(-1+4 \beta ) \left(2+k^2 \beta \right) \eta },}\\
 {\varepsilon\frac{\left(k+2 \sqrt{2+\left(-8+k^2\right) \beta }\right) \sqrt{2+\beta  \left(-8+k \left(k+4 k \beta -4 \sqrt{2+\left(-8+k^2\right) \beta
}\right)\right)}}{(-1+4 \beta ) \left(2+k^2 \beta \right)},}\\
 {\varepsilon\frac{\sqrt{2+\beta  \left(-8+k \left(k+4 k \beta -4 \sqrt{2+\left(-8+k^2\right) \beta }\right)\right)} \left(k (-2+8 \beta -\eta )-2 \sqrt{2+\left(-8+k^2\right)
\beta } \eta \right)}{(-1+4 \beta ) \left(2+k^2 \beta \right) \eta }}\).\\ For some choices of parameters, say  $\beta=1, \eta=1$, or $\beta=1, \eta=2$, they are saddle. 
	
\item $P_{10}^\varepsilon$: $(X_\phi, Y_\phi, S_1, S_2,Y)=$\noindent\( {\left(\varepsilon\frac{\sqrt{4+2 \beta  \left(-8+k \left(k+4 k \beta -4 \sqrt{2+\left(-8+k^2\right) \beta }\right)\right)}}{2+k^2 \beta },\right.}\\
 {\varepsilon\frac{\left(2 k \beta +\sqrt{2+\left(-8+k^2\right) \beta }\right) \sqrt{2+\beta  \left(-8+k \left(k+4 k \beta -4 \sqrt{2+\left(-8+k^2\right)
\beta }\right)\right)}}{(-1+4 \beta ) \left(2+k^2 \beta \right)},}\\
 {\left. \varepsilon\frac{\left(k+2 \sqrt{2+\left(-8+k^2\right) \beta }\right) \sqrt{2+\beta  \left(-8+k \left(k+4 k \beta -4 \sqrt{2+\left(-8+k^2\right)
\beta }\right)\right)}}{(-1+4 \beta ) \left(2+k^2 \beta \right)},0,\frac{1}{\eta }\right)}\).   \\
Exists for
\begin{enumerate}
\item $\eta \geq 1,  k>2 \sqrt{2},  \beta >\frac{1}{4}$, or 
\item $\eta \geq 1,  k=-2 \sqrt{2},  \beta >\frac{1}{4}$, or
\item $\eta \geq 1,  k=2 \sqrt{2},  \beta >\frac{1}{4}$, or 
\item $\eta \geq 1,  k<-2 \sqrt{2},  \beta >\frac{1}{4}$, or 
\item $\eta \geq 1,  k=0,  \beta <\frac{1}{4}$, or
\item $\eta \geq 1,  k=-2 \sqrt{2},  \beta  <-\frac{1}{4}$, or
\item $\eta \geq 1,  k=2 \sqrt{2},  \beta <-\frac{1}{4}$, or
	\item $\eta \geq 1,  k>2 \sqrt{2}, -\frac{2}{k^2}<\beta <\frac{1}{4}$, or
	\item $\eta \geq 1,  k<-2 \sqrt{2},  -\frac{2}{k^2}<\beta <\frac{1}{4}$, or
	\item $\eta \geq 1,  k>2 \sqrt{2},  -\frac{2}{k^2-8}\leq \beta <-\frac{2}{k^2}$, or
	\item $\eta \geq 1,  k<-2 \sqrt{2}, -\frac{2}{k^2-8}\leq \beta <-\frac{2}{k^2}$, or
	\item $\eta \geq 1,  k=-2 \sqrt{2},  -\frac{1}{4}<\beta <\frac{1}{4}$, or
	\item $\eta \geq 1,  k=2 \sqrt{2},  -\frac{1}{4}<\beta <\frac{1}{4}$, or
	\item $\eta \geq 1,  0<k<2 \sqrt{2},  \beta  <-\frac{2}{k^2}$, or
	\item $\eta \geq 1,  -2 \sqrt{2}<k<0,  \beta <-\frac{2}{k^2}$, or
	\item $\eta \geq 1,  0<k<2 \sqrt{2},  \frac{1}{4}<\beta \leq -\frac{2}{k^2-8}$, or
	\item $\eta \geq 1,  -2 \sqrt{2}<k<0,  \frac{1}{4}<\beta \leq  -\frac{2}{k^2-8}$, or
	\item $\eta \geq 1,  0<k<2 \sqrt{2},  -\frac{2}{k^2}<\beta <\frac{1}{4}$, or 
	\item $\eta \geq 1,  -2 \sqrt{2}<k<0,  -\frac{2}{k^2}<\beta <\frac{1}{4}$.
\end{enumerate}
  The eigenvalues are\\\noindent\( {0,0,\varepsilon\frac{2 \left(2 k \beta -\sqrt{2+\left(-8+k^2\right) \beta }\right) \sqrt{2+\beta  \left(-8+k \left(k+4 k \beta +4 \sqrt{2+\left(-8+k^2\right)
\beta }\right)\right)}}{(-1+4 \beta ) \left(2+k^2 \beta \right) \eta },}\\
 {\varepsilon\frac{\left(k-2 \sqrt{2+\left(-8+k^2\right) \beta }\right) \sqrt{2+\beta  \left(-8+k \left(k+4 k \beta +4 \sqrt{2+\left(-8+k^2\right) \beta
}\right)\right)}}{(-1+4 \beta ) \left(2+k^2 \beta \right)},}\\
 {\varepsilon\frac{\sqrt{2+\beta  \left(-8+k \left(k+4 k \beta +4 \sqrt{2+\left(-8+k^2\right) \beta }\right)\right)} \left(k (-2+8 \beta -\eta )+2 \sqrt{2+\left(-8+k^2\right)
\beta } \eta \right)}{(-1+4 \beta ) \left(2+k^2 \beta \right) \eta }}\). \\
$P_{10}^+$ (resp. $P_{10}^-$) is nonhyperbolic with a 3D unstable (resp. stable) manifold for 
\begin{enumerate}
\item $ -2 \sqrt{2}<k<0 ,  \beta <-\frac{2}{k^2} ,  1<\eta <4 \sqrt{\frac{\beta  k^4-8 \beta  k^2+2 k^2}{\left(k^2-8\right)^2}}-\frac{2 k^2}{k^2-8}$, or 
\item $2 \sqrt{\frac{2}{3}}<k<2 \sqrt{2} ,  \frac{9 k^2-8}{16 k^2}<\beta \leq -\frac{2}{k^2-8} ,  1<\eta <-\frac{2
   k^2}{k^2-8}-4 \sqrt{\frac{\beta  k^4-8 \beta  k^2+2 k^2}{\left(k^2-8\right)^2}}$, or 
\item $k=2 \sqrt{2} ,  \beta >\frac{1}{2} , 
   1<\eta <4 \beta -1$, or
\item $k>2 \sqrt{2} ,  \beta >\frac{9 k^2-8}{16 k^2} ,  1<\eta <4 \sqrt{\frac{\beta  k^4-8 \beta  k^2+2
   k^2}{\left(k^2-8\right)^2}}-\frac{2 k^2}{k^2-8}$. 
\end{enumerate}
$P_{10}^+$ (resp. $P_{10}^-$) nonhyperbolic with a 3D stable (resp. unstable) manifold for 
\begin{enumerate}
\item $ k=-2 \sqrt{\frac{2}{5}} ,  -\frac{5}{4}<\beta <\frac{1}{4} ,  1<\eta <\frac{1}{2} \sqrt{5-16 \beta }+\frac{1}{2} $, or 
\item $k=-2
   \sqrt{\frac{2}{3}} ,  \frac{1}{4}<\beta <\frac{3}{8} ,  1<\eta <\sqrt{3-8 \beta }+1 $, or
\item $ k=-2 \sqrt{\frac{2}{3}} , 
   -\frac{3}{4}<\beta <\frac{1}{4} ,  1<\eta <\sqrt{3-8 \beta }+1 $, or 
\item $ -2 \sqrt{\frac{2}{3}}<k<-2 \sqrt{\frac{2}{5}} , 
   \frac{1}{4}<\beta <\frac{9 k^2-8}{16 k^2} ,  1<\eta <4 \sqrt{\frac{\beta  k^4-8 \beta  k^2+2 k^2}{\left(k^2-8\right)^2}}-\frac{2
   k^2}{k^2-8} $, or 
\item $ -2 \sqrt{2}<k<-2 \sqrt{\frac{2}{3}} ,  \beta =-\frac{2}{k^2-8} ,  1<\eta <8 \beta -2 $, or 
\item $ -2 \sqrt{\frac{2}{5}}<k<0 ,  -\frac{2}{k^2}<\beta <\frac{9 k^2-8}{16 k^2} ,  1<\eta <4 \sqrt{\frac{\beta  k^4-8 \beta  k^2+2
   k^2}{\left(k^2-8\right)^2}}-\frac{2 k^2}{k^2-8} $, or 
\item $-2 \sqrt{2}<k<-2 \sqrt{\frac{2}{3}} ,  \frac{1}{4}<\beta
   <-\frac{2}{k^2-8} ,  1<\eta <4 \sqrt{\frac{\beta  k^4-8 \beta  k^2+2 k^2}{\left(k^2-8\right)^2}}-\frac{2 k^2}{k^2-8} $, or 
\item $ -2
   \sqrt{2}<k<-2 \sqrt{\frac{2}{3}} ,  -\frac{2}{k^2}<\beta <\frac{1}{4} ,  1<\eta <4 \sqrt{\frac{\beta  k^4-8 \beta  k^2+2
   k^2}{\left(k^2-8\right)^2}}-\frac{2 k^2}{k^2-8} $, or 
\item $ -2 \sqrt{\frac{2}{3}}<k<-2 \sqrt{\frac{2}{5}} ,  -\frac{2}{k^2}<\beta
   <\frac{1}{4} ,  1<\eta <4 \sqrt{\frac{\beta  k^4-8 \beta  k^2+2 k^2}{\left(k^2-8\right)^2}}-\frac{2 k^2}{k^2-8} $.
\end{enumerate}

\item $P_{11}^\varepsilon$: $(X_\phi, Y_\phi, S_1, S_2,Y)=$\noindent\( {\left(\varepsilon\frac{\sqrt{4+2 \beta  \left(-8+k \left(k+4 k \beta +4 \sqrt{2+\left(-8+k^2\right) \beta }\right)\right)}}{2+k^2 \beta },\right.}\\
 {\varepsilon\frac{\left(2 k \beta -\sqrt{2+\left(-8+k^2\right) \beta }\right) \sqrt{2+\beta  \left(-8+k \left(k+4 k \beta +4 \sqrt{2+\left(-8+k^2\right)
\beta }\right)\right)}}{(-1+4 \beta ) \left(2+k^2 \beta \right)},}\\
 {\left. \varepsilon\frac{\left(k-2 \sqrt{2+\left(-8+k^2\right) \beta }\right) \sqrt{2+\beta  \left(-8+k \left(k+4 k \beta +4 \sqrt{2+\left(-8+k^2\right)
\beta }\right)\right)}}{(-1+4 \beta ) \left(2+k^2 \beta \right)},0,\frac{1}{\eta }\right)}\). 
Exists for
\begin{enumerate}
\item $\eta =1,  k<-2 \sqrt{\frac{2}{3}},  \beta =\frac{9 k^2-8}{16 k^2}$, or 
\item $\eta =1,  k>2 \sqrt{\frac{2}{3}},  \beta =\frac{9 k^2-8}{16 k^2}$, or 
\item $\eta =1,  0<k\leq 2 \sqrt{\frac{2}{3}},  \beta \leq \frac{9 k^2-8}{16 k^2}$, or
\item $\eta =1,  k<-2 \sqrt{\frac{2}{3}},  \beta \leq \frac{1}{k^2}$, or
\item $\eta =1,  k>2 \sqrt{\frac{2}{3}},  \beta \leq  \frac{1}{k^2}$, or 
\item $\eta =1,  -2 \sqrt{\frac{2}{3}}\leq k<0,  \beta \leq \frac{9 k^2-8}{16 k^2}$, or 
\item $\eta >1,  k>2 \sqrt{2} \sqrt{\frac{\eta }{\eta +2}},  \beta =\frac{-8 \eta ^2+\eta ^2 k^2+4 \eta  k^2+4 k^2}{16 k^2}$, or 
\item $\eta >1,  k<-2 \sqrt{2} \sqrt{\frac{\eta }{\eta +2}},  \beta =\frac{-8 \eta ^2+\eta ^2 k^2+4 \eta  k^2+4 k^2}{16 k^2}$, or 
\item $\eta >1,  0<k<2 \sqrt{2} \sqrt{\frac{\eta }{\eta +2}},  \beta \leq \frac{-8 \eta ^2+\eta ^2 k^2+4 \eta  k^2+4 k^2}{16 k^2}$, or
\item $\eta >1,  -2 \sqrt{2} \sqrt{\frac{\eta }{\eta +2}}<k<0,  \beta \leq \frac{-8 \eta ^2+\eta ^2 k^2+4 \eta  k^2+4 k^2}{16 k^2}$, or 
\item $\eta >1,  k\geq 2 \sqrt{2} \sqrt{\frac{\eta }{\eta +2}},  \beta \leq \frac{\eta }{k^2}$, or 
\item $\eta >1,  k\leq -2 \sqrt{2} \sqrt{\frac{\eta }{\eta +2}},  \beta \leq \frac{\eta }{k^2}$.
\end{enumerate}

 The eigenvalues are\\
\noindent\( {0,0,\frac{2 \left(2 k \beta -\sqrt{2+\left(-8+k^2\right) \beta }\right) \sqrt{2+\beta  \left(-8+k \left(k+4 k \beta +4 \sqrt{2+\left(-8+k^2\right)
\beta }\right)\right)}}{(-1+4 \beta ) \left(2+k^2 \beta \right) \eta },}\\
 {\frac{\left(k-2 \sqrt{2+\left(-8+k^2\right) \beta }\right) \sqrt{2+\beta  \left(-8+k \left(k+4 k \beta +4 \sqrt{2+\left(-8+k^2\right) \beta
}\right)\right)}}{(-1+4 \beta ) \left(2+k^2 \beta \right)},}\\
 {\frac{\sqrt{2+\beta  \left(-8+k \left(k+4 k \beta +4 \sqrt{2+\left(-8+k^2\right) \beta }\right)\right)} \left(k (-2+8 \beta -\eta )+2 \sqrt{2+\left(-8+k^2\right)
\beta } \eta \right)}{(-1+4 \beta ) \left(2+k^2 \beta \right) \eta }}\).
\\ $P_{11}^+$ (resp. $P_{11}^-$) is nonhyperbolic with a 3D stable (resp. unstable) manifold for 
\begin{enumerate}
\item $ \beta >\frac{1}{2} ,  k=-2 \sqrt{2} ,  1<\eta <4 \beta -1 $, or 
\item $ \beta \geq \frac{9}{16} ,  k<-2 \sqrt{2} ,  1<\eta <4
   \sqrt{\frac{\beta  k^4-8 \beta  k^2+2 k^2}{\left(k^2-8\right)^2}}-\frac{2 k^2}{k^2-8} $, or 
\item $ \beta <-\frac{1}{4} ,  \sqrt{2}
   \sqrt{-\frac{1}{\beta }}<k<2 \sqrt{2} ,  1<\eta <4 \sqrt{\frac{\beta  k^4-8 \beta  k^2+2 k^2}{\left(k^2-8\right)^2}}-\frac{2
   k^2}{k^2-8} $, or 
\item $ \frac{1}{2}<\beta <\frac{9}{16} ,  -2 \sqrt{2} \sqrt{-\frac{1}{16 \beta -9}}<k<-2 \sqrt{2} ,  1<\eta <4
   \sqrt{\frac{\beta  k^4-8 \beta  k^2+2 k^2}{\left(k^2-8\right)^2}}-\frac{2 k^2}{k^2-8} $, or 
\item $ \beta >\frac{1}{2} ,  -2
   \sqrt{2}<k\leq -\sqrt{2} \sqrt{\frac{4 \beta -1}{\beta }} ,  1<\eta <-\frac{2 k^2}{k^2-8}-4 \sqrt{\frac{\beta  k^4-8 \beta  k^2+2
   k^2}{\left(k^2-8\right)^2}} $, or 
\item $ \frac{3}{8}<\beta \leq  \frac{1}{2} ,  -2 \sqrt{2} \sqrt{-\frac{1}{16 \beta -9}}<k\leq
   -\sqrt{2} \sqrt{\frac{4 \beta -1}{\beta }} ,  1<\eta <-\frac{2 k^2}{k^2-8}-4 \sqrt{\frac{\beta  k^4-8 \beta  k^2+2
   k^2}{\left(k^2-8\right)^2}} $.
\end{enumerate}
 $P_{11}^+$ (resp. $P_{11}^-$)  is nonhyperbolic with a 3D unstable (resp. stable) manifold for 
	\begin{enumerate}
	\item $\beta >\frac{3}{8} ,  k=\sqrt{2} \sqrt{\frac{4 \beta -1}{\beta }} ,  1<\eta <8 \beta -2$, or 
	\item $\beta >\frac{3}{8} , 
   \sqrt{2} \sqrt{\frac{4 \beta -1}{\beta }}<k<2 \sqrt{2} ,  1<\eta <4 \sqrt{\frac{\beta  k^4-8 \beta  k^2+2
   k^2}{\left(k^2-8\right)^2}}-\frac{2 k^2}{k^2-8}$, or 
	\item $-\frac{1}{4}<\beta <\frac{1}{4} ,  2 \sqrt{2} \sqrt{-\frac{1}{16 \beta
   -9}}<k<2 \sqrt{2} ,  1<\eta <4 \sqrt{\frac{\beta  k^4-8 \beta  k^2+2 k^2}{\left(k^2-8\right)^2}}-\frac{2 k^2}{k^2-8}$, or
	\item $\frac{1}{4}<\beta \leq \frac{3}{8} ,  2 \sqrt{2} \sqrt{-\frac{1}{16 \beta -9}}<k<2 \sqrt{2} ,  1<\eta <4 \sqrt{\frac{\beta  k^4-8
   \beta  k^2+2 k^2}{\left(k^2-8\right)^2}}-\frac{2 k^2}{k^2-8}$, or 
	\item $\beta \leq -\frac{1}{4} ,  2 \sqrt{2} \sqrt{-\frac{1}{16
   \beta -9}}<k<\sqrt{2} \sqrt{-\frac{1}{\beta }} ,  1<\eta <4 \sqrt{\frac{\beta  k^4-8 \beta  k^2+2 k^2}{\left(k^2-8\right)^2}}-\frac{2
   k^2}{k^2-8}$.
	\end{enumerate}

\end{enumerate}

\section{Equilibrium points for Model 4}
\label{App_A2}

The details
of the stability of the relevant equilibrium point (curves/surfaces of equilibrium point) of the dynamical system \eqref{EQS-EQS} are as follows. 

\begin{enumerate}

 \item $L_1^\varepsilon$: $(X_\phi, Y_\phi, S_1, S_2,Y)=\Big(X_{\phi}^*,Y_{\phi}^*,S_1^*,1-\beta  {S_1^*}^2-{X_{\phi}^*}^2,0\Big)$.  Always exists. The eigenvalues are $0,0,0,-\frac{{S_1^*}}{n+1},-{S_1^*}$. Nonhyperbolic with a 2D unstable manifold (resp. stable) for $S_1^*<0$ (resp. $S_1^*>0$).

 \item $L_2^\varepsilon$: $(X_\phi, Y_\phi, S_1, S_2,Y)=(\varepsilon, \varepsilon, 0, 0, Y_0)$.  Exists for $1\leq Y_0\leq 1$. The eigenvalues are\\$0,0,2 Y_0\varepsilon,4 Y_0\varepsilon,\left(4-\sqrt{2} k\right) Y_0\varepsilon$. 
 $L_2^+$ (resp. $L_2^-$) is nonhyperbolic with a 3D unstable (resp. stable) manifold for $Y_0>0, k<2\sqrt{2}$. Otherwise the unstable (resp. stable) manifold is 2D.

 \item $L_3^\varepsilon$: $(X_\phi, Y_\phi, S_1, S_2,Y)=(-\varepsilon,\varepsilon,0,0,Y_0)$.  Exists for $-1\leq Y_0 \leq 1$. The eigenvalues are \\$0,0,2 Y_0\varepsilon,4 Y_0\varepsilon,\left(\sqrt{2} k+4\right) Y_0\varepsilon$.
 $L_3^+$ (resp. $L_3^-$) is nonhyperbolic with a 3D unstable (resp. stable) manifold for $Y_0>0, k>-2\sqrt{2}$. Otherwise the unstable (resp. stable) manifold is 2D.

 \item $L_4^\varepsilon$: $(X_\phi, Y_\phi, S_1, S_2,Y)=(0,\varepsilon,0,1,Y_0)$. Exists for $-1\leq Y_0\leq 1$. The eigenvalues are \\$0,-Y_0\varepsilon,-Y_0\varepsilon, 2 Y_0\varepsilon, 2 Y_0\varepsilon$. They are nonhyperbolic and behaves as saddle.

 \item $L_5^\varepsilon$: $(X_\phi, Y_\phi, S_1, S_2,Y)=\Big(\varepsilon{X_{\phi}^*}, \varepsilon, \varepsilon\frac{\sqrt{1-{X_{\phi}^*}^2}}{\sqrt{\beta }},0,1\Big)$. Exists for $\beta<0, {X_{\phi}^*}^2\geq 1$ or $\beta>0, {X_{\phi}^*}^2\leq 1$. The eigenvalues are $0,\frac{\sqrt{1-{X_{\phi}^*}^2}}{\sqrt{\beta } (n+1)}\varepsilon, 2\varepsilon,\left(-\sqrt{2} k {X_{\phi}^*}-\frac{2 \sqrt{1-{X_{\phi}^*}^2}}{\sqrt{\beta }}+4\right)\varepsilon,\left(4-\frac{3 \sqrt{1-{X_{\phi}^*}^2}}{\sqrt{\beta }}\right) \varepsilon$. 
  $L_5^+$ (resp. $L_5^-$) is nonhyperbolic, with a 4D unstable (resp. stable) manifold for
\begin{enumerate}
\item $ k\in \mathbb{R},  {X_\phi^*}=0,  \beta >\frac{9}{16},  n> 0 $, or 
\item $ -1<{X_\phi^*}<0,  k>\frac{2
   \sqrt{2}}{3 {X_\phi^*}},  \beta >\frac{9}{16} \left(1- {X_\phi^*}^2\right),  n> 0$, or 
\item $0<{X_\phi^*}<1,  k\leq \frac{2 \sqrt{2}}{3 {X_\phi^*}},  \beta >\frac{9}{16} \left(1- {X_\phi^*}^2\right),  n> 0$, or
\item $-1<{X_\phi^*}<0,  \frac{2 \sqrt{2}}{{X_\phi^*}}<k\leq \frac{2 \sqrt{2}}{3 {X_\phi^*}},  \beta >\frac{2-2
   {X_\phi^*}^2}{k^2 {X_\phi^*}^2-4 \sqrt{2} k {X_\phi^*}+8},  n> 0$, or
\item	$0<{X_\phi^*}<1,  \frac{2
   \sqrt{2}}{3 {X_\phi^*}}<k<\frac{2 \sqrt{2}}{{X_\phi^*}},  \beta >\frac{2-2 {X_\phi^*}^2}{k^2 {X_\phi^*}^2-4 \sqrt{2}
   k {X_\phi^*}+8},  n> 0$. 
\end{enumerate}

 \item $L_6^\varepsilon$: $(X_\phi, Y_\phi, S_1, S_2,Y)=\Big(\varepsilon{X_{\phi}^*},\varepsilon,-\varepsilon\frac{\sqrt{1-{X_{\phi}^*}^2}}{\sqrt{\beta }},0,1\Big)$.  Exists for $\beta<0, {X_{\phi}^*}^2\geq 1$ or $\beta>0, {X_{\phi}^*}^2\leq 1$. The eigenvalues are\\$0,-\frac{\sqrt{1-{X_{\phi}^*}^2}}{\sqrt{\beta } (n+1)}\varepsilon, 2\varepsilon, \left(-\sqrt{2} k {X_{\phi}^*}+\frac{2 \sqrt{1-{X_{\phi}^*}^2}}{\sqrt{\beta }}+4\right)\varepsilon, \left(\frac{3 \sqrt{1-{X_{\phi}^*}^2}}{\sqrt{\beta }}+4\right)\varepsilon$. They are nonhyperbolic and behaves as saddle.

\item $P_1^\varepsilon$: $(X_\phi, Y_\phi, S_1, S_2,Y)=\Big(\varepsilon,\frac{2 \sqrt{2}}{k}\varepsilon,0,0,0\Big)$.  Exists for $k\leq -2 \sqrt{2}$, or $k\geq 2 \sqrt{2}$. The eigenvalues are $0,0,0,0,0$.  They are nonhyperbolic.

 \item $P_2^\varepsilon$: $(X_\phi, Y_\phi, S_1, S_2,Y)=\Big(0,\varepsilon,\frac{2}{3}\varepsilon,1-\frac{8 \beta }{9},1\Big)$. Exists for $0\leq \beta \leq \frac{9}{8}$. The eigenvalues are\\$\frac{2}{3 (n+1)}\varepsilon, -\varepsilon,\frac{2}{3}\varepsilon, \frac{1}{6} \left(-\sqrt{64 \beta -63}-3\right)\varepsilon, \frac{1}{6} \left(\sqrt{64 \beta -63}-3\right)\varepsilon$. They are saddle.

 \item $P_3^\varepsilon$: $(X_\phi, Y_\phi, S_1, S_2,Y)=(0,\varepsilon,0,1,1)$. The eigenvalues are $0, -\varepsilon, -\varepsilon, 2\varepsilon, 2\varepsilon$.     They are nonhyperbolic and behaves as saddle.
 
 \item $P_4^\varepsilon$: $(X_\phi, Y_\phi, S_1, S_2,Y)=\Big(0,\varepsilon,\frac{\varepsilon}{\sqrt{\beta }},0,1\Big)$. Exists for $\beta>0$. The eigenvalues are\\$0,\frac{1}{\sqrt{\beta } (n+1)}\varepsilon, \frac{2 \left(2 \sqrt{\beta }-1\right)}{\sqrt{\beta }}\varepsilon, \frac{4 \sqrt{\beta }-3}{\sqrt{\beta }}\varepsilon, 2 \varepsilon$.  $P_4^+$ (resp. $P_4^-$) is nonhyperbolic with a 4D unstable (resp. stable) manifold for $n> 0, \beta >\frac{9}{16}$. 

\item $P_5^\varepsilon$: $(X_\phi, Y_\phi, S_1, S_2,Y)=\Big(0,\varepsilon, \frac{3\varepsilon}{4 \beta },0,1\Big)$.  Exists for  $\beta <0$, or $\beta \geq \frac{9}{16}$. The eigenvalues are\\$\frac{3}{4 \beta  (n+1)}\varepsilon, \frac{3}{4 \beta }\varepsilon,-\frac{8 \beta -9}{4 \beta }\varepsilon, -\frac{16 \beta -9}{8 \beta }\varepsilon, -\frac{16 \beta -9}{8 \beta }\varepsilon$. 
   $P_5^+$ (resp. $P_5^-$) is a sink (resp. a source) for $\beta <0, n> 0$ or a saddle otherwise. 

 \item $P_6^\varepsilon$: $(X_\phi, Y_\phi, S_1, S_2,Y)=\Big(0,\varepsilon,-\frac{\varepsilon}{\sqrt{\beta }},0,1\Big)$. Exists for $\beta>0$. The eigenvalues are\\$0,-\frac{1}{\sqrt{\beta } (n+1)}\varepsilon, \frac{2 \left(2 \sqrt{\beta }+1\right)}{\sqrt{\beta }}\varepsilon, \frac{4 \sqrt{\beta }+3}{\sqrt{\beta }}\varepsilon, 2\varepsilon$. They are nonhyperbolic and behaves as saddle.

 \item $P_7^\varepsilon$: $(X_\phi, Y_\phi, S_1, S_2,Y)=\Big(\frac{k \sqrt{\beta }}{\sqrt{\beta  k^2+2}}\varepsilon, \frac{2 \sqrt{2} \sqrt{\beta }}{\sqrt{\beta  k^2+2}}\varepsilon, \frac{\sqrt{2}}{\sqrt{\beta } \sqrt{\beta  k^2+2}}\varepsilon,0,1\Big)$. Exists for
\begin{enumerate}
\item $k\leq -2 \sqrt{2},  \beta \geq 0$, or 
\item $-2 \sqrt{2}<k<2 \sqrt{2},  0\leq \beta \leq -\frac{4}{2 k^2-16}$, or
\item $k\geq 2 \sqrt{2},  \beta \geq 0$. 
\end{enumerate}
The eigenvalues are $\frac{\sqrt{2} \left(\beta  \left(k^2-4\right)+2\right)}{\sqrt{\beta } \sqrt{\beta  k^2+2}}\varepsilon, \frac{\sqrt{2}}{\sqrt{\beta } (n+1) \sqrt{\beta  k^2+2}}\varepsilon, \frac{\sqrt{2} \left(\beta  k^2-1\right)}{\sqrt{\beta } \sqrt{\beta 
   k^2+2}}\varepsilon, \frac{\beta  \left(k^2-8\right)+2}{\sqrt{2} \sqrt{\beta } \sqrt{\beta  k^2+2}}\varepsilon, \frac{\beta  \left(k^2-8\right)+2}{\sqrt{2} \sqrt{\beta } \sqrt{\beta  k^2+2}}\varepsilon$. 
	 \\
	$P_7^+$ (resp. $P_7^-$) is a source (resp. a sink) for 
	\begin{enumerate}
	\item $k\leq -2 \sqrt{2},  \beta >\frac{1}{k^2},  n>0$, or 
	\item $-2 \sqrt{2}<k<-2 \sqrt{\frac{2}{3}},  \frac{1}{k^2}<\beta
   <-\frac{2}{k^2-8},  n>0$, or 
	\item $2 \sqrt{\frac{2}{3}}<k<2 \sqrt{2},  \frac{1}{k^2}<\beta <-\frac{2}{k^2-8}, 
   n>0$, or 
	\item $k\geq 2 \sqrt{2},  \beta >\frac{1}{k^2},  n>0$.
	\end{enumerate}
It is a saddle otherwise.

\item $P_8^\varepsilon$: $(X_\phi, Y_\phi, S_1, S_2,Y)=\left(\frac{k \beta \varepsilon}{\sqrt{\left(\frac{\beta  k^2}{2}+1\right)\left(\left(k^2-2\right) \beta +2\right)}}, \varepsilon\sqrt{\frac{{\beta  k^2+2}}{{\left(k^2-2\right) \beta +2}}}, \frac{2\varepsilon}{\sqrt{\left(\beta  k^2+2\right) \left(\left(k^2-2\right) \beta
   +2\right)}}, \frac{\left(k^2-4\right) \beta +2}{\left(k^2-2\right) \beta +2},1\right)$.  Exists for
	\begin{enumerate}
	\item $k<0,  -\frac{2}{k^2}\leq \beta \leq 0$, or 
	\item $k=0,  \beta \leq 0$, or 
	\item $k>0,  -\frac{2}{k^2}\leq \beta \leq 0$.
	\end{enumerate}
The eigenvalues are\\$\frac{2 \beta  k^2-2}{\sqrt{\beta  k^2+2} \sqrt{\beta  \left(k^2-2\right)+2}}\varepsilon, -\frac{\sqrt{\beta  k^2+2}}{\sqrt{\beta 
   \left(k^2-2\right)+2}}\varepsilon, \frac{2}{(n+1) \sqrt{\beta  k^2+2} \sqrt{\beta  \left(k^2-2\right)+2}}\varepsilon$,\\
	$\frac{-\sqrt{\beta  k^2+2}-\sqrt{\beta 
   \left(-\left(7 k^2-32\right)\right)-14}}{2 \sqrt{\beta  \left(k^2-2\right)+2}}\varepsilon, \frac{\sqrt{\beta  \left(-\left(7
   k^2-32\right)\right)-14}-\sqrt{\beta  k^2+2}}{2 \sqrt{\beta  \left(k^2-2\right)+2}}\varepsilon$.  They are saddle.

\item $P_9^\varepsilon$: $(X_\phi, Y_\phi, S_1, S_2,Y)=\Big(\frac{3 k \varepsilon}{\sqrt{k^2 \left(16 \beta  \left(2 k^2 \beta -1\right)+27\right)-16}},\frac{2 \left(2 \beta  k^2+1\right)\varepsilon}{\sqrt{16 \beta ^2 k^4+\left(\frac{27}{2}-8 \beta \right) k^2-8}},\frac{3 k^2 \varepsilon}{\sqrt{16 \beta ^2
   k^4+\left(\frac{27}{2}-8 \beta \right) k^2-8}},0,1\Big)$. Exists for 
	\begin{enumerate}
	\item $k<-2 \sqrt{\frac{2}{3}}, \beta =\frac{9 k^2-8}{16 k^2}$, or 
	\item $k>2 \sqrt{\frac{2}{3}}, \beta =\frac{9 k^2-8}{16 k^2}$, or 
	\item $k<-2 \sqrt{\frac{2}{3}}, \beta \leq \frac{1}{k^2}$, or 
	\item $k>2 \sqrt{\frac{2}{3}}, \beta \leq \frac{1}{k^2}$, or 
	\item $0<k\leq 2 \sqrt{\frac{2}{3}}, \beta \leq \frac{9 k^2-8}{16 k^2}$, or 
	\item $-2\sqrt{\frac{2}{3}}\leq k<0, \beta \leq \frac{9 k^2-8}{16 k^2}$.
	\end{enumerate}
		For $\beta=1$ the eigenvalues are\\$\frac{3 k^2 \varepsilon}{ (n+1)\sqrt{16 k^4+\frac{11 k^2}{2}-8}}, \frac{-7 k^2-8}{\sqrt{64 k^4+22 k^2-32}}\varepsilon, \frac{-7 k^2-i \sqrt{7 k^2+8} \sqrt{17 k^2-32}-8}{2 \sqrt{64 k^4+22 k^2-32}}\varepsilon, \frac{-7 k^2+i \sqrt{7 k^2+8} \sqrt{17
   k^2-32}-8}{2 \sqrt{64 k^4+22 k^2-32}}\varepsilon, \frac{k^2-4}{\sqrt{16 k^4+\frac{11 k^2}{2}-8}}\varepsilon$. 	For $\beta=1$ they are saddle.

\item $P_{10}^\varepsilon$: $(X_\phi, Y_\phi, S_1, S_2,Y)=\Big(\varepsilon\frac{\sqrt{2 \beta  \left(k \left(4 \beta  k+k-4 \sqrt{\left(k^2-8\right) \beta +2}\right)-8\right)+4}}{\beta  k^2+2}$,\\$\varepsilon\frac{\left(2 k \beta +\sqrt{\left(k^2-8\right) \beta +2}\right) \sqrt{\beta  \left(k \left(4
   \beta  k+k-4 \sqrt{\left(k^2-8\right) \beta +2}\right)-8\right)+2}}{(4 \beta -1) \left(\beta  k^2+2\right)}$,\\$\varepsilon\frac{\left(k+2 \sqrt{\left(k^2-8\right) \beta +2}\right) \sqrt{\beta  \left(k \left(4 \beta  k+k-4
   \sqrt{\left(k^2-8\right) \beta +2}\right)-8\right)+2}}{(4 \beta -1) \left(\beta  k^2+2\right)},0,1\Big)$. 
	Exists for 
	\begin{enumerate}
	\item $k=0,  \beta <\frac{1}{4}$, or \item $k=2 \sqrt{2},  -\frac{1}{4}<\beta <\frac{1}{4}$, or 
	\item $k=2 \sqrt{2},  \beta <-\frac{1}{4}$, or 
	\item $k=2 \sqrt{2},  \beta >\frac{1}{4}$, or 
	\item $k=-2 \sqrt{2},  -\frac{1}{4}<\beta <\frac{1}{4}$, or 
	\item $k=-2 \sqrt{2},  \beta <-\frac{1}{4}$, or 
	\item $k=-2 \sqrt{2},  \beta >\frac{1}{4}$, or
  \item $0<k<2 \sqrt{2},  -\frac{2}{k^2}<\beta <\frac{1}{4}$, or 
	\item $0<k<2 \sqrt{2},  \beta <-\frac{2}{k^2}$, or
  \item $0<k<2 \sqrt{2},  \frac{1}{4}<\beta \leq -\frac{2}{k^2-8}$, or 
	\item $-2 \sqrt{2}<k<0,  -\frac{2}{k^2}<\beta <\frac{1}{4}$, or 
	\item $-2 \sqrt{2}<k<0,  \beta <-\frac{2}{k^2}$, or 
	\item $-2 \sqrt{2}<k<0,  \frac{1}{4}<\beta \leq -\frac{2}{k^2-8}$, or 
	\item $k<-2 \sqrt{2},  -\frac{2}{k^2}<\beta <\frac{1}{4}$, or 
	\item $k<-2 \sqrt{2},  -\frac{2}{k^2-8}\leq \beta <-\frac{2}{k^2}$, or 
	\item $k<-2 \sqrt{2},  \beta >\frac{1}{4}$, or 
	\item $k>2 \sqrt{2},  -\frac{2}{k^2}<\beta <\frac{1}{4}$, or 
	\item $k>2 \sqrt{2},  -\frac{2}{k^2-8}\leq \beta <-\frac{2}{k^2}$, or 
	\item $k>2 \sqrt{2},  \beta >\frac{1}{4}$.
	\end{enumerate}
 The eigenvalues are \\
	$0,0,\frac{2 \left(\sqrt{\beta  \left(k^2-8\right)+2}+2 \beta  k\right) \sqrt{\beta  \left(k \left(-4 \sqrt{\beta  \left(k^2-8\right)+2}+4 \beta  k+k\right)-8\right)+2}}{(4 \beta -1) \left(\beta 
   k^2+2\right)}\varepsilon$, \\
	$\frac{\left(2 \sqrt{\beta  \left(k^2-8\right)+2}+k\right) \sqrt{\beta  \left(k \left(-4 \sqrt{\beta  \left(k^2-8\right)+2}+4 \beta  k+k\right)-8\right)+2}}{(4 \beta -1) (n+1) \left(\beta 
   k^2+2\right)}\varepsilon$,\\ $\frac{\left((8 \beta -3) k-2 \sqrt{\beta  \left(k^2-8\right)+2}\right) \sqrt{\beta  \left(k \left(-4 \sqrt{\beta  \left(k^2-8\right)+2}+4 \beta  k+k\right)-8\right)+2}}{(4 \beta -1) \left(\beta 
   k^2+2\right)}\varepsilon$.  \\
	$P_{10}^+$ (resp. $P_{10}^-$) has a 3D stable (resp. unstable) manifold for
	\begin{enumerate}
	\item $-2 \sqrt{2}<k<-2 \sqrt{\frac{2}{3}},  n\geq 0,  \frac{1}{4}<\beta \leq -\frac{2}{k^2-8}$, or 
	\item $-2 \sqrt{2}<k<-2 \sqrt{\frac{2}{3}},  n\geq 0,  -\frac{2}{k^2}<\beta <\frac{1}{4}$, or
  \item $k=-2 \sqrt{\frac{2}{3}},  n\geq 0,  -\frac{3}{4}<\beta <\frac{1}{4}$, or 
	\item $k=-2 \sqrt{\frac{2}{3}},  n\geq 0,  \frac{1}{4}<\beta <\frac{3}{8}$, or \item $-2 \sqrt{\frac{2}{3}}<k<-2
   \sqrt{\frac{2}{5}},  n\geq 0,  -\frac{2}{k^2}<\beta <\frac{1}{4}$, or 
	\item $-2 \sqrt{\frac{2}{3}}<k<-2 \sqrt{\frac{2}{5}},  n\geq 0,  \frac{1}{4}<\beta <\frac{9 k^2-8}{16 k^2}$, or \item $k=-2
   \sqrt{\frac{2}{5}},  n\geq 0,  -\frac{5}{4}<\beta <\frac{1}{4}$, or 
	\item $-2 \sqrt{\frac{2}{5}}<k<-\frac{2 \sqrt{2}}{3},  n\geq 0,  -\frac{2}{k^2}<\beta <\frac{9 k^2-8}{16 k^2}$, or
  \item $k=-\frac{2 \sqrt{2}}{3},  n\geq 0,  -\frac{9}{4}<\beta <0$, or 
	\item $-\frac{2 \sqrt{2}}{3}<k<0,  n\geq 0,  -\frac{2}{k^2}<\beta <\frac{9 k^2-8}{16 k^2}$.
	\end{enumerate}
$P_{10}^+$ (resp. $P_{10}^-$) has a 3D unstable (resp. stable) manifold for
  \begin{enumerate}
	\item $-2 \sqrt{2}<k<0,  n\geq 0,  \beta <-\frac{2}{k^2}$, or 
	\item $2 \sqrt{\frac{2}{3}}<k<2 \sqrt{2},  n\geq 0,  \frac{9 k^2-8}{16 k^2}<\beta \leq -\frac{2}{k^2-8}$, or 
	\item $k\geq 2 \sqrt{2}, n\geq 0,  \beta >\frac{9 k^2-8}{16 k^2}$.
	\end{enumerate}
	
\item $P_{11}^\varepsilon$: $(X_\phi, Y_\phi, S_1, S_2,Y)=\Big(\varepsilon\frac{\sqrt{2 \beta  \left(k \left(4 \beta  k+k+4 \sqrt{\left(k^2-8\right) \beta +2}\right)-8\right)+4}}{\beta  k^2+2}$,\\
$\varepsilon\frac{\left(2 k \beta -\sqrt{\left(k^2-8\right) \beta +2}\right) \sqrt{\beta  \left(k \left(4
   \beta  k+k+4 \sqrt{\left(k^2-8\right) \beta +2}\right)-8\right)+2}}{(4 \beta -1) \left(\beta  k^2+2\right)}$,\\$\varepsilon\frac{\left(k-2 \sqrt{\left(k^2-8\right) \beta +2}\right) \sqrt{\beta  \left(k \left(4 \beta  k+k+4
   \sqrt{\left(k^2-8\right) \beta +2}\right)-8\right)+2}}{(4 \beta -1) \left(\beta  k^2+2\right)},0,1\Big)$. 
	Exists for
	\begin{enumerate}
	\item $k=0,  \beta <\frac{1}{4}$, or 
	\item $k=2 \sqrt{2},  -\frac{1}{4}<\beta <\frac{1}{4}$, or 
	\item $k=2 \sqrt{2},  \beta  <-\frac{1}{4}$, or 
	\item $k=2 \sqrt{2},  \beta >\frac{1}{4}$, or 
	\item $k=-2 \sqrt{2},  -\frac{1}{4}<\beta <\frac{1}{4}$, or 
	\item $k=-2 \sqrt{2},  \beta <-\frac{1}{4}$, or 
	\item $k=-2 \sqrt{2},  \beta >\frac{1}{4}$, or
  \item $0<k<2 \sqrt{2},  -\frac{2}{k^2}<\beta <\frac{1}{4}$, or 
	\item $0<k<2 \sqrt{2},  \beta <-\frac{2}{k^2}$, or
  \item $0<k<2 \sqrt{2},  \frac{1}{4}<\beta \leq -\frac{2}{k^2-8}$, or 
	\item $-2 \sqrt{2}<k<0,  -\frac{2}{k^2}<\beta <\frac{1}{4}$, or 
	\item $-2 \sqrt{2}<k<0,  \beta <-\frac{2}{k^2}$, or 
	\item $-2 \sqrt{2}<k<0,  \frac{1}{4}<\beta \leq -\frac{2}{k^2-8}$, or 
	\item $k<-2 \sqrt{2},  -\frac{2}{k^2}<\beta <\frac{1}{4}$, or 
	\item $k<-2 \sqrt{2},  -\frac{2}{k^2-8}\leq \beta <-\frac{2}{k^2}$, or 
	\item $k<-2 \sqrt{2},  \beta >\frac{1}{4}$, or 
	\item $k>2 \sqrt{2},  -\frac{2}{k^2}<\beta <\frac{1}{4}$, or 
	\item $k>2 \sqrt{2},  -\frac{2}{k^2-8}\leq \beta <-\frac{2}{k^2}$, or 
	\item $k>2 \sqrt{2},  \beta >\frac{1}{4}$.
	\end{enumerate}
	The eigenvalues are\\
	$0,0,\frac{\left(k-2 \sqrt{\beta  \left(k^2-8\right)+2}\right) \sqrt{\beta  \left(k \left(4 \sqrt{\beta  \left(k^2-8\right)+2}+4 \beta  k+k\right)-8\right)+2}}{(4 \beta -1) (n+1) \left(\beta  k^2+2\right)}\varepsilon$,\\ $\frac{2
   \left(2 \beta  k-\sqrt{\beta  \left(k^2-8\right)+2}\right) \sqrt{\beta  \left(k \left(4 \sqrt{\beta  \left(k^2-8\right)+2}+4 \beta  k+k\right)-8\right)+2}}{(4 \beta -1) \left(\beta  k^2+2\right)}\varepsilon$,\\ $\frac{\left(2 \sqrt{\beta
    \left(k^2-8\right)+2}+(8 \beta -3) k\right) \sqrt{\beta  \left(k \left(4 \sqrt{\beta  \left(k^2-8\right)+2}+4 \beta  k+k\right)-8\right)+2}}{(4 \beta -1) \left(\beta  k^2+2\right)}\varepsilon$.\\  
		$P_{11}^+$ (resp. $P_{11}^-$) has a 3D stable (resp. unstable) manifold for 
		\begin{enumerate}
		\item $k\leq -2 \sqrt{2},  n\geq 0,  \beta >\frac{9 k^2-8}{16 k^2}$, or 
		\item $-2 \sqrt{2}<k<-2 \sqrt{\frac{2}{3}},  n\geq 0,  \frac{9 k^2-8}{16 k^2}<\beta \leq -\frac{2}{k^2-8}$, or 
		\item $0<k<2\sqrt{2},  n\geq 0,  \beta <-\frac{2}{k^2}$. 
		\end{enumerate}
		$P_{11}^+$ (resp. $P_{11}^-$) has a 3D unstable (resp. stable) manifold for 
		\begin{enumerate}
		\item $k=\frac{2 \sqrt{2}}{3},  n\geq 0,  -\frac{9}{4}<\beta <0$, or 
		\item $2 \sqrt{\frac{2}{5}}<k<2 \sqrt{2},  n\geq 0,  -\frac{2}{k^2}<\beta <\frac{1}{4}$, or 
		\item $0<k<\frac{2 \sqrt{2}}{3},  n\geq   0,  -\frac{2}{k^2}<\beta <\frac{9 k^2-8}{16 k^2}$, or 
	  \item $\frac{2 \sqrt{2}}{3}<k\leq 2 \sqrt{\frac{2}{5}},  n\geq 0,  -\frac{2}{k^2}<\beta <\frac{9 k^2-8}{16 k^2}$, or 
	  \item $2  \sqrt{\frac{2}{5}}<k\leq 2 \sqrt{\frac{2}{3}},  n\geq 0,  \frac{1}{4}<\beta <\frac{9 k^2-8}{16 k^2}$, or 
	  \item $2 \sqrt{\frac{2}{3}}<k<2 \sqrt{2},  n\geq 0,  \frac{1}{4}<\beta \leq -\frac{2}{k^2-8}$.
	  \end{enumerate}
\end{enumerate}

\end{document}